\documentclass{emulateapj}
\usepackage{aas_macros}
\usepackage{apjfonts}
\usepackage{graphicx}
\usepackage{amsmath}
\usepackage{color}

\newcommand{\pdel}[2]%
{\frac{\partial{#2}}{\partial{#1}}}

\newcommand{\pddel}[2]%
{\frac{\partial^2{#2}}{\partial{#1}^2}}

\begin{document}

\title{Three-dimensional Hydrodynamic Core-Collapse Supernova Simulations 
for an $11.2 M_{\odot}$ Star with Spectral Neutrino Transport}

\author{Tomoya Takiwaki\altaffilmark{1}, Kei Kotake\altaffilmark{1,2}, 
and Yudai Suwa\altaffilmark{3}}
\affil{\altaffilmark{1}Center for Computational Astrophysics, National
Astronomical Observatory of Japan, 2-21-1, Osawa, Mitaka, Tokyo,
181-8588, Japan}
\affil{\altaffilmark{2}Division of Theoretical Astronomy, National Astronomical Observatory of Japan, 2-21-1, Osawa, Mitaka, Tokyo, 181-8588, Japan}
\affil{$^3$Yukawa Institute for Theoretical Physics, Kyoto
  University, Oiwake-cho, Kitashirakawa, Sakyo-ku, Kyoto, 606-8502,
  Japan}
\begin{abstract}
We present numerical results on three-dimensional 
(3D) hydrodynamic core-collapse simulations of 
an $11.2 M_{\odot}$ star. By comparing one-(1D) and 
two-dimensional(2D) results with those of 3D, we study how the increasing
 spacial multi-dimensionality affects the postbounce supernova dynamics. 
 The calculations were performed with an energy-dependent treatment of 
the neutrino transport that is solved by
 the isotropic diffusion source approximation scheme.
In agreement with previous study, our 1D model does not produce explosions for 
 the 11.2 $M_{\odot}$ star, while the neutrino-driven 
revival of the stalled bounce shock is obtained both in the 2D and 3D models.
 The standing accretion-shock instability (SASI) is observed in 
 the 3D models, in which the dominant mode of the SASI is bipolar ($\ell=2$) with 
  its saturation amplitudes being slightly smaller than 2D. 
By performing a tracer-particle analysis, we show that 
the maximum residency time of material 
in the gain region becomes longer in 3D due to non-axisymmetric flow motions
 than in 2D,
 which is one of advantageous aspects of 3D models to obtain neutrino-driven explosions.
 Our results show that convective matter motions below the gain radius become 
much more violent in 3D than in 2D, making the neutrino luminosity larger
 for 3D. Nevertheless the emitted neutrino 
 energies are made smaller due to the enhanced cooling. Our results indicate 
whether these advantages for driving 3D explosions  
could or could not overwhelm
 the disadvantages is sensitive to the employed numerical resolutions.
 An encouraging finding is that the shock expansion tends to become 
 more energetic for models with finer resolutions.
To draw a robust conclusion, 3D simulations 
 with much more higher numerical resolutions and also with more advanced treatment of 
 neutrino transport as well as of gravity are needed, which could be 
 hopefully practicable by utilizing forthcoming Petaflops-class supercomputers.  
 \end{abstract}


\maketitle 
\section{Introduction}
 Core-collapse supernovae have long drawn the attention of astrophysicists 
because they have 
many aspects playing important roles
in astrophysics. They are the mother of neutron stars and black
holes; they play an important role for acceleration of cosmic rays; they influence
galactic dynamics triggering further star formation; they are
 gigantic emitters of neutrinos and gravitational waves.
 They are also a major site for nucleosynthesis, so, naturally, any attempt 
to address human origins may need to begin
with an understanding of core-collapse supernovae. 

Ever since the first numerical simulation \citep{colgate}, 
the neutrino-heating mechanism,
 in which a stalled bounce shock is revived by neutrino energy deposition to trigger
  explosions \citep{wils85,bethe85,bethe},
 has been the working hypothesis of supernova theorists for these $\sim$ 45 years.
 However, one important lesson we have learned from 
\citet{rampp00,lieb01,thom03,sumi05} who implemented the best input physics and 
numerics to date, is that the mechanism fails to blow up 
canonical massive stars in spherical symmetric (1D) simulations.
Pushed by mounting supernova observations of the blast morphology
 (e.g., \citet{wang01,maeda08,tanaka}, and references therein), 
it is now almost certain that the
breaking of the spherical symmetry is the key to solve the supernova problem.
 So far a number of multidimensional (multi-D) hydrodynamic simulations have shown
 that hydrodynamic motions associated with convective overturn (e.g., 
\citet{herant,burr95,jankamueller96,frye02,fryer04a}) and the 
Standing-Accretion-Shock-Instability (SASI, e.g., 
\citet{blon03,sche04,scheck06,Ohni05,ohnishi07,thierry,iwakami1,iwakami2,
  murphy08,rodrigo09,rodrigo09_2}, and
 references therein) can help the onset of the neutrino-driven explosion. 


In fact, the neutrino-driven explosions have been obtained in the
following state-of-the-art two-dimensional (2D) simulations (e.g., table 1 in 
 \citet{kota11_rev}).  Using
the MuDBaTH code which includes one of the best available neutrino
transfer approximations, \cite{buras06} firstly reported explosions for
a non-rotating low-mass ($11.2 M_\odot$) progenitor of \cite{woos02},
and then for a $15 M_{\odot}$ progenitor of \citet{woos95} with a
moderately rapid rotation imposed \citep{marek}.  By implementing a
multi-group flux-limited diffusion algorithm to the CHIMERA code
\cite[e.g.,][]{bruenn}, \citet{yakunin} obtained explosions for a
non-rotating progenitors of \citet{woos02} 
in the mass range of $12 M_{\odot}$ and 25$M_{\odot}$.  More recently, \cite{suwa} pointed out that a
stronger explosion is obtained for a rapidly rotating $13M_\odot$
progenitor of \cite{nomo88} compared to the corresponding non-rotating
model, in which the isotropic diffusion source approximation (IDSA)
for the spectral neutrino transport \citep{idsa} is implemented in
the ZEUS code.

This success, however, is opening further new questions. 
First of all, the explosion energies obtained in these 2D simulations
are typically underpowered 
by one or two orders of magnitudes to explain the canonical supernova kinetic energy 
($\sim 10^{51}$ erg).
Moreover, the softer nuclear equation of state (EOS), such as of the
\citet{latt91} (LS) EOS with an incompressibility at nuclear
densities, $K$, of 180 MeV, is employed in those simulations.  On top
of a striking evidence that favors a stiffer EOS based on the
nuclear experimental data ($K=240 \pm 20$ MeV, \citet{shlo06}), 
the soft EOS may not account
for the recently observed massive neutron star of $\sim 2 M_{\odot}$
\citep{demo10}\footnote{The maximum mass for the LS180 EOS is about $1.8 M_{\odot}$ 
 (e.g., \citet{oconnor,kiuc08}).}. 
 Using a stiffer EOS, the explosion energy may be
even lower as inferred from \citet{marek} who did not obtain the
neutrino-driven explosion for their model with $K=263$ MeV\footnote{On the other hand,
 they obtained 2D explosions for Shen EOS ($K=281$MeV, H.-T. Janka, private 
communication).}.
 What is then missing furthermore ?  The neutrino-driven mechanism would be 
 assisted by other candidate mechanisms such as the acoustic mechanism (e.g.,
 \citet{burr06}) or the magnetohydrodynamic mechanism (e.g., 
\citet{kota04b,taki04,taki09,burr07,fogli_B,martin11,taki_kota}, see also \citet{kota06} for collective references 
therein). 
We may get the answer by taking into account new 
ingredients, such as exotic physics 
in the core of the protoneutron star (e.g., \citet{takahara88,sage09}), viscous heating by the
magnetorotational instability \citep{thomp05,masa11}, or energy
dissipation via Alfv\'en waves \citep{suzu08}.

But before seeking alternative scenarios, it may be of primary importance to investigate 
how the explosion criteria extensively studied so far in 2D simulations 
could or could not be changed in 3D simulations.
 \citet{nordhaus} is the first to argue that 
the critical neutrino luminosity for producing neutrino-driven explosions 
becomes smaller in 3D than 2D. They employed 
 the CASTRO code with an adaptive mesh refinement technique, by which unprecedentedly 
high resolution 3D calculations were made possible. Since it is generally
  computationally expensive to solve the neutrino transport in 3D, 
 they employed a light-bulb scheme (e.g., \citet{jankamueller96}) 
to trigger explosions, in which the heating and cooling by neutrinos 
  are treated by a parametric manner. 
Since the light-bulb scheme can capture fundamental properties of 
 neutrino-driven explosions (albeit on the qualitative grounds), it is one of 
the most prevailing approximation adopted in recent 3D models
(e.g., \citet{iwakami1,iwakami2,annop}). A number of 
 important findings have been reported recently in these simulations,
 including a potential role of non-axisymmetric SASI flows in generating spins 
(\citet{annop,rantsiou}, see also \citet{blo07,fern}) and magnetic 
fields \citep{endeve} of pulsars, stochastic nature of 
 gravitational-wave (e.g., \citet{kotake09,kotake11,ewald11}) and 
 neutrino emission (e.g., \citet{kneller}).

To go up the ladders beyond the light-bulb scheme, we explore in this study possible 
3D effects 
in the supernova mechanism by performing 3D, multigroup, radiation-hydrodynamic 
core-collapse simulations.
For the multigroup transport, the IDSA scheme is implemented, 
which can be done rather in a 
 straightforward manner by extending our 2D modules \citep{suwa,suwa11} to 3D.
 This can be made possible because we apply the so-called
ray-by-ray approach (e.g., \citet{buras06}) 
in which the neutrino transport is solved along a
given radial direction assuming that the hydrodynamic medium for the
direction is spherically symmetric. From a technical point of view, it is worth
 mentioning that the ray-by-ray treatment is highly efficient in parallization\footnote{along each radial ray}
 on present supercomputers, most of which 
 employ the message-passing-interface (MPI) routines.
 We focus here on the evolution of an $11.2 M_{\odot}$ star of \citet{woos02}.
 We first choose 
 such a lighter progenitor star, not only because we follow
 a tradition in 2D literatures (e.g., \citet{buras06,burr06}), but also because 
 the neutrino-driven
 shock revival for the progenitor 
was reported to occur rather earlier after bounce in 2D models by \citet{buras06}.
 So we anticipate that the cost of  3D simulations would not be too expensive for 
 the progenitor.
 By comparing with our 1D and 2D results, we study how the increasing 
multi-dimensionality could affect
 the postbounce supernova dynamics. 

The paper opens with descriptions of the initial models and the
numerical methods (Section 2).  The main results are shown in Section 3.
We summarize our results and discuss their implications in Section 4.

\begin{figure*}[htbp]
    \centering
    \includegraphics[width=.35\linewidth]{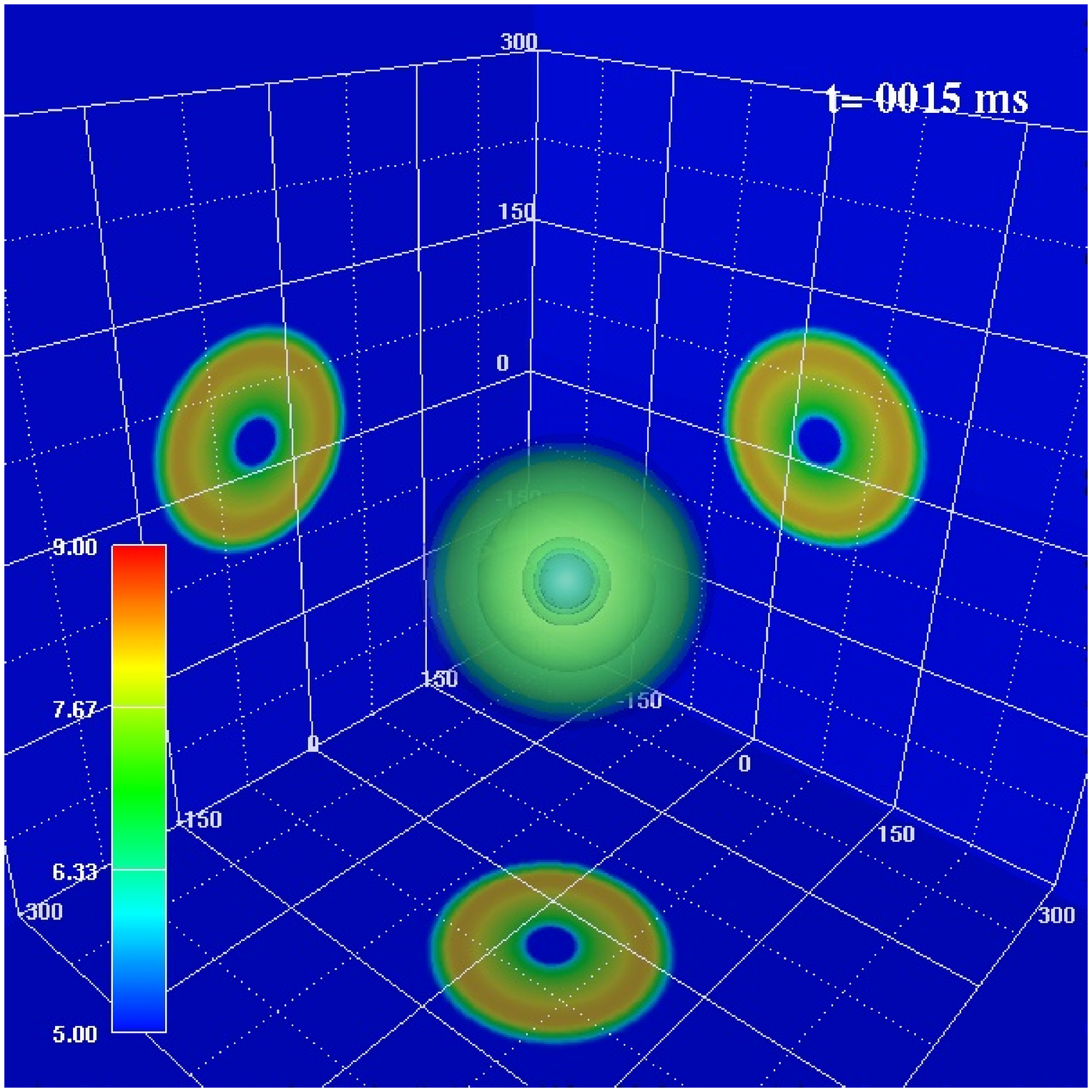}
    \includegraphics[width=.35\linewidth]{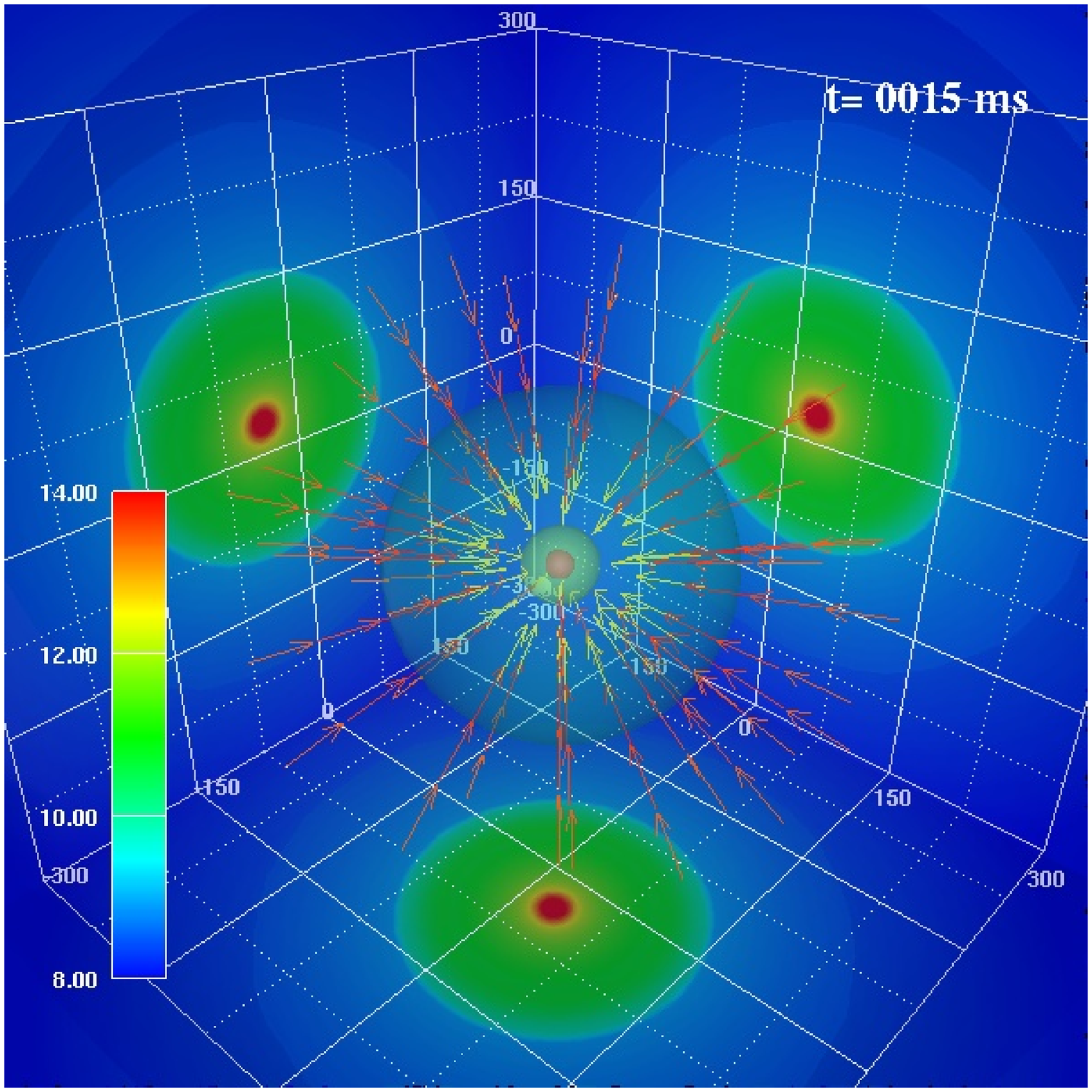}\\
    \includegraphics[width=.35\linewidth]{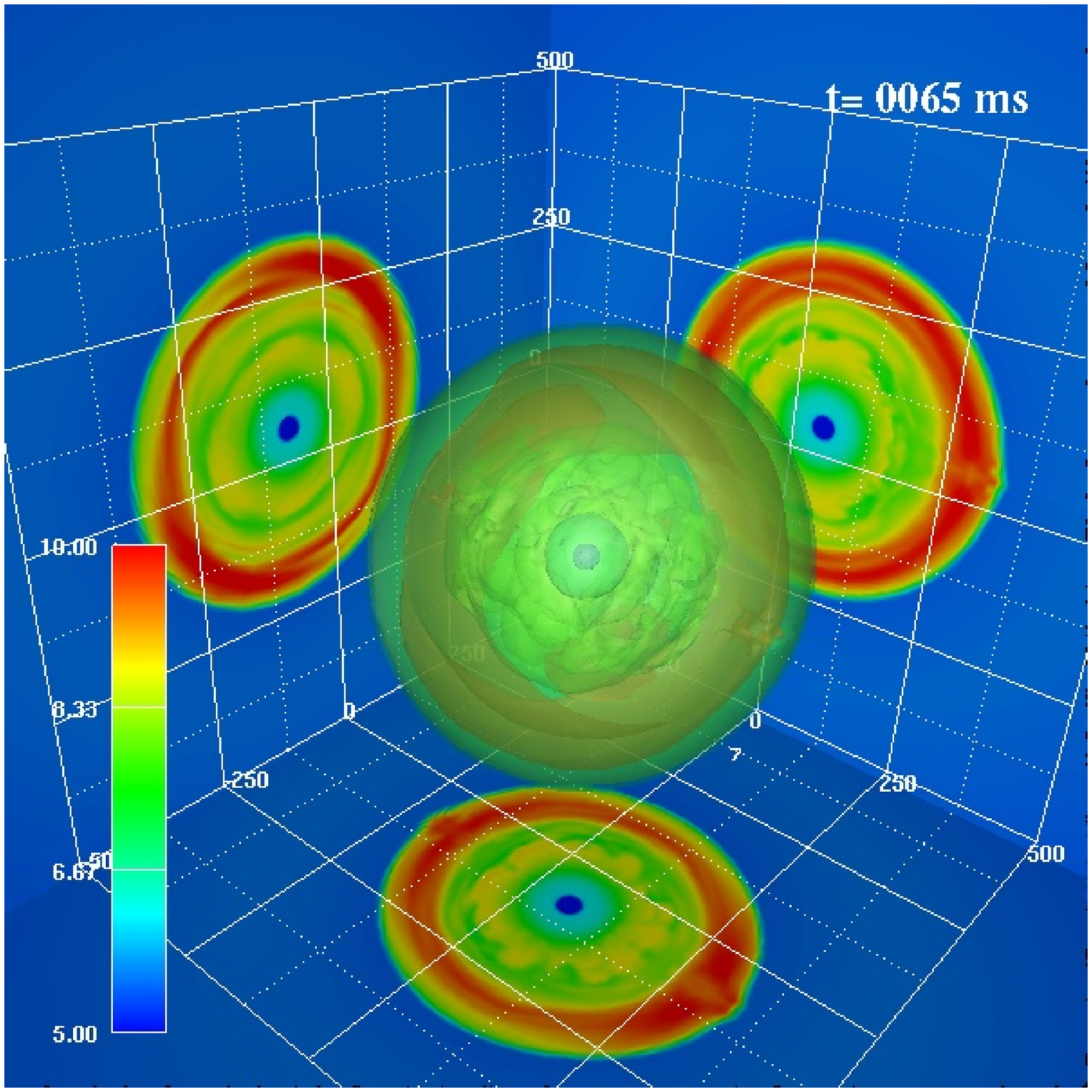}
    \includegraphics[width=.35\linewidth]{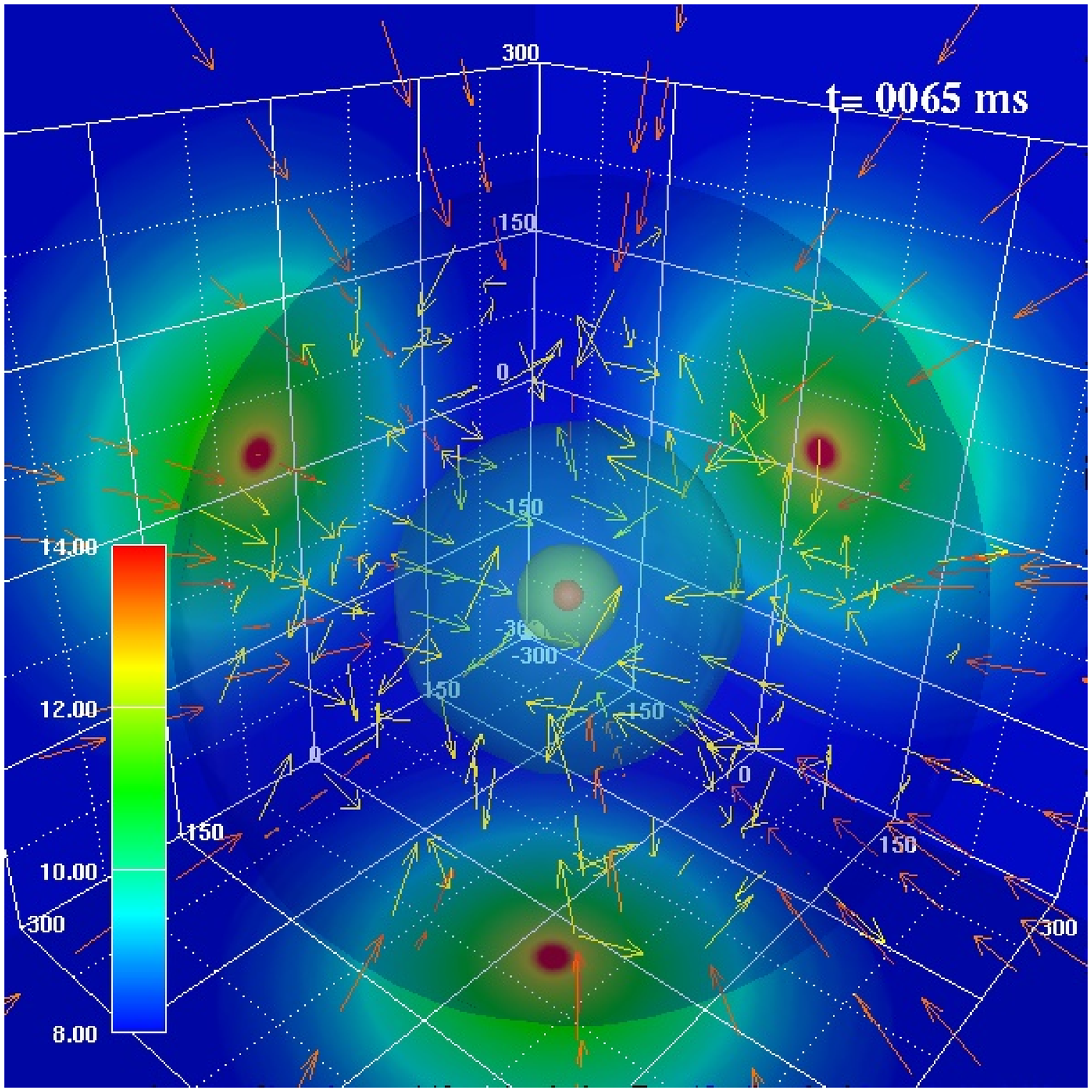}\\
    \includegraphics[width=.35\linewidth]{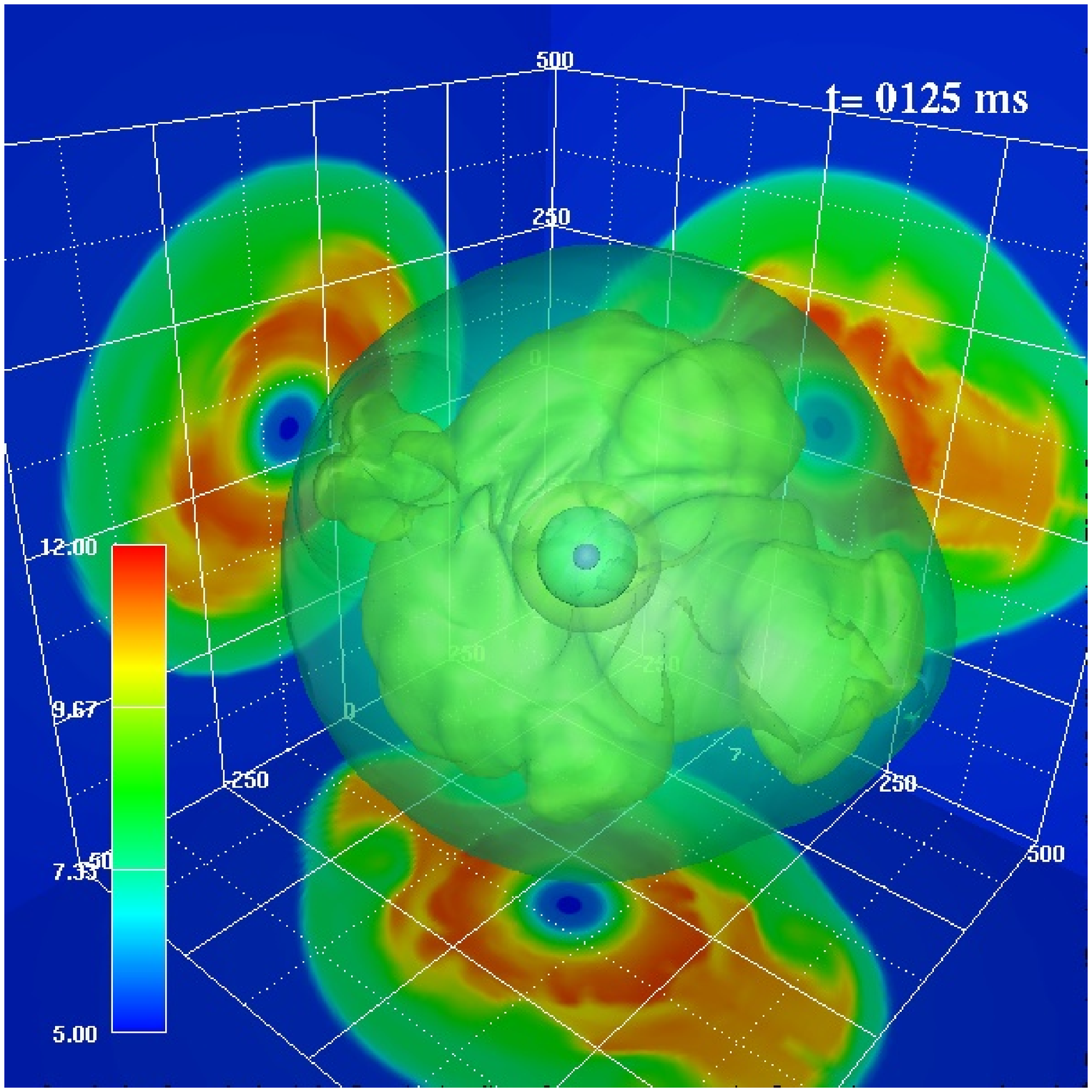}
    \includegraphics[width=.35\linewidth]{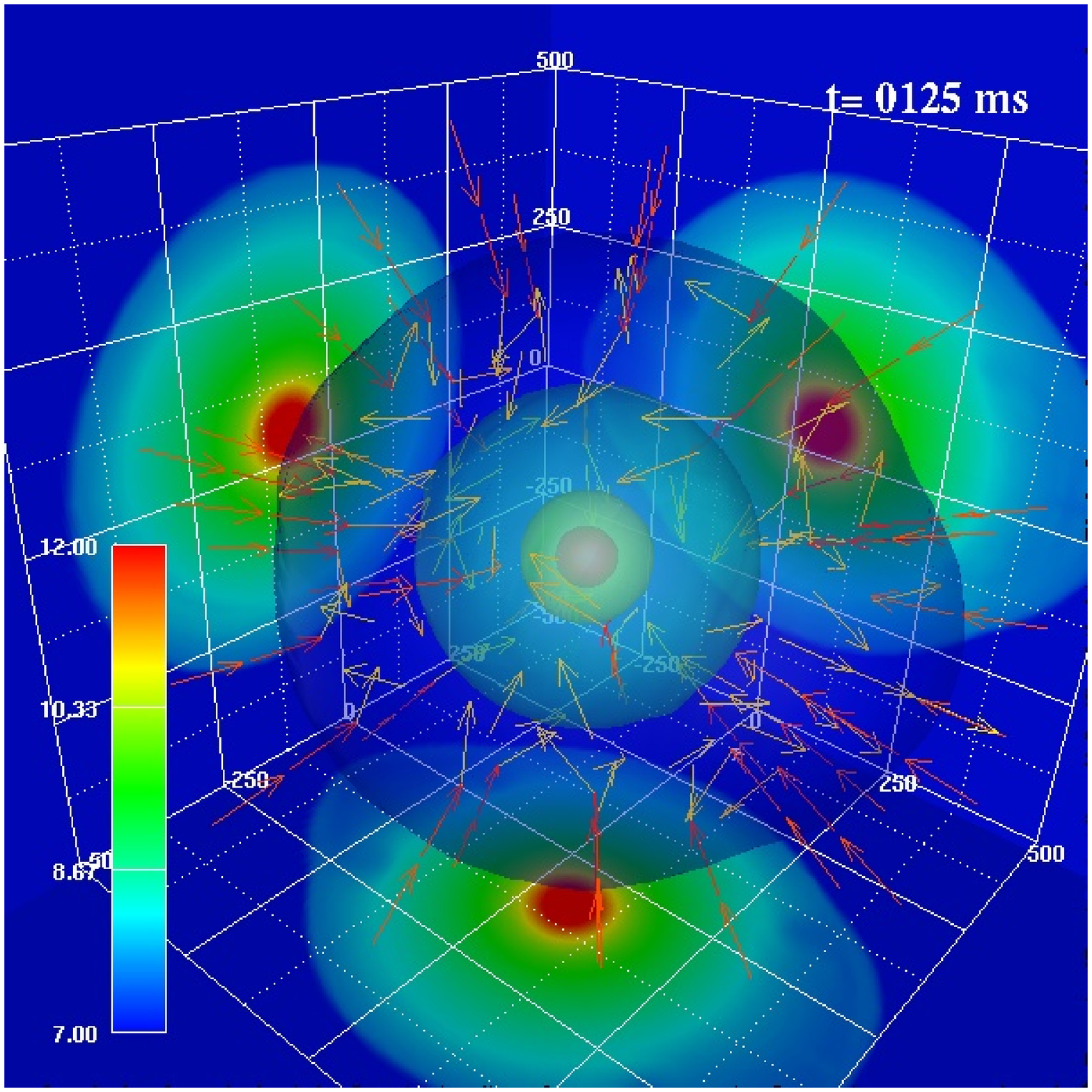}\\
 \caption{Three dimensional plots of entropy per baryon (left panels) 
and logarithmic density
 (right panels, in unit of ${\rm g}/{\rm cm}^3$) for three snapshots (top; $t = 15$ ms, middle; $t=65$ ms, and 
 bottom; $t=125$ ms measured after bounce ($t\equiv 0$)) of our 3D model. 
 In the right panels,  velocities are indicated by arrows.
The contours on the cross sections in the 
$x=0$ (back right), $y=0$ (back bottom), and $z=0$ (back left) planes are, 
respectively projected on the sidewalls of the graphs. For each snapshot, 
 the linear scale is indicated along the axis in unit of km.}
\label{f1}
\end{figure*}

\begin{figure*}[htbp]
    \centering
    \includegraphics[width=.35\linewidth]{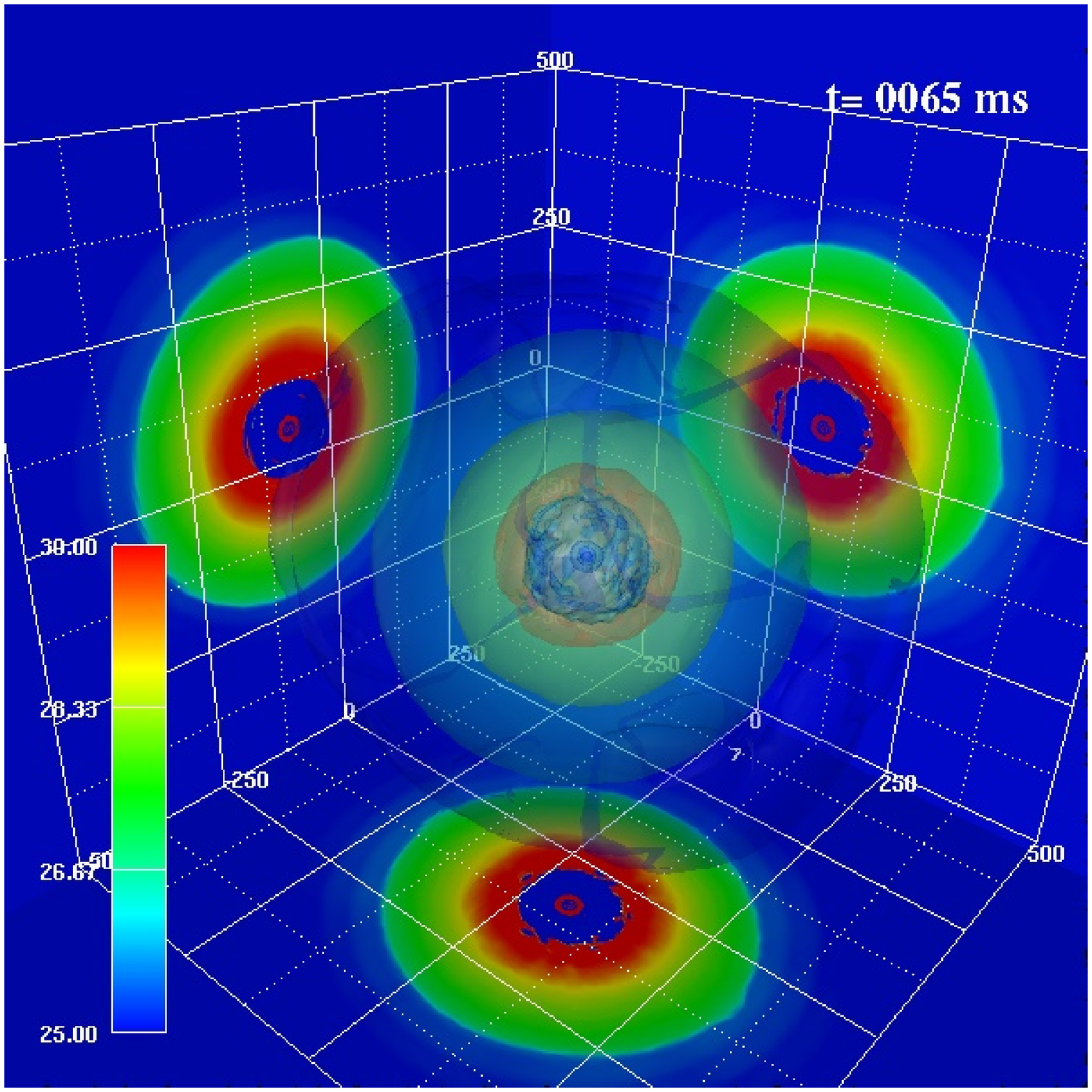}
    \includegraphics[width=.35\linewidth]{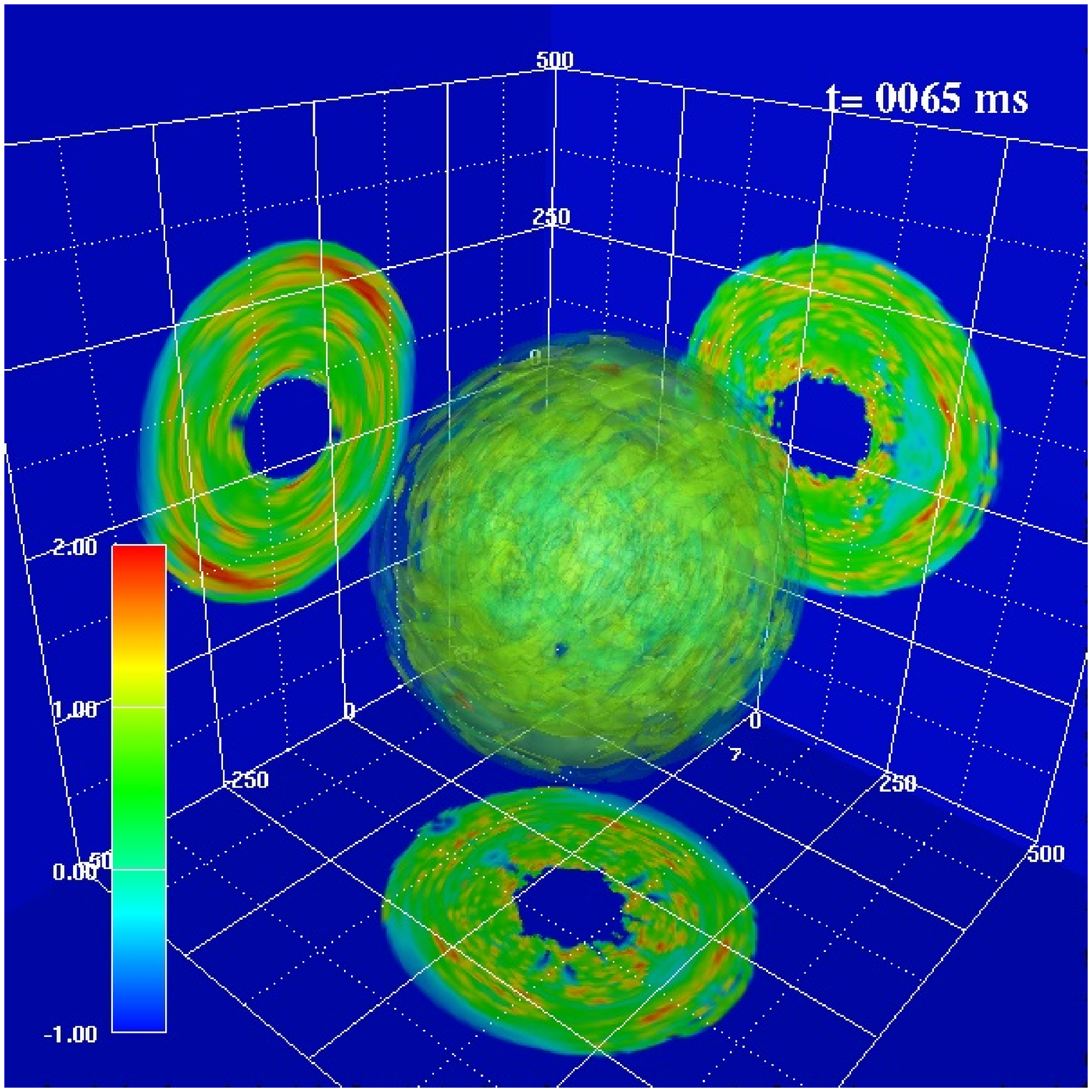}\\
    \includegraphics[width=.35\linewidth]{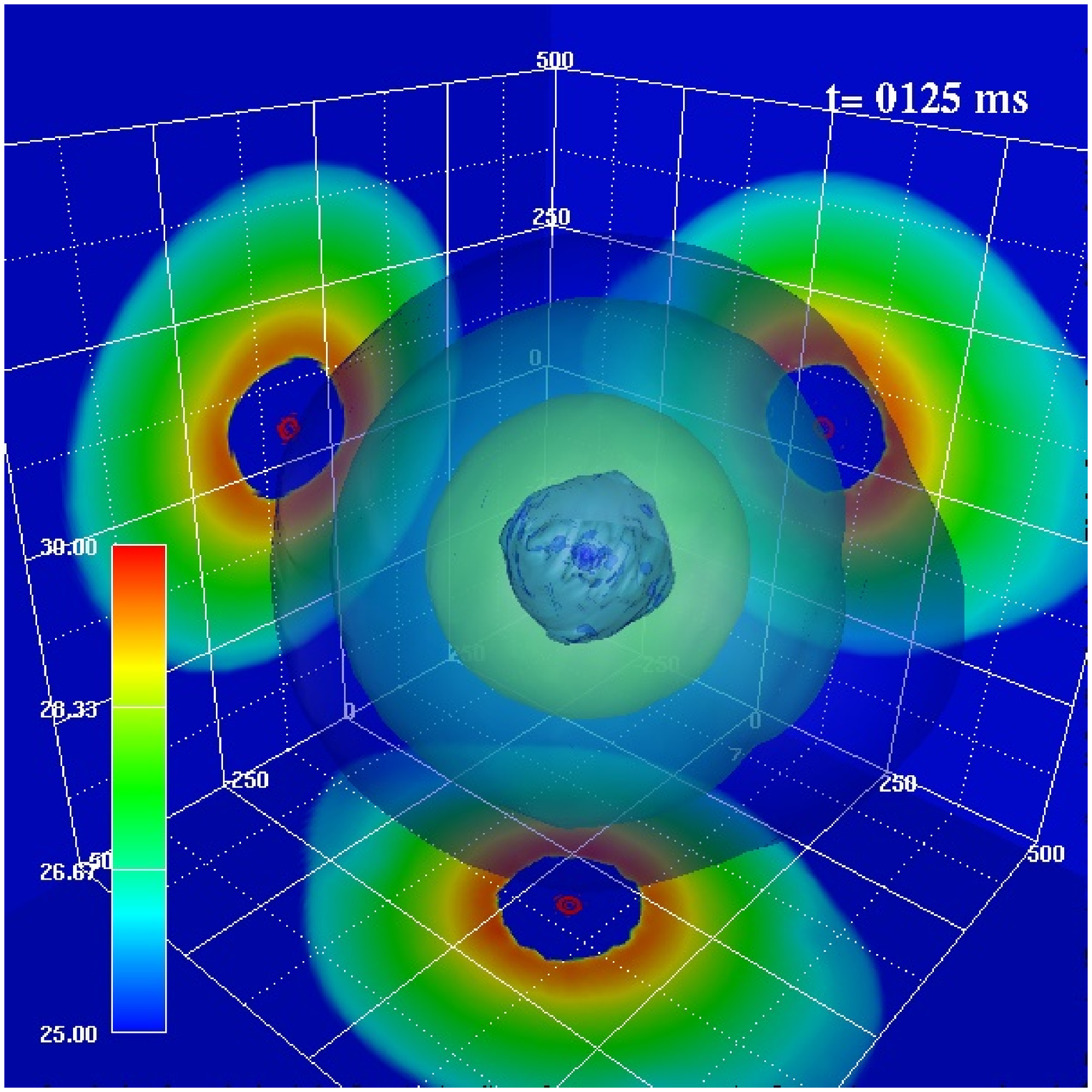}
    \includegraphics[width=.35\linewidth]{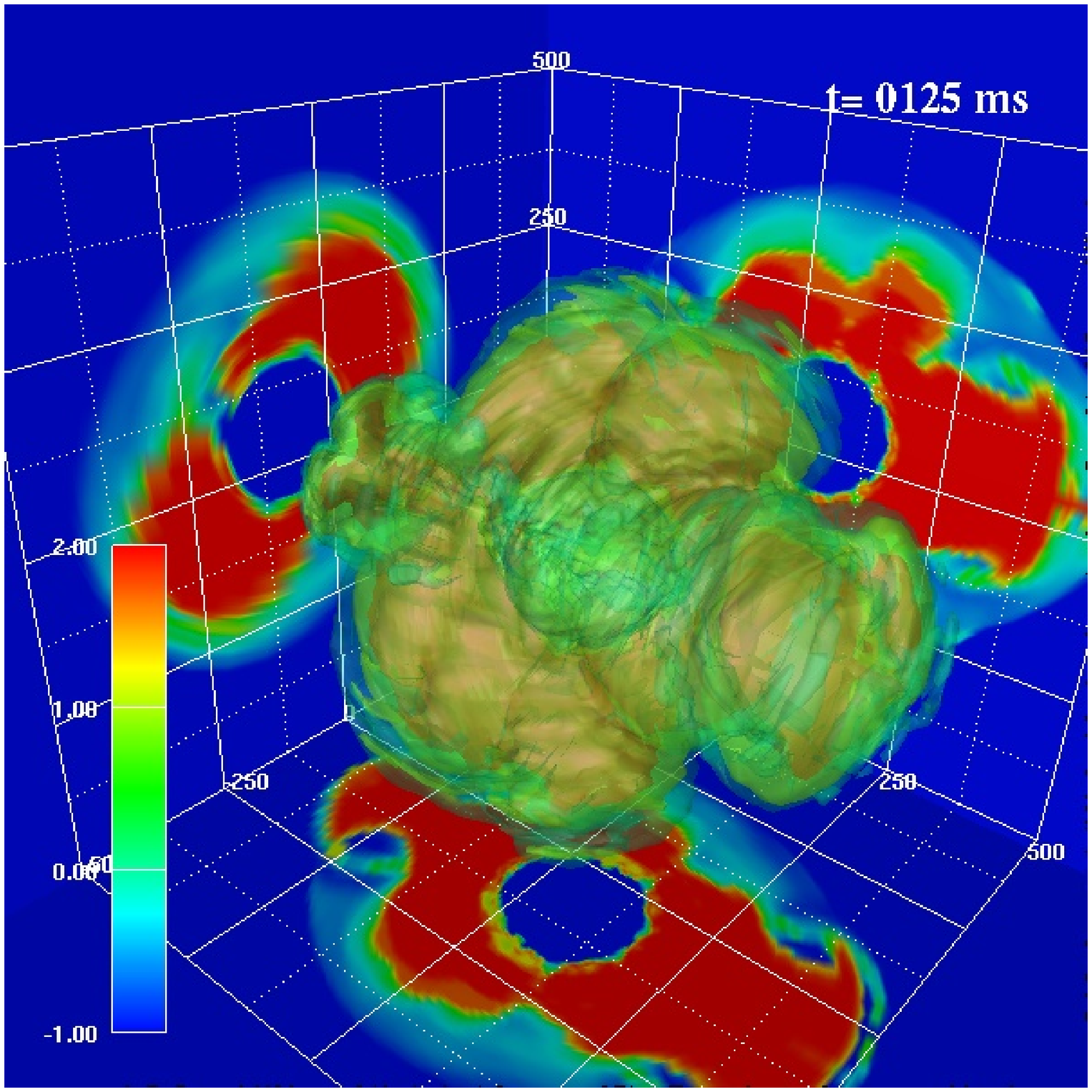}\\
 \caption{Same as Figure 1 but for the net neutrino heating rate (left panels,
 logarithmic in unit of erg/cm$^3$/s and $\tau_\mathrm{adv}/\tau_\mathrm{heat}$
 (right panels, see text for details), which is the ratio of the advection to the neutrino heating timescale. The gain region (colored 
 by red in the top left panel) is shown to be formed at around $t=65$ ms after bounce,
 which coincides with the epoch approximately when the neutrino-driven shock revival 
 initiates in our 3D model. 
The condition of $\tau_\mathrm{adv}/\tau_\mathrm{heat}\gtrsim1$ is satisfied 
behind the aspherical shock, the low-mode deformation of which is characterized by 
  the SASI (bottom right panel).}
\label{f2}
\end{figure*}

\section{Numerical Methods and Initial Models \label{sec2}}

 The basic evolution equations for our 3D simulations are
written as,
\begin{equation}
  \frac{\mathrm{d}\rho}{\mathrm{d}t}+\rho\nabla\cdot\mathbf{v}=0,
\end{equation}
\begin{equation}
  \rho \frac{\mathrm{d}\mathbf{v}}{\mathrm{d}t}=-\nabla P -\rho \nabla \Phi,
\end{equation}
\begin{equation}
\frac{\mathrm{\partial}e^*}{\mathrm{\partial}t}+
\nabla \cdot
\left[\left(e^* + P\right) \mathbf{v}\right]= -\rho \mathbf{v} \cdot \nabla  \Phi+ Q_{E}
\label{eq:e},
\end{equation}
\begin{equation}
\frac{{\mathrm{d}Y_e}}{\mathrm{d}t} = \Gamma_N,
\label{eq:ye}
\end{equation}
\begin{equation}
  \bigtriangleup{\Phi} = 4\pi G \rho,
\end{equation}
where $\rho, \mathbf{v}, P, \mathbf{v}, e^*, \Phi$, are density, fluid
velocity, gas pressure including the radiation pressure of neutrinos,
total energy density, gravitational potential, respectively. 
$\frac{{\mathrm{d}}}{\mathrm{d}t}$ denotes the Lagrangian derivative.
As for the hydro-solver, we employ the
ZEUS-MP code \citep{hayes} which has been modified for core-collapse
simulations \citep[e.g.,][]{iwakami1,iwakami2}. $Q_{E}$
and $\Gamma_{N}$ (in Equations (\ref{eq:e}) and (\ref{eq:ye})) represent
the change of energy and electron fraction ($Y_e$) due to the
interactions with neutrinos. To estimate these quantities, we
employ the IDSA scheme \citep{idsa}. 
The IDSA scheme
splits the neutrino distribution into two components, both of which
are solved using separate numerical techniques. 
 Although the current IDSA scheme
does not yet include heavy lepton neutrinos ($\nu_x$)
 and the inelastic neutrino scattering
with electrons, these simplifications
 save a significant amount of computational time compared to the canonical Boltzmann 
solvers (see \cite{idsa} for more details). As already mentioned, we employ 
the ray-by-ray approximation, by which the 3D radiation transport is reduced 
essentially to the 1D problem.
Following the prescription in
\cite{mueller}, we improve the accuracy of the total energy
conservation by using a conservation form in Equation (\ref{eq:e}),
instead of solving the evolution of internal energy as originally
designed in the ZEUS code. A Poisson equation (in Equation (5)) can be 
 solved either by the ICCG\footnote{Incomplete Cholesky Conjugate Gradient} method in the original ZEUS-MP code or 
by the multi-domain spectral method developed in the Lorene code \citep{lorene}. 
For the calculations presented here,
 the monopole approximation is employed.
 
The computational grid is comprised of 300 logarithmically spaced,
radial zones to cover from the center up to 5000 km  and 64 polar ($\theta$) and 32
azimuthal ($\phi$) uniform mesh points for our 3D model, 
which are used to cover the whole solid angle. To vary numerical resolutions,
 we run one more 3D model that has one-half of the mesh numbers in the 
$\phi$ direction ($n_{\phi}$=16), while fixing the mesh numbers in other directions.
 Both in 2D and 3D 
 models, we take the same mesh numbers in the polar direction ($n_{\theta}$=64),
 so that we 
 could see how the dynamics could change due to the additional degree of freedom 
in the $\phi$ direction. For the spectral transport, we use 20 logarithmically 
spaced energy bins reaching from 3 to 300 MeV. For our non-rotating progenitor, 
 the dynamics of collapsing iron core 
 proceeds totally spherically till the stall
 of the bounce shock. 
 To save the computational
 time, we start our 2D and 3D simulation by remapping the 1D data after the stall of 
the bounce shock to the multi-D grids. 
To induce non-spherical instability, we add random velocity 
perturbations at less than 1 $\%$ of the unperturbed radial velocity.

\section{Results}
In the following (section \ref{part1}), 
we first outline hydrodynamic features in our 3D model.
 Then in sections \ref{part2} and 3.3, 
we move on to discuss how 3D effects impact on the explosion dynamics by 
 comparing with the 1D and 2D results.

\subsection{3D dynamics from core-collapse through postbounce turbulence till explosion}\label{part1}

Figure \ref{f1} shows three snapshots, which 
are helpful to characterize hydrodynamic features in the 3D model.
  Top panel is for $t=15$ ms after bounce, showing that the bounce shock stalls 
(indicated by inward arrows in the top right panel) at a radius of $150$ km. Note that 
 colors of the velocity arrows are taken to change from yellow to red as 
 the absolute values become larger. By looking carefully at the top right panel, 
  matter flows in supersonically (indicated by reddish arrows) 
in the standing shock (the central transparent sphere), and then advects
 subsonically (indicated by yellowish arrows) to 
 the protoneutron star (PNS or the unshocked core, the central bluish region 
in the top left panel).
 As seen, the entropy (left panel) and density (right panel) configurations are 
essentially spherical at this epoch.\footnote{It is approximately 5 ms after we 
remap the 1D data to the 3D grids.}

 The middle panels shows an epoch ($t=65$ ms)
 when the neutrino-driven convection is already active.
 From the right panel, turbulent motions can be 
 seen (arrows in random directions) inside the standing shock, which is
 indicated by the boundary between red and yellow arrows.
 The entropy behind the standing shock becomes high by the neutrino-heating 
 (reddish regions in the left panel). The size of the neutrino-heated hot 
 bubble becomes larger in a non-axisymmetric way later on, which is indicated by  
 smaller structures encompassed by the stalled shock (i.e., inside the central 
greenish sphere in the left panel). 

 The bottom panels ($t=125$ ms) show the epoch when the revived shock is expanding 
 aspherically, which is indicated by the outgoing yellowish arrows 
in the right panel. The asphericity of 
 the expanding shocks could be more clearly visible by the sidewall panels.
From the entropy distribution (left panel), the expanding shock is 
shown to touch a radius of $\sim 500$ km (the projected back bottom panel). 
 Inside the expanding shock (enclosed by the greenish membrane in the 
 left panel), the bumpy structures of the hot bubbles are seen. In contrast to
 these smaller asphericities, the deformation of the shock surface is mild, 
which is a consequence of the SASI as will be discussed in section \ref{part2}.

Figure \ref{f2} shows the net neutrino heating rate (left panels) and the 
 ratio of the residency timescale to the neutrino-heating timescale (right panels) 
for the 
two snapshots in Figure \ref{f1} 
(at $t=65$ ms (top panels) and $t=125$ ms (bottom panels)).
 Here the residency timescale and the neutrino heating timescale\footnote{Originally 
 the ratio of the advection to the neutrino-heating timescale is known as a 
 useful quantity to diagnose the success ($\tau_\mathrm{adv}/\tau_\mathrm{heat}\gtrsim 
1$, i.e., the neutrino-heating timescale is shorter than the advection timescale of 
material in the gain region) or failure 
($\tau_\mathrm{adv}/\tau_\mathrm{heat}\lesssim 1$) of the 
neutrino-driven explosion (e.g., \citet{goshy,jank01,thomp05}).} 
are locally defined as,
\begin{eqnarray}
 t_{\mathrm{res}}\left(r,\theta,\phi\right) &=&
\begin{cases}
 \frac{r-r_{\mathrm{gain}}\left(\theta,\phi\right)}{-v_r}
 & {\rm for}~v_r < 0, \\
\frac{r_{\mathrm{shock}}\left(\theta,\phi\right)-r}{v_r} 
& {\rm for}~v_r > 0,
\end{cases}\\
 t_{\mathrm{heat}}\left(r,\theta,\phi\right) &=&
 \frac{|e_{\mathrm{bind}}|}{Q^{+}_{\nu,~{\rm total}}},
\label{residency}
\end{eqnarray}
where $r_{\rm gain}$ is the gain radius that depends on $\theta$ and $\phi$, 
$r_{\rm shock}$
 is the shock radius, and $v_r$ is 
 the radial velocity. We take the above criteria
 in order to estimate the residency timescale
 for material with positive radial velocities ($v_r > 0$) behind the shock.\footnote{We take the word of ``residency'' timescale from \citet{murphy08}, which we feel 
 more appropriate than the canonical ``advection'' timescale 
especially when we need to estimate the timescale regarding the postshock
 material with positive radial velocities.} The heating timescale can be rather 
straightforwardly defined by dividing the local binding energy 
 ($e_{\mathrm{bind}}=\frac{1}{2}\rho\mathbf{v}^2+e-\rho\Phi$ [erg/cm$^3$]) in 
 the gain layer
 by the net neutrino-heating rate (:$Q^{+}_{\nu,~{\rm total}}$ [erg/cm$^3$/s]). 

\begin{figure}[htbp]
    \centering
    \includegraphics[width=.80\linewidth]{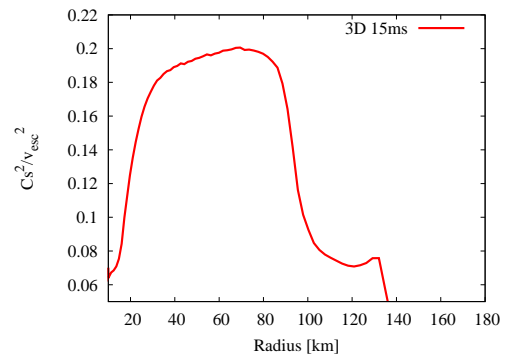}
 \caption{Ratio of sound speed ($c_s$) and escape velocity ($v_{\rm esc}$) 
squared as a function of 
 radius in our 3D model. This snapshot roughly coincides with 
the epoch when the neutrino-heating 
begins to revive the stalled bounce into explosion. This result is in agreement
 with the ante-sonic condition $(c_s^2/v_{\rm esc}^2 \sim 0.2) $ for producing explosions \citep{pejcha}.}
\label{f3}
\end{figure}

\begin{figure*}[htbp]
    \centering
    \includegraphics[width=.35\linewidth]{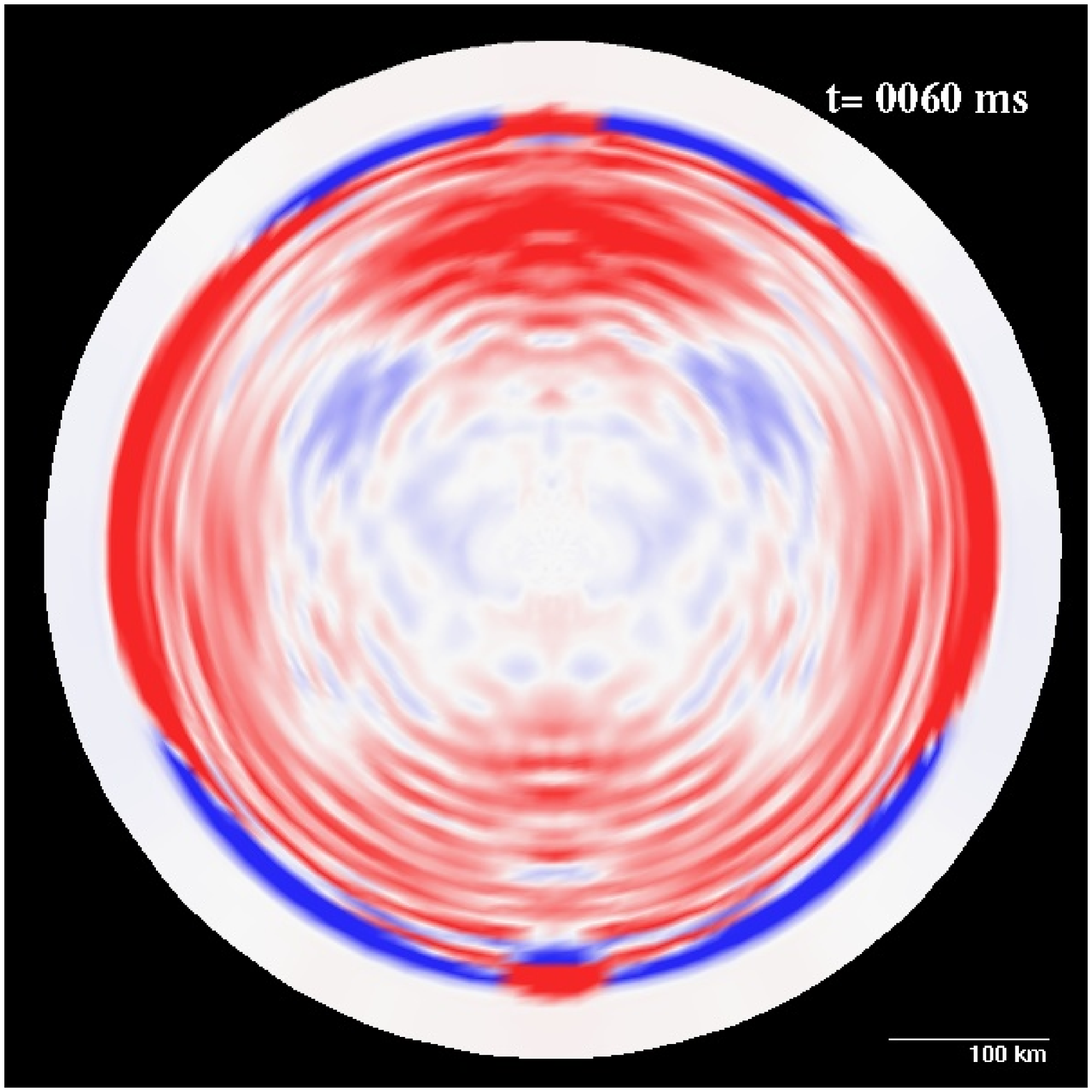}
    \includegraphics[width=.35\linewidth]{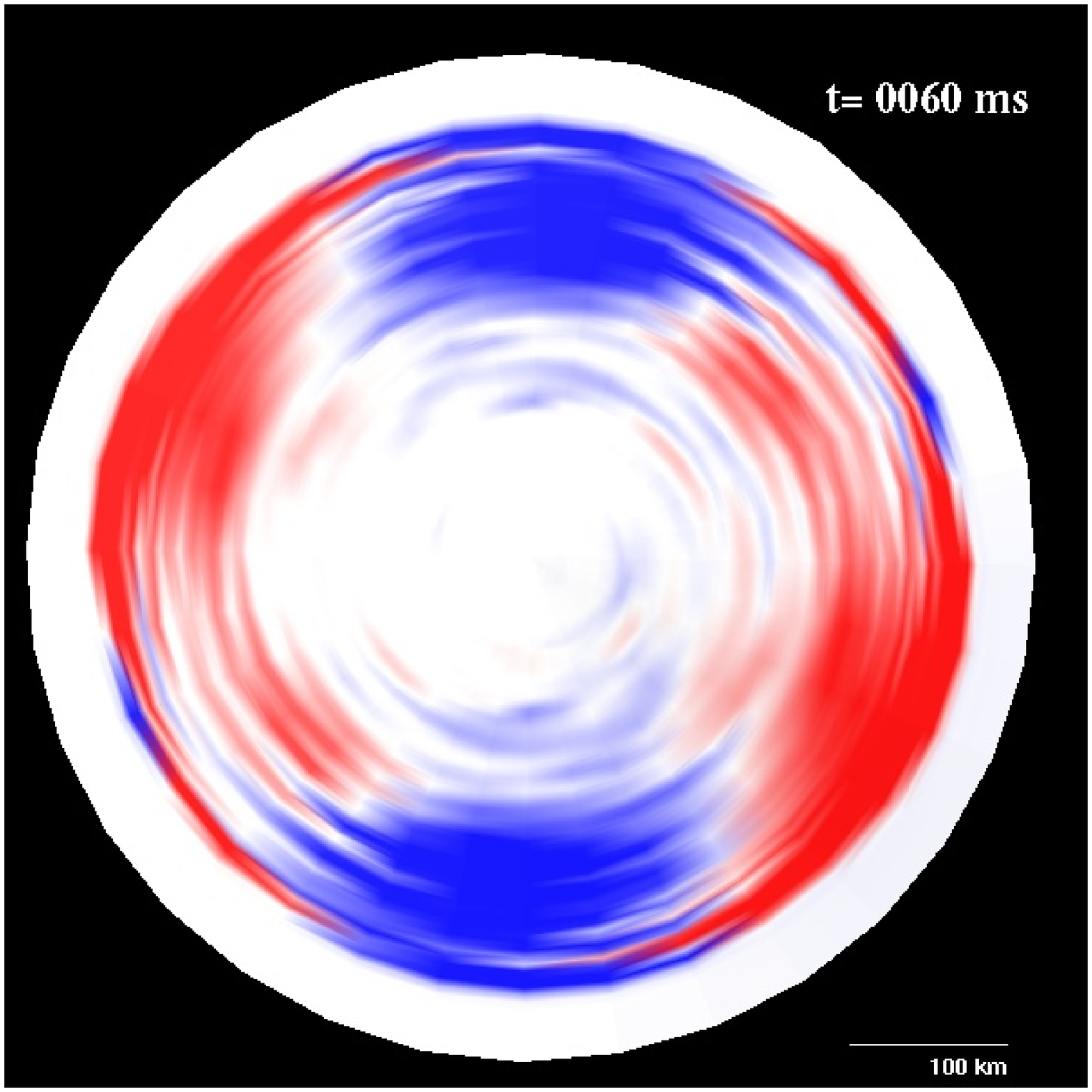}\\
    \includegraphics[width=.35\linewidth]{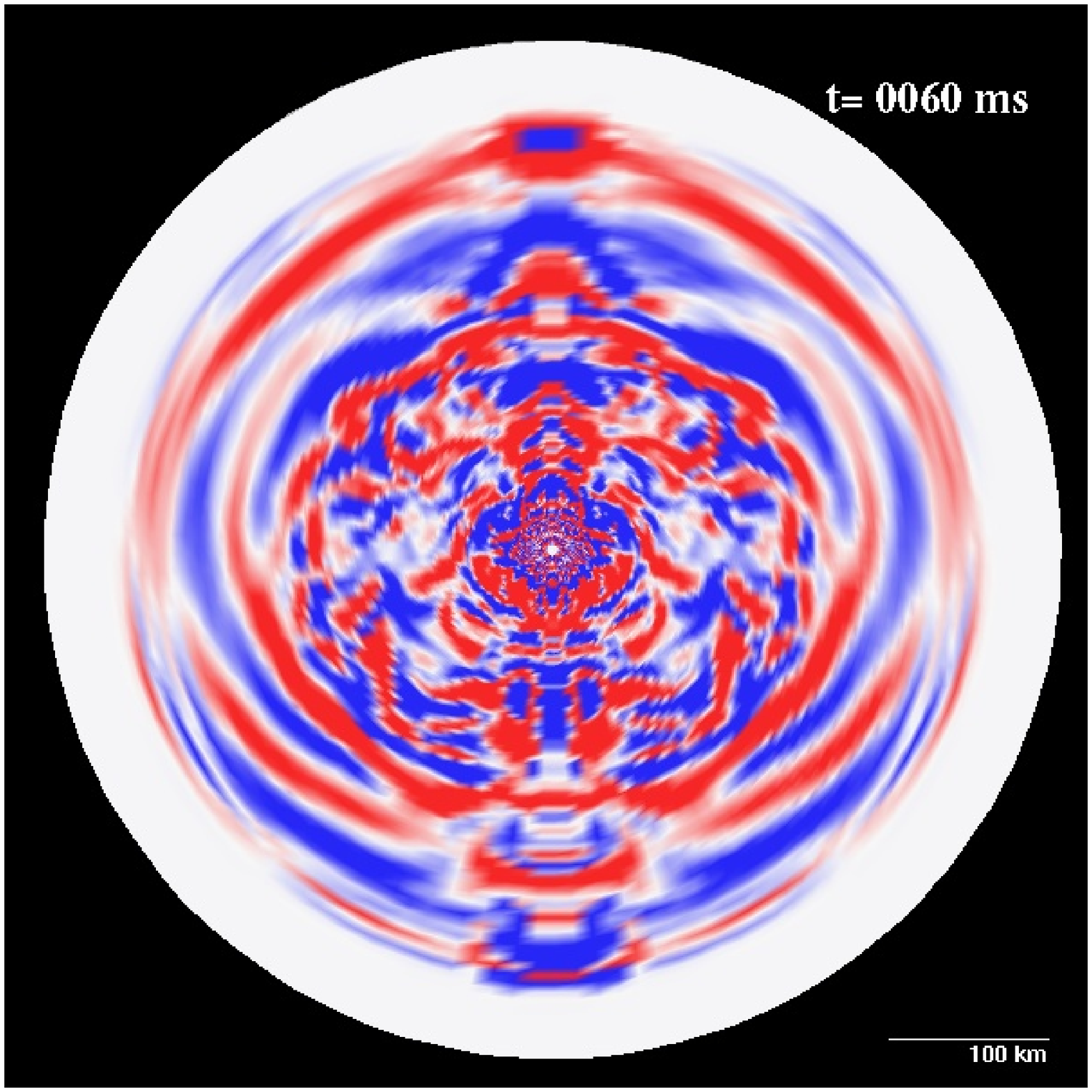}
    \includegraphics[width=.35\linewidth]{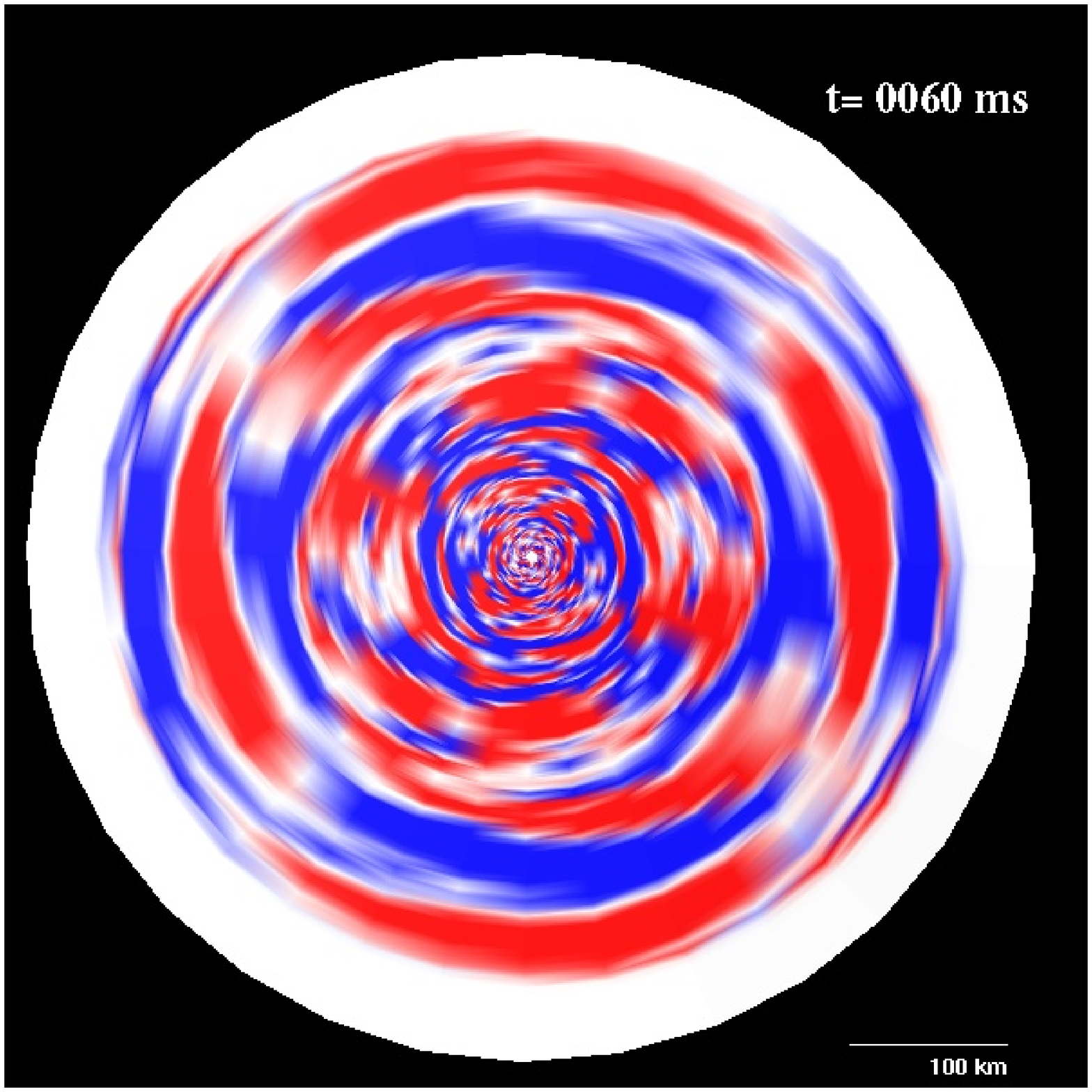}
 \caption{Top panels show distributions of the 
pressure perturbation ($\Delta p/\langle p \rangle $, see text for definition).
 Left and right panel corresponds to
 a partial cutaway in the $YZ$ and $XY$ plane, respectively.
 Bottom panels show the vorticity distributions 
 ($ \left(\nabla \times \mathbf{v}\right)_{\perp}$ with $\perp$ 
 being $\phi$ or $\theta$ in the left and right panel, respectively).
 The circle that screens between the regions colored 
by blue and red and the whitish region outside corresponds to the surface of the 
stalled shock. Note that the positive and negative values are colored by red and 
 blue, respectively (e.g.,
 $+|\Delta p/\langle p \rangle |$ (red) or $-|\Delta p/\langle p \rangle |$ (blue) for the pressure perturbation).
  The linear scale and 
 the time of these snapshots of the 3D model ($t=60$ ms after bounce) are 
indicated in the top right, and bottom right 
edge of each plot.}
\label{f4}
\end{figure*}

At $t=65$ ms after bounce (top panels), the gain region is clearly formed (reddish
 region in the left panel), where the ratio of the two timescales 
exceeds unity (yellowish region in the right panel). At $t=125$ ms after bounce 
 (bottom panels), the ratio reaches about 2 (reddish region in the bottom right panel)
 behind the shock (compare the bottom right panel in Figure \ref{f1}), which presents 
 an evidence that the shock-revival is driven by the neutrino-heating mechanism.
 Recently \citet{pejcha} proposed an alternative definition of the onset time
 of the explosion, which is the so-called ante-sonic condition. From Figure \ref{f3},
 it can be seen that the criteria that the ratio of sound speed and escape velocity
 squared $\sim 0.2$, is also satisfied in our 3D model 
when the neutrino-driven explosion sets in.

Figure \ref{f4} shows distributions of 
pressure perturbation (top) and vorticity (bottom) at 
$t=60$ ms after bounce. 
 Here the pressure perturbation is 
 estimated by $\Delta p/\langle p \rangle $, with $\Delta p$ representing the deviation 
from the angle average pressure (:$\langle p \rangle$) at a given
position. Here we define the angle average of variable $A$ as
\begin{equation}
\langle A \rangle  = \frac{\int \mathrm{d}\Omega A}{4\pi}.
\end{equation}
 The positive and negative deviations are colored by red and blue, respectively (e.g.,
 $+|\Delta p/\langle p \rangle |$ (red) or $-|\Delta p/\langle p\rangle |$ (blue)).
The left and 
 right panels are for an equatorial ($\theta = \pi/2$, and $\phi =0$) 
and a polar observer ($\theta = 0$), respectively. 
 In each plot, the circle that screens between 
the colored region and the whitish region outside corresponds 
to the surface of the stalled shock. From the top left panel, it is shown 
that the pressure waves (colored by red or blue) propagate outwards up to behind the 
 stalled shock in a concentric fashion.  
Seen from the polar direction (top
 right), the color pattern at this 
 snapshot indicates the dominance of $\ell =2$ and $m=2$ modes 
 in the pressure perturbation, which is related to the growth of SASI as we will 
 discuss in section \ref{part2}.

 The vorticity distributions seen from the equator (bottom left panel)
 show 
 that the red and blue stripes appear alternatively behind the stalled 
shock. Seen from the pole (bottom right), the vorticity waves are shown to be spinning 
 around the polar axis (the origin of the figure), which would 
 be related to the growth of the spiral SASI modes. These fundamental features of 
 the acoustic-vorticity feedbacks are 
 akin to the ones obtained in 
 \citet{sato_f} who studied extensively the properties of SASI by their 
  idealized numerical simulations.
 Our results might provide a
 supporting evidence that the advective-acoustic cycle (e.g., \citet{fog1,fog2,fog3})
does work also in 3D simulations. 
 
\begin{figure}[htbp]
    \centering
    \includegraphics[width=.90\linewidth]{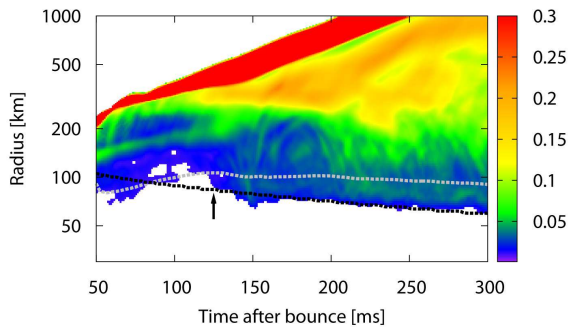}
 \caption{Entropy dispersion ($\sigma_s$) in spacetime diagrams for the 3D model (see text for 
 the definition).
The dotted gray line represents
 the position of the gain radius, while the dotted black line shows the position of
 the PNS surface. The black arrow inserted at around $125$ ms represents the 
 epoch when the position where the entropy production takes place, shifts from 
the gain radius (dotted gray line) to the PNS surface (dotted black line).}
\label{f5}
\end{figure}

Figure \ref{f5} shows the spacetime diagrams of entropy dispersion ($\sigma_s$) 
for the 3D model.
  Note that the dispersion of quantity $A$ with 
 respect to angular variation is 
 defined by 
\begin{equation}
\sigma_{A} = \sqrt{\int \mathrm{d}\Omega \left(A- \langle A \rangle \right)^2/(4\pi)},
\end{equation}
 where $\langle A \rangle $ represents the angle average (Equation (8)).
 It is rather uncertain where the entropy production actually 
 takes place in the supernova core in the context of the 
advective-acoustic cycle  (e.g., \citet{sato_f}). 
The primary position is the surface of the PNS, where the advecting material receives 
faster deceleration by the walls of the PNS due to the localized gravitational 
potential (e.g., \citet{blon03}). In addition, the infalling material 
 could also receive faster deceleration just outside the gain radius,
 where the neutrino-heating becomes maximum. Our 3D results tell that both of the 
 two candidates are relevant indeed. As seen from Figure 5, the position where 
 the entropy production takes place, roughly coincides with the gain radius 
(the dotted grey line) 
before $125$ ms after bounce (indicated by the upward arrow). Later on, the position is 
 shown to transit to the surface of the PNS surface (the dotted black line).

Until now, we have focused on the postbounce dynamics only for our 3D model.
 From the next sections, we move 
 on to look into more detail how they differ from the 1D and 2D results.

\subsection{Blast morphology and explosion dynamics}\label{part2}

\begin{figure*}[htbp]
    \centering
    \includegraphics[width=.31\linewidth]{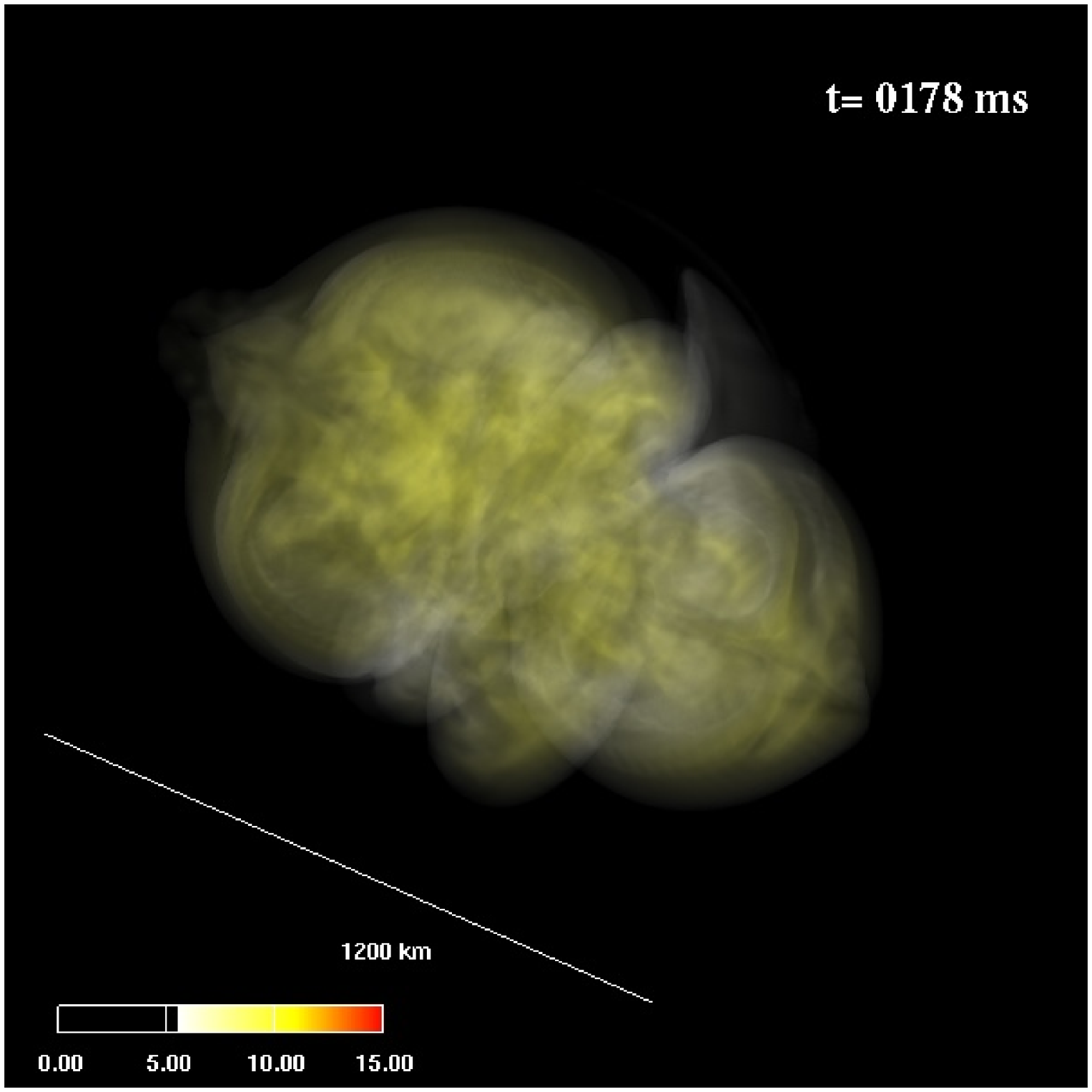}
    \includegraphics[width=.31\linewidth]{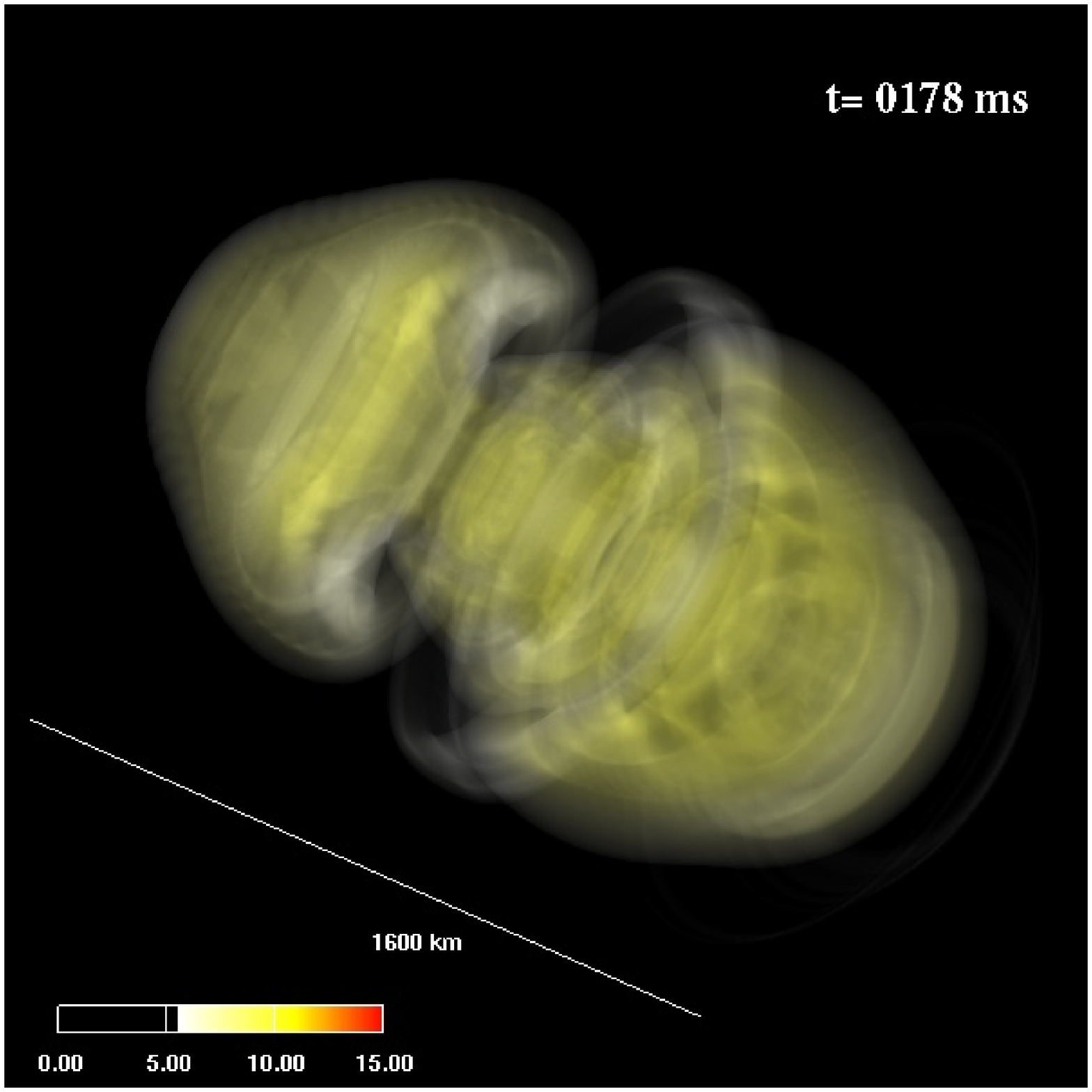}
    \includegraphics[width=.31\linewidth]{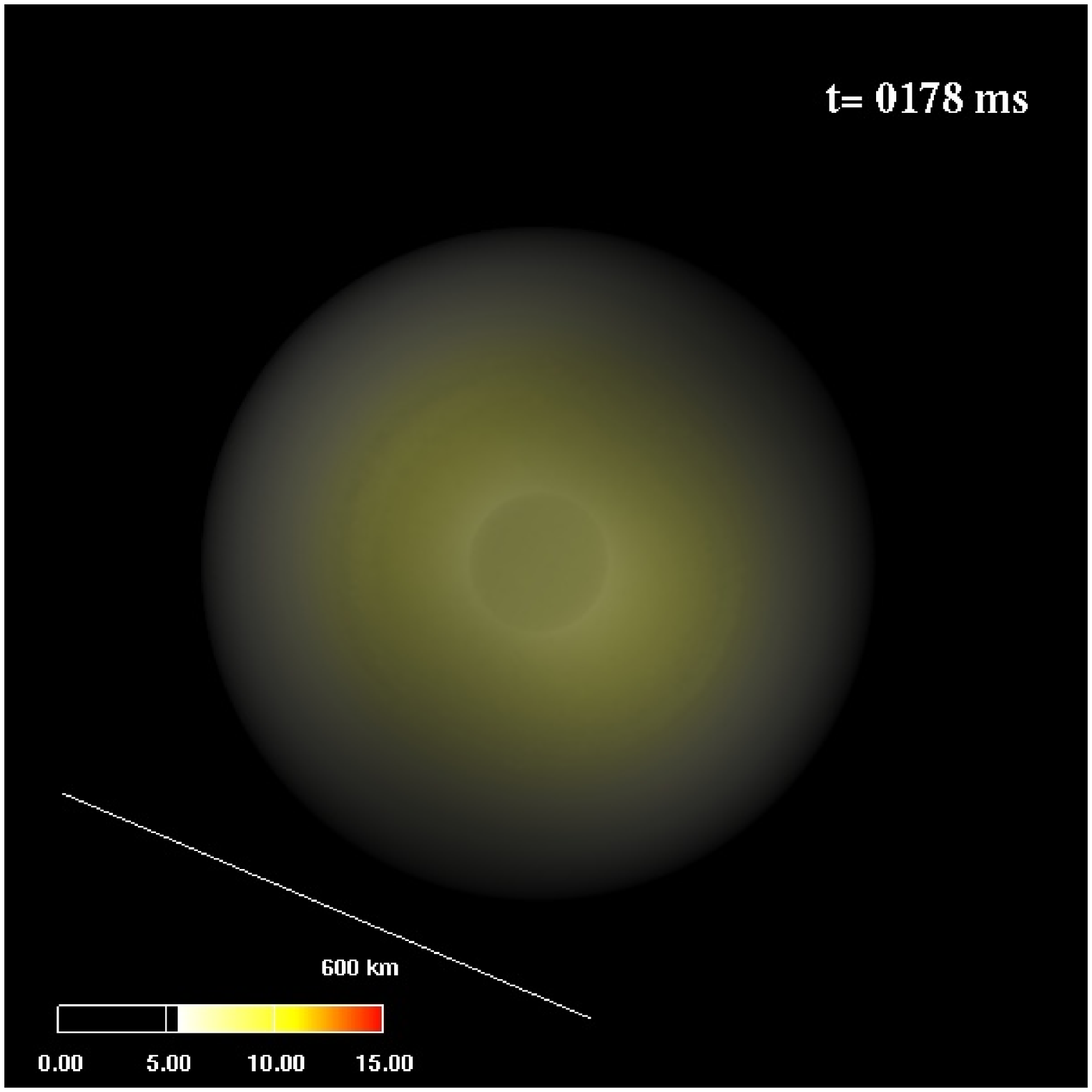}
 \caption{Volume rendering of entropy showing the blast morphology 
 in our 3D (left), 2D (middle), and 1D (right) model (at $t=178$ ms after bounce),
 respectively. The linear scale is indicated in each panel.}
\label{f6}
\end{figure*}

\begin{figure*}[htbp]
    \centering
    \includegraphics[width=.35\linewidth]{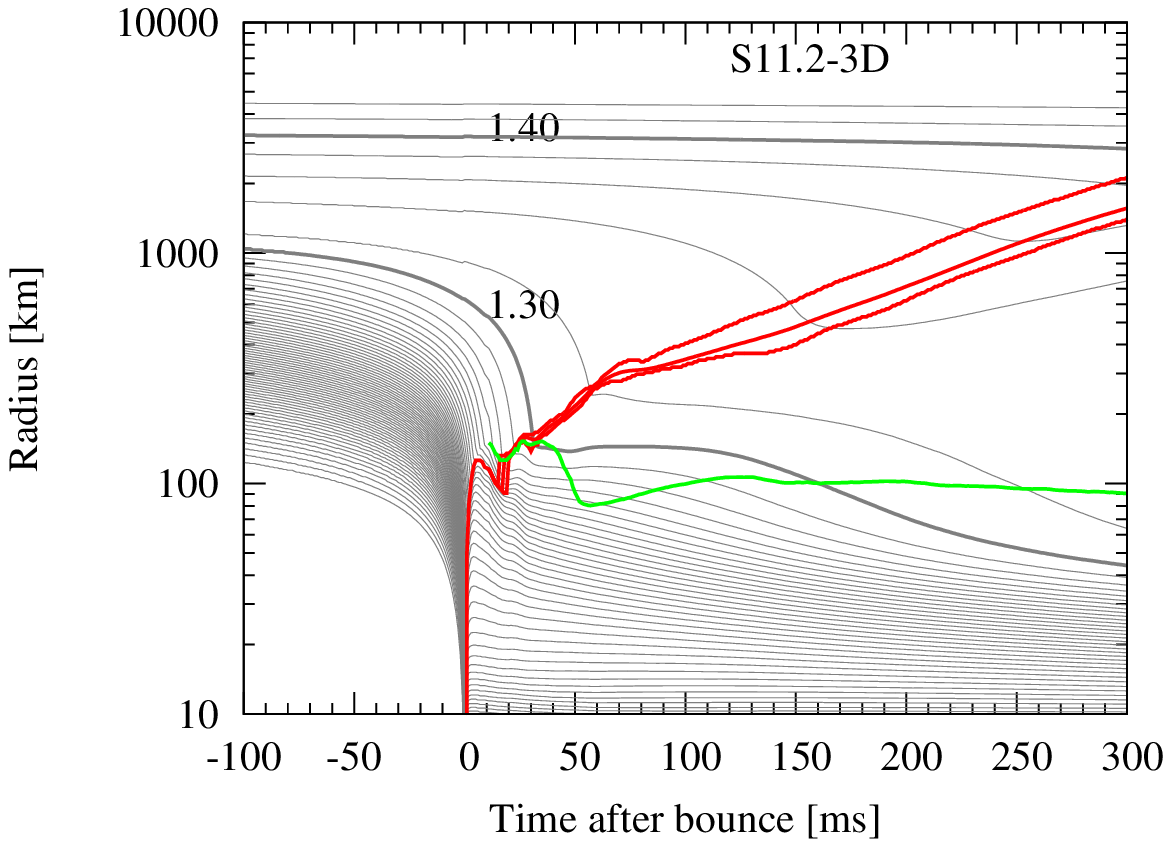}
    \includegraphics[width=.35\linewidth]{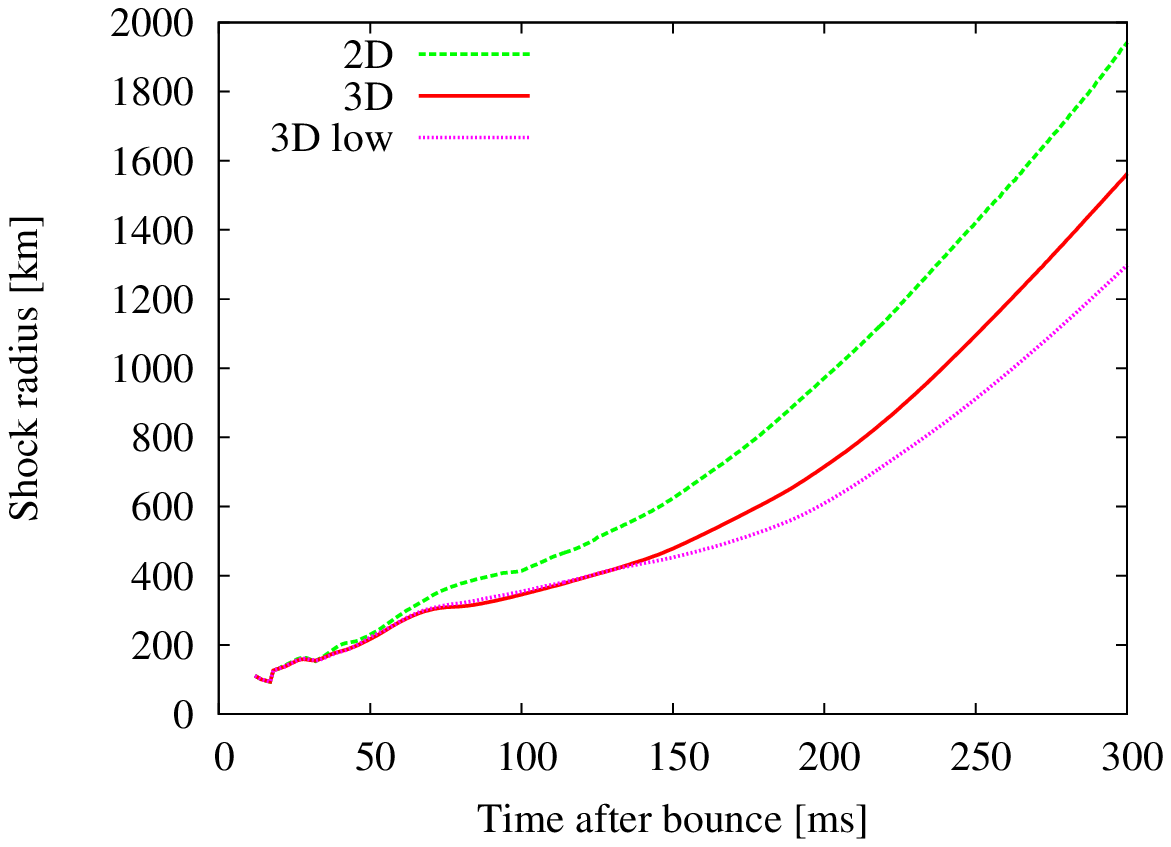}
 \caption{Time evolution of the 3D model, visualized by mass shell trajectories 
in thin gray lines (left panel).
  Thick red lines show the position of shock waves, noting that the maximum (top),
 average (middle), and the minimum (bottom) shock position are shown, respectively. 
 The green line represents the shock position of the 1D model. "1.30" and 
 "1.40" indicates
 the mass in unit of $M_{\odot}$ enclosed inside the mass-shell. Right panel shows 
 the evolution of average shock radii for the 2D (green line), 
 and 3D (red line) models. The "3D low" (pink line) corresponds to the low resolution 3D 
model, in which the mesh numbers are taken to be half of the standard model
 (see Section 2).}
\label{f7}
\end{figure*}

Figure \ref{f6} shows the blast morphology for our 3D (left panel), 2D (middle), 
 and 1D (right) model, respectively. In the 2D model (middle panel),
 the morphology is symmetric around the coordinate symmetry axis.\footnote{Note that the 
 polar axis is tilted (about $\pi/4$) both in the left and middle panel.}
In contrast, non-axisymmetric structures are clearly shown in the 3D model (left panel).
 The direction of explosion is rather closely aligned with the polar axis 
in the 3D model.
 Owing to the use of the spherical coordinates, we cannot omit the possibility that 
the polar axis still gives a special direction in our 3D simulations.
  However we suspect that the 
alignment might be just an accident, because the axis-free 3D explosions were obtained 
 in a number of parametric 3D explosion 
models by using the same hydro-code (e.g., \citet{iwakami1,iwakami2,kotake09,kotake11}). 
  To clearly witness the stochastic natures concerning the explosion direction,
 we may need to investigate a number of 
  3D models by changing initial perturbations and numerical resolutions
 systematically, which we think as an important extension of this study 
(Takiwaki et al. in preparation).

 The left panel of Figure \ref{f7} shows mass-shell trajectories 
 for the 3D (red lines) and 1D model (green line), respectively.
 At around 300 ms after bounce, the average shock radius for the 3D model exceeds
 1000 km in radius. On the other hand, an explosion is not obtained 
 for the 1D model, which is in agreement with \citet{buras06}. 
The right panel of Figure \ref{f7} shows 
 a comparison of the average shock radius vs. postbounce time. In the 2D model, 
the shock expands rather 
continuously after bounce. This trend is qualitatively consistent with the 2D result by
 \citet{buras06} (see their Figure 15 for model 
s112\underline{\hspace{0.4em}}128\underline{\hspace{0.4em}}f), however
 the average shock of our 2D model expands much faster than theirs.
 We suspect that all of the neglected effects in this work including 
 general relativistic effects, inelastic neutrino-electron scattering, and cooling 
by heavy-lepton neutrinos, could give a more optimistic condition to produce explosions.
 Apparently these ingredients should be appropriately implemented, which we hope 
 to be practicable in the next-generation 3D simulations.

 Comparing the shock evolution between our 2D (green line in 
 the right panel of Figure \ref{f7}) and 3D model (red line), 
the shock is shown to expand much faster for 2D.
 The pink line labeled by "3D low" is for the low resolution 3D 
model, in which the mesh numbers are taken to be half of the standard model
 (see Section 2). Comparing with our standard 3D model (red line),
  the shock expansion becomes less energetic for the low resolution model
 (later than $\sim 150$ ms). Above results indicate that explosions are easiest 
to obtain in 2D, followed in order by 3D, and 3D (low).
  At first sight, this may look contradicted with the finding by 
 \citet{nordhaus} who pointed out that explosions could be more easily obtained 
 in 3D than 2D. In the following section, we proceed to discuss what is the 
reason of the discrepancy more in detail.

\subsection{Comparison between 2D and 3D}\label{part3}
In this section, we move on to illuminate the key differences between our 2D and 3D models.
 For the purpose, we highlight the SASI (section \ref{331}) and convective activities 
(section \ref{333}),  the residency (section \ref{332}) and neutrino-heating timescales 
 (section \ref{334}), respectively.

\begin{figure*}[htbp]
    \centering
    \includegraphics[width=.40\linewidth]{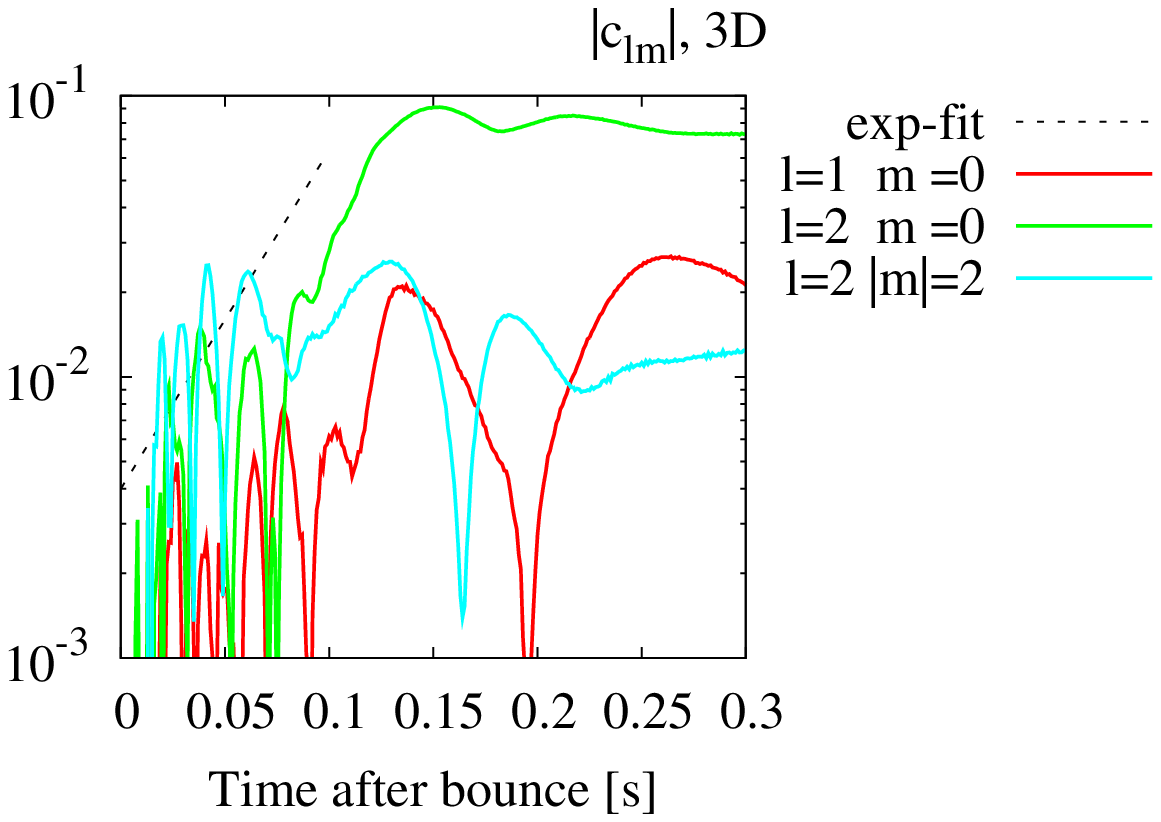}
    \includegraphics[width=.40\linewidth]{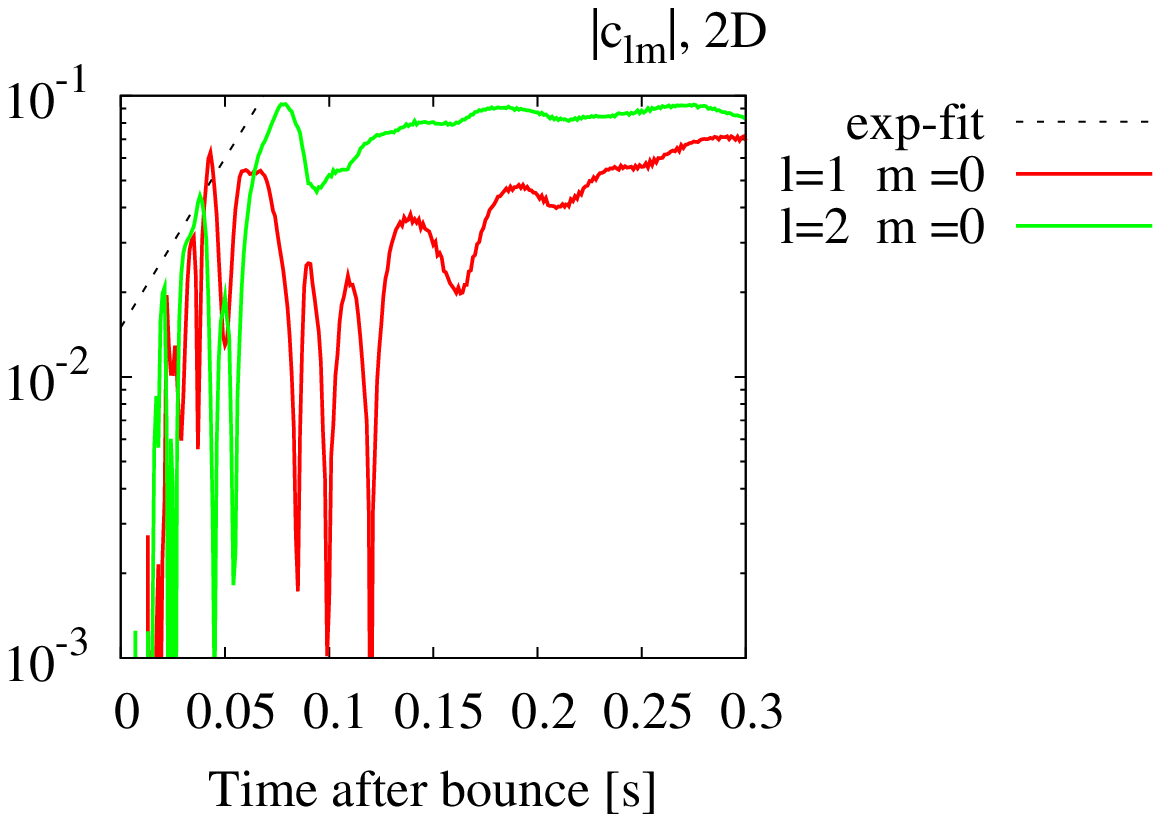}\\
    \includegraphics[width=.40\linewidth]{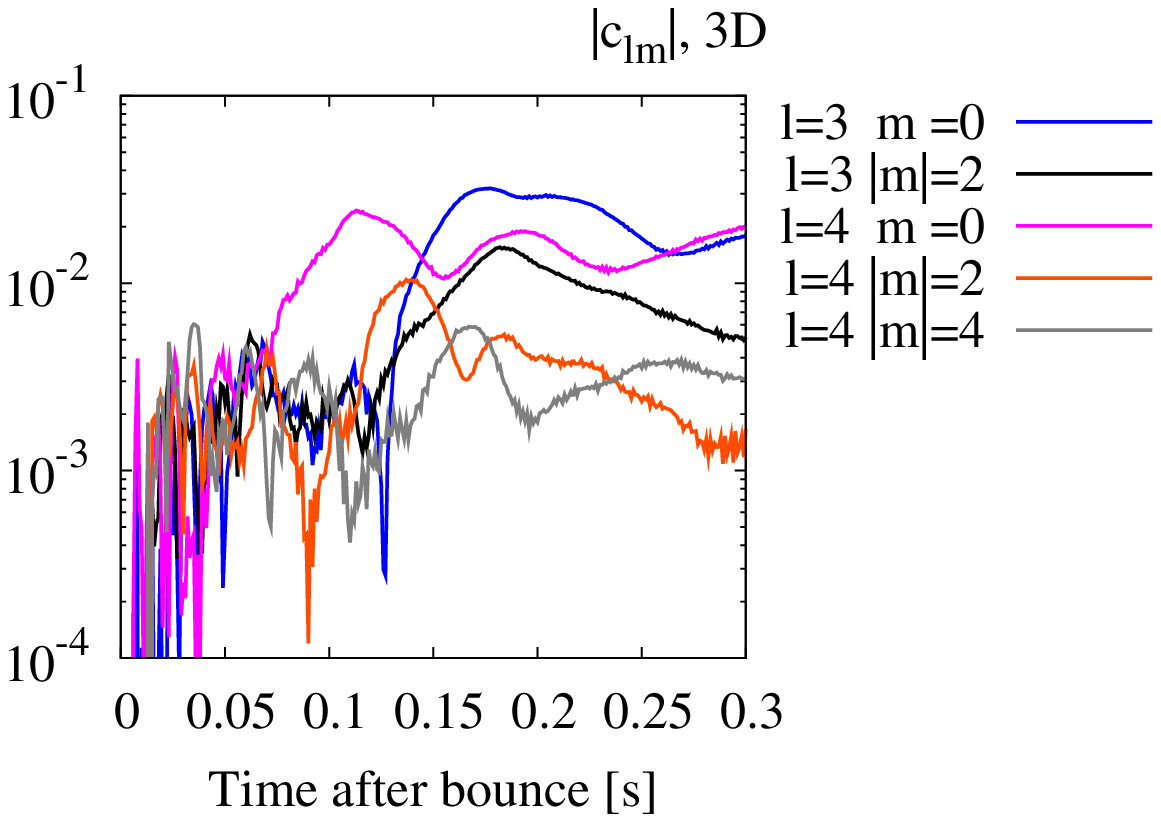}
    \includegraphics[width=.40\linewidth]{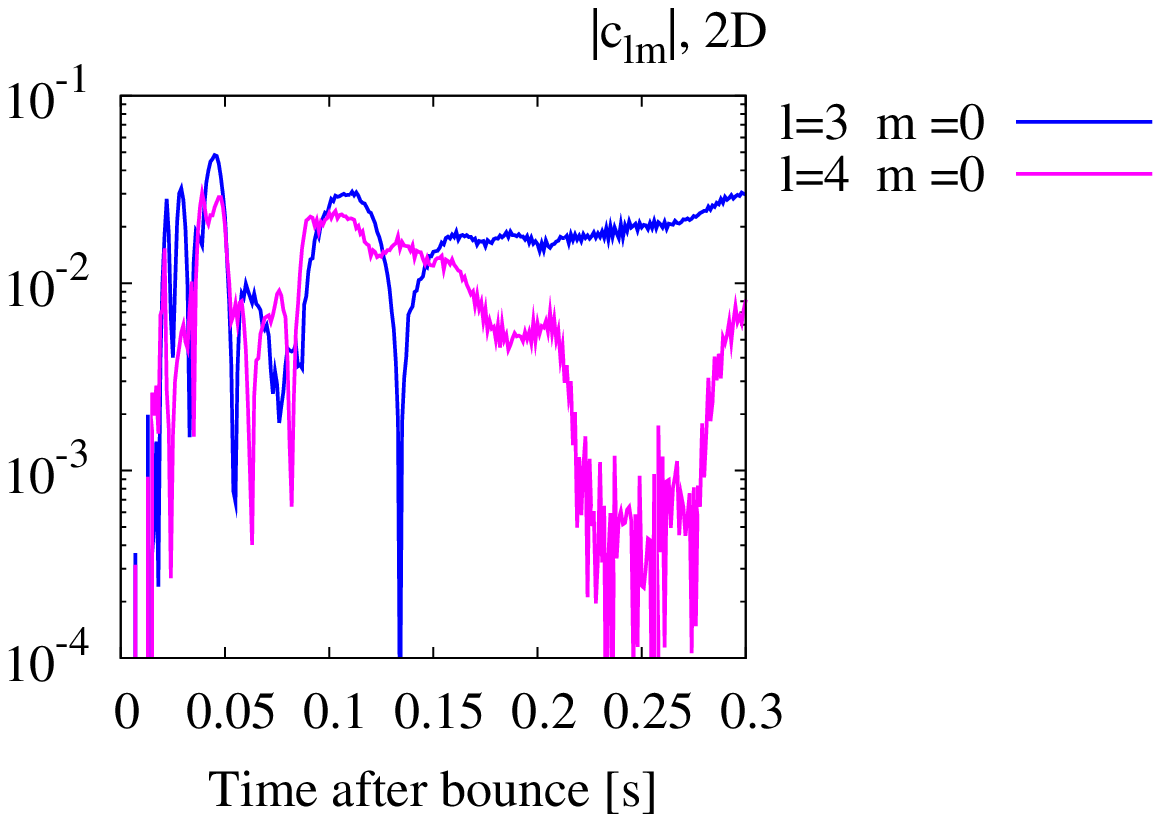}\\
 \caption{Time evolution of the normalized amplitudes $|c_{lm}/c_{00}|$
 in our 3D (left panels) and 2D (right panels) model, respectively. Lower
 and higher modes are selected in top and bottom panels. Note that the time in 
 the figure is measured after bounce. 
 The black dotted lines labeled by "exp-fit" in the top two panels 
 indicate the linear growth rate of the SASI 
 (see text for more details).
}
\label{f8}
\end{figure*}

\subsubsection{SASI activities in 2D and 3D}\label{331}
To compare the SASI activities in 2D and 3D, 
we first perform the mode analysis of the shock wave.
The deformation of the shock surface can be expanded as a linear combination of the
 spherical harmonics components $Y_{lm} (\theta, \phi)$:
\begin{equation}
R_S(\theta, \phi) = \sum^{\infty}_{l=0} \sum^{l}_{m=-l} c_{lm} \, Y_{l m}(\theta, \phi),
\nonumber
\label{eq:eq10}
\end{equation}
where $Y_{lm}$ is expressed by the associated Legendre polynomial $P_{lm}$ and 
a constant $K_{lm}$ given as
\begin{equation}
Y_{lm} = K_{lm} P_{lm}(\cos \theta) \, e^{im\phi},
\nonumber
\end{equation}
\begin{equation}
K_{lm} = \sqrt{\frac{2l+1}{4\pi}\frac{(l-m)!}{(l+m)!}}.
\nonumber
\end{equation}
Here the expansion coefficients read,
\begin{equation}
c_{lm}=\int^{2\pi}_0 \! \! \! \! d\phi  \! \int^{\pi}_0 \! \! d\theta \, \sin \theta \, R_S(\theta, \phi) \, Y_{lm}^{*} (\theta, \phi),
\label{coefficient}
\end{equation}
where the superscript * denotes complex conjugation.

Figure \ref{f8} shows the time evolution of the expansion coefficients (Equation
 (\ref{coefficient})) for the 3D (left panel) and 2D model (right panel), respectively.
 As can be seen, the amplitude of each mode grows exponentially 
until $\sim$ 100-150 ms after bounce, which corresponds to the linear
 SASI phase.

 The top panels show that the mode of ($\ell,m$)=(2,0) (green line) 
is dominant when the SASI enters to the saturation phase, which is common both 
 in our 2D and 3D models.
 The epoch when the SASI shifts from the linear to non-linear phase 
 is much delayed for 3D ($t \gtrsim 150$ ms) than 2D ($t\gtrsim 80$ ms),
 which was also seen in the parametric 3D models by 
\citet{iwakami1} (e.g., their Figure 9). These transition timescales 
 are also consistent with \citet{buras06} who employed the same progenitor model
 as ours in their 2D simulations in which detailed neutrino transport was solved
(see their Figure 22).
 It is also worth mentioning that the timescale seems rather insensitive
 to the employed progenitor. In fact, Figure 5 in \citet{marek} shows the
 transition timescale to be around 150 ms for a 15 $M_{\odot}$ progenitor
 model.
 In the bottom panels of Figure \ref{f8},
  the saturation levels of the even modes of ($\ell,|m|$)= (4,0), (4,2), (4,4) in
 3D are shown to become much larger than those in 2D (pink line), while the odd mode of ($\ell,m$)= (3,0) is much 
the same. 

Based on the pioneering work by \citet{houck92},
 the linear growth rate of the SASI in core-collapse case 
 was presented by \citet{scheck08}. They pointed out
 that the cycle efficiency (:$Q$) which
  represents how many times the average radius expands 
 compared to the original position per a unit oscillation frequency (:$\omega_{\rm osc}$)
 of the SASI, is an important quantity to characterize the linear growth rate.
 From Figure 8, $Q$ and $\omega_{\rm osc}^{-1}$
 in our simulation are approximately estimated 
 to be 2 and 25 ms, respectively. Note that these values are in agreement with the 
ones obtained in 2D simulations by \citet{scheck08} (e.g., their Figure 17).
 From the two quantities, the linear growth rate 
can be straightforwardly estimated as $\exp(ln(Q)\,t\, \omega_{\rm osc}$), 
 which is shown in the top panels of Figure 8 as black-dotted lines.
 As can be seen, the growth rates observed both in our 3D (top left panel in 
 Figure 8) and 2D simulations (top right) are close to the linear 
 growth rate, which seems a rather generic trend for the 
 low-modes ($\ell = 1,2$) of the SASI. Note here that 
the normalized amplitude of the shock (the value in the vertical axis of Figure 8) 
 shortly after bounce is actually so small as $10^{-6}$ to $10^{-5}$.
 Deduced from the linear growth rate above, the amplification due to the SASI 
in the linear phase is at most $\sim 20$ within 100 ms after bounce.
 On the other hand, the amplitudes do increase more than about $10^{-1}/10^{-5} \sim
 10^4$ for the epoch. So the SASI is not the only agent for the shock deformation.
 In fact the amplitudes are observed to sharply increases from $10^{-5}$ to $\sim 
 5 \times 10^{-3}$ within 10 ms after bounce, which is predominantly 
 driven by the Rayleigh-Taylor instability behind the shock.
 To summarize, our results suggest that the rapid deformation after bounce
 is triggered by the Rayleigh-Taylor instability and 
the subsequent deformation with much milder growth rates is predominantly 
determined by the SASI.

 Concerning the saturation levels of the dominant mode of ($\ell,m$)=(2,0) between
 2D and 3D,
 it is slightly larger for 2D than 3D (top panels, green line). The dominance of 
 this bipolar mode can be also seen in the blast morphology (Figure 6). 
The second-order mode of 
($\ell,m$)=(1,0) is shown to be much smaller for 3D than 2D (red lines in 
 the top panels in Figure \ref{f8}).
 This is qualitatively consistent with \citet{nordhaus} who did not observe the 
 dominance of the ($\ell,m$)=(1,0) mode in their 3D models.
 Note that this agreement might be just by chance.
 In \citet{iwakami1}, they observed the 
 dominance of ($\ell,m$)=(1,0) mode in the saturation phase (see their Figure 12).
 From 3D results reported so far \citep{iwakami1,iwakami2,annop,fern},
 it seems almost certain that the low modes ($\ell$=1,2) are 
dominant in the 3D SASI-aided neutrino-driven explosions. 
However it might be rather uncertain
which of the two modes  ($\ell$=1 or 2) becomes dominant. This may reflect
 the fact that the explosion dynamics in 3D proceeds totally stochastically.

\begin{figure*}[htbp]
    \centering
    \includegraphics[width=.35\linewidth]{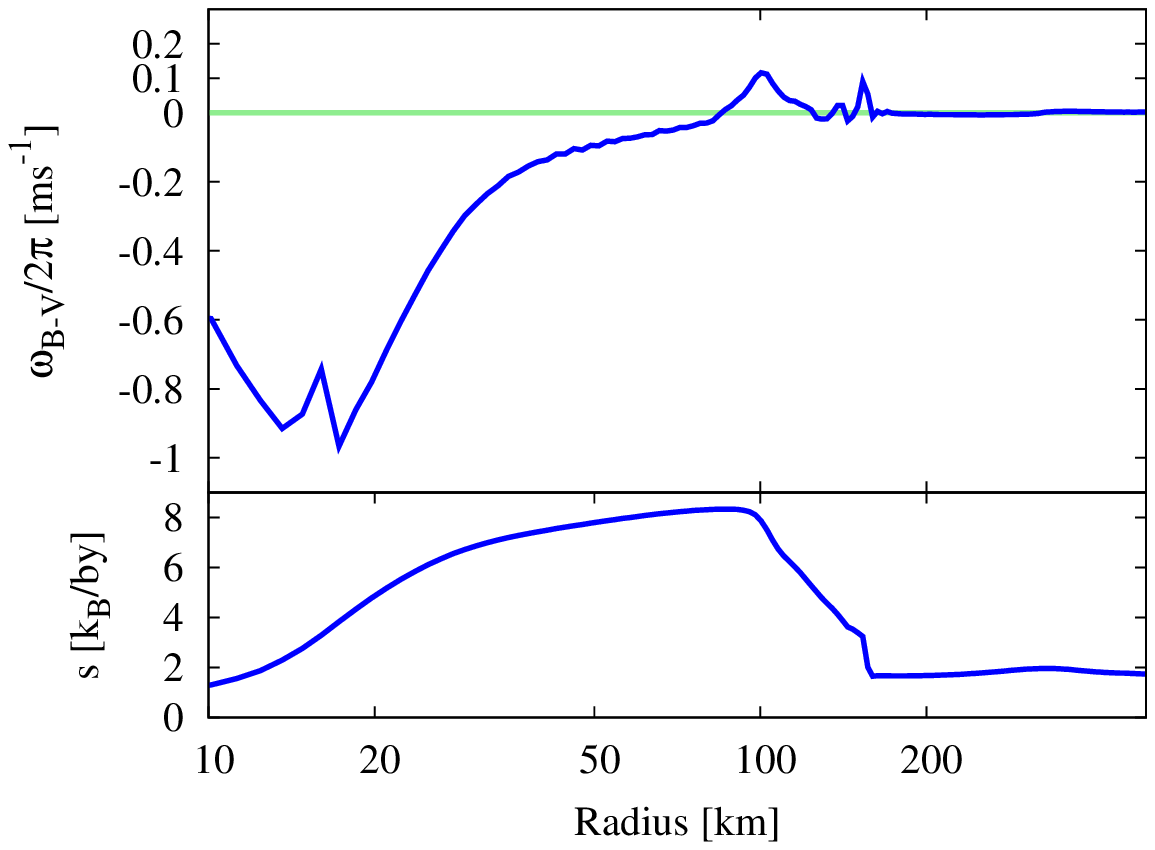}
    \includegraphics[width=.35\linewidth]{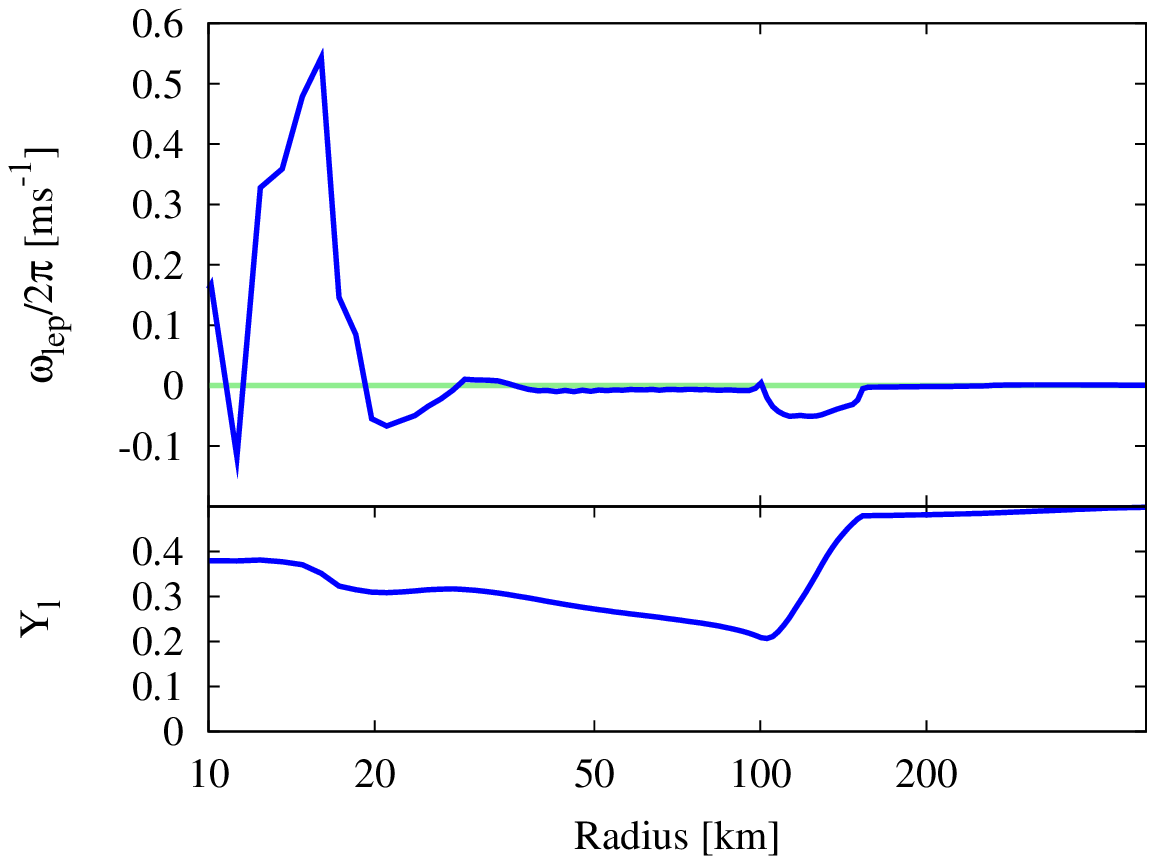}
\caption{Profile of the Brunt-V\"ais\"al\"a (B-V) frequency (left top panel) and 
 entropy (left bottom pane) at 10 ms after bounce for our 3D model. The right panel 
 shows the B-V frequency contributed only from the lepton gradient.
 Note that angle-averaged 
quantities are plotted here. The horizontal green line in the top panels 
is shown just for the reference
 of $\omega_{\rm B-V} = 0$.
 }
 \label{f10a}
\end{figure*}

\begin{figure*}[htbp]
    \centering
    \includegraphics[width=.35\linewidth]{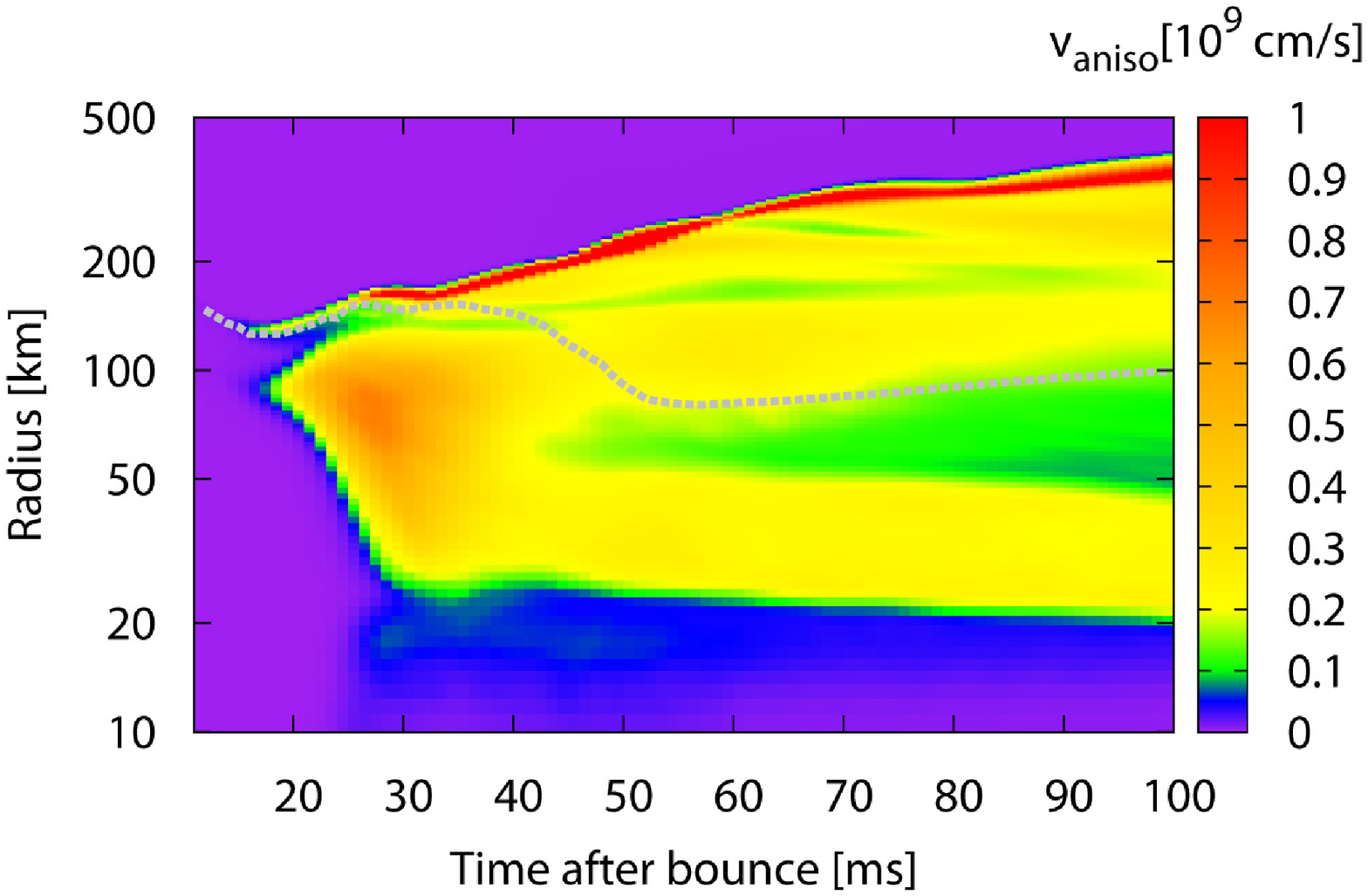}
    \includegraphics[width=.35\linewidth]{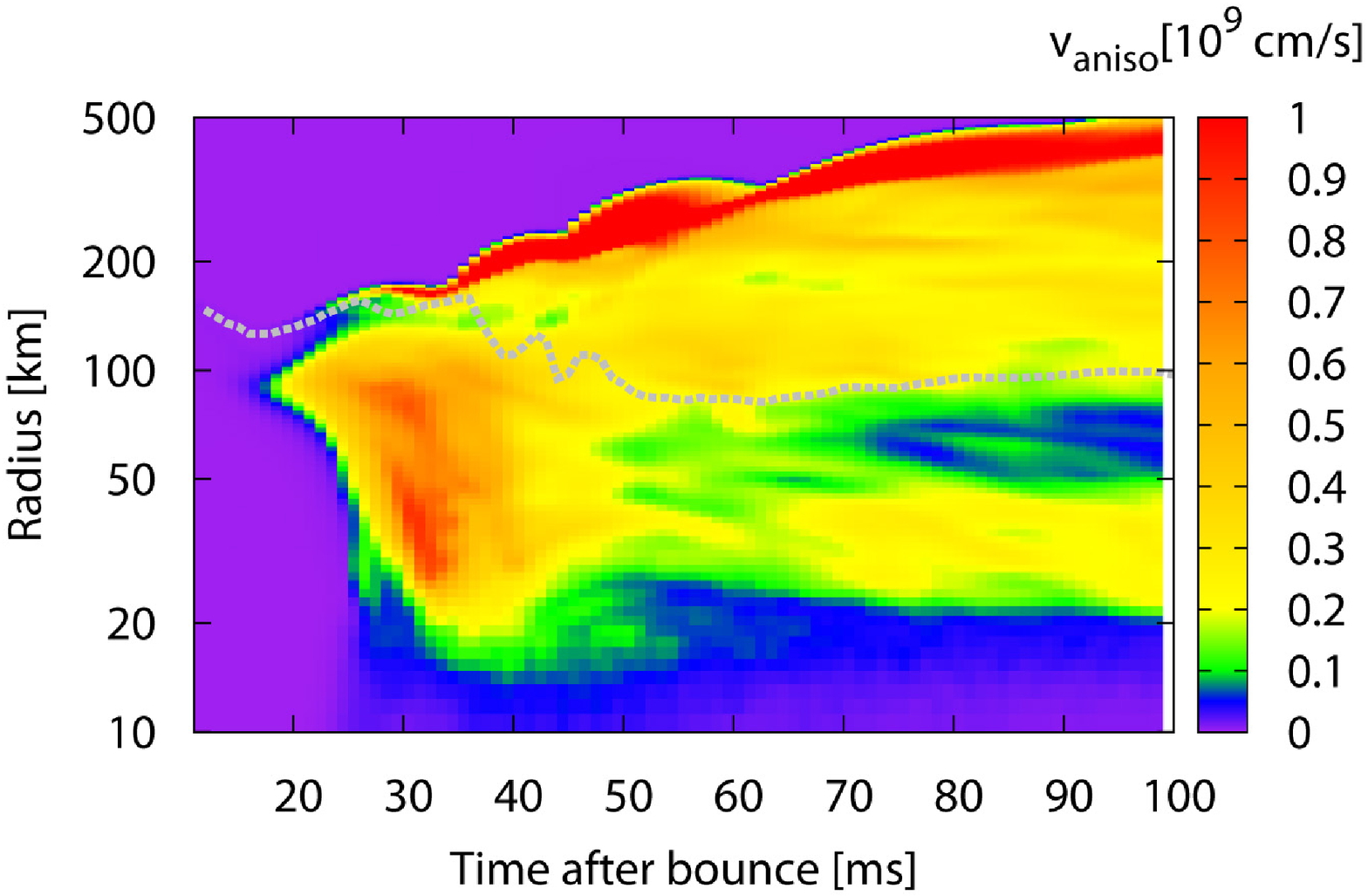}\\
    \includegraphics[width=.35\linewidth]{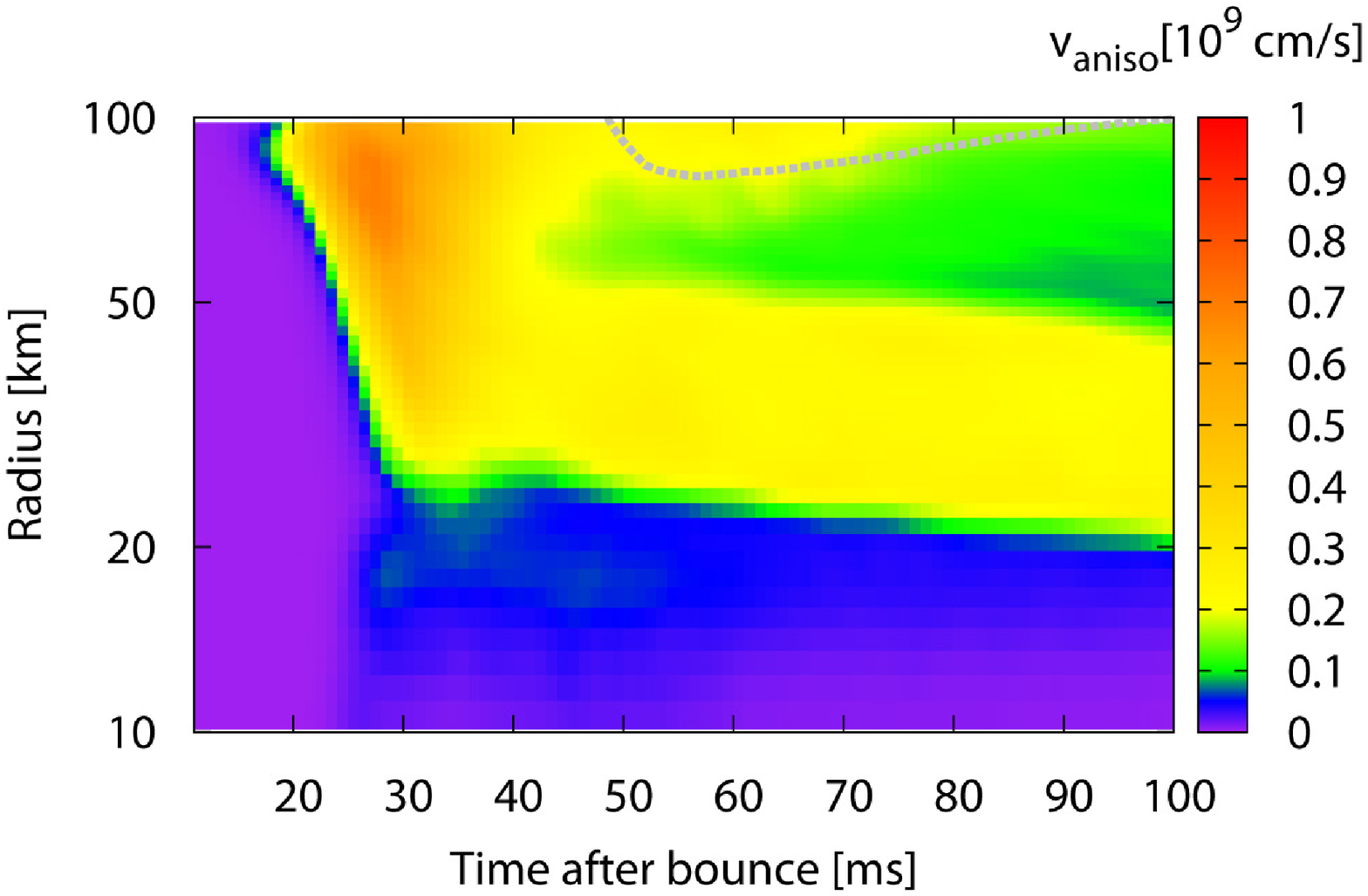}
    \includegraphics[width=.35\linewidth]{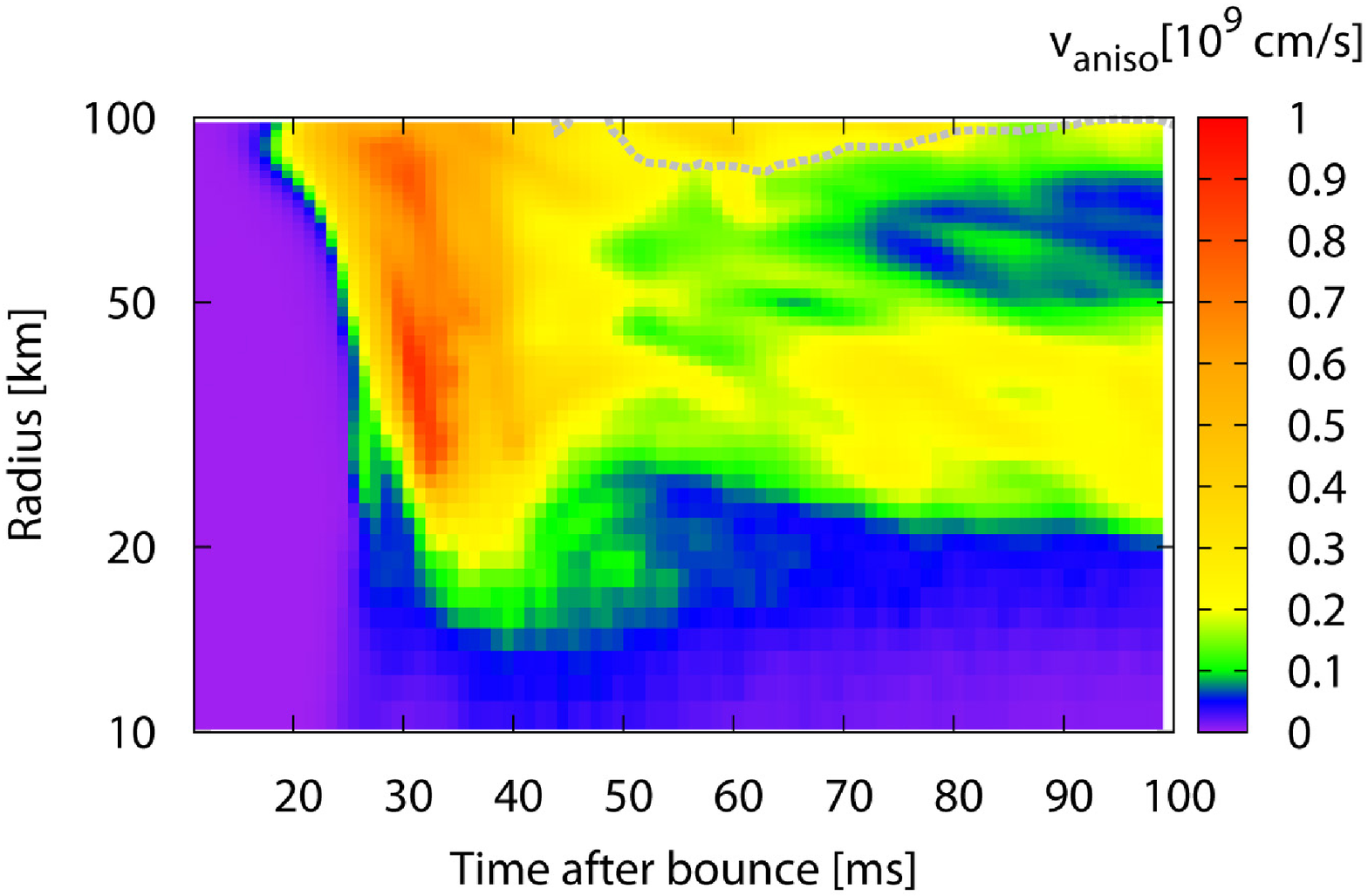}\\
 \caption{Evolution of convective activities in our 3D (left) and 2D (right) model,
 respectively. The color-scale is for anisotropic velocity
 (see text for more details), in which higher anisotropy (colored by yellow to red) 
 is related to active convective overturns.
 The dotted gray line represents the gain radius. The bottom panels
 are just zoom up of the top panels focusing on the central region.}
\label{f10b}
\end{figure*}

\subsubsection{Convective activities in 2D and 3D} \label{333}

To discuss convective activities, we compute the Brunt-V\"ais\"al\"a (B-V) frequency 
 which is defined as (e.g., \citet{buras06}), 
\begin{equation}
\omega_{\mathrm{B-V}} =
\mathrm{sgn}\left(C_{\mathrm{L}}\right)\sqrt{
\left|\frac{C_{\mathrm{L}}}{\rho}\frac{\mathrm{d}\Phi}{ \mathrm{d} r} \right|}
\end{equation}
 with $ d \Phi/ d r$ being the local gravitational acceleration. $C_{L}$ is the 
 Ledoux-criterion, which is given by
\begin{equation}
C_{\mathrm{L}}= - \left(\frac{\partial \rho}{\partial P}\right)_{s,Y_l}
\left[
\left(\frac{\partial P}{\partial s}\right)_{\rho,Y_l}
\left(\frac{\mathrm{d} s}{\mathrm{d} r}\right)
+
\left(\frac{\partial P}{\partial Y_l}\right)_{\rho,s}
\left(\frac{\mathrm{d} Y_l}{\mathrm{d} r}\right)
\right],
\end{equation}
with $Y_l$ being the lepton fraction.
 It predicts instability in static layers if $C_{\rm L} > 0$.
 The B-V  frequency denotes the linear growth rate of fluctuations,
 if it is positive (instability). If it is negative (stable), it denotes the negative 
 of the oscillation frequency of stable modes.

The left panel of Figure \ref{f10a}
 shows the profile of the B-V frequency for our 3D model at 10 ms 
after bounce. The negative entropy gradient (bottom left panel) 
between the gain radius ($\sim 100$ km in radius) 
and the stalled shock ($\sim 160$ km) 
 makes there convectively unstable. The region in the vicinity of the PNS 
 ($10 -20$ km in radius, bottom right panel) has a negative lepton gradient, 
 which can make there convectively unstable (the top right panel). However this 
 region turns out to be convectively stable due to positive entropy gradient (compare
 bottom left panel). Both in our 2D and 3D models, the convectively unstable regions 
 persist only behind the stalled shock triggered by the negative 
 entropy gradient.
 
Figure \ref{f10b} shows evolution of convective activities for the 3D (left) and 
2D (right) model, respectively. To measure the strength of convective activities,
 we define the anisotropic velocity as,
\begin{equation}
 v_{\mathrm{aniso}}=
\sqrt{\langle\rho \left((v_r-\langle v_r \rangle)^2+v_\theta^2+v_\phi^2\right)\rangle/
\langle\rho\rangle}.
\end{equation}
 By this definition, higher anisotropy comes from greater deviation 
 in the radial motions ($v_r - \langle v_r \rangle$) or larger 
 non-radial ($v_\theta,~v_{\phi}$) motions.

 From the top left panel of Figure \ref{f10b}, 
the initial formation of convectively unstable 
 regions is shown to be around 15 ms after bounce (seen as a 
 sudden formation of the non-zero $v_{\rm aniso}$). Subsequently, the convectively 
 unstable regions are shown to advect to the center. At around $20-30$ km in radius,
 the anisotropic velocities are strongly suppressed (seen as a change from
 yellowish to bluish region at $\sim$ 30 ms after bounce) 
due to the stabilizing positive entropy gradient (see
 the left panel of Figure \ref{f10a}). As a result, the convective overturns
 are shown to persistently stay in the regions above the PNS ($\sim 20-30$ km in radius)
 and below $\sim 50$ km in radius. This is seen as a (horizontal) yellow stripe 
in the bottom left panel.
 Since the infalling velocities below the gain region (dotted gray line)
 are so high that the convectively unstable material cannot stay there for long.
 This may be 
 the reason that the anisotropic velocity becomes relatively low (seen as 
 greenish in the bottom left panel) between the gain
 radius ($\lesssim 100$ km) and the upper position of the yellow strip 
($\sim 50$ km). These overall trends obtained in the 3D model are common to 2D
 (right panels). In 2D, a more drastic overshooting of the convectively unstable material
 to the convectively stable region is seen (compare the bottom panels for 20 - 40 ms 
 after bounce). The area of the brought-in convectively unstable region 
(equivalently the yellowish stripe) is shown to be larger for 3D than 2D. 
 Such a vigorous convective overturn in our 3D model becomes
 essential in analyzing the neutrino-heating timescales
 later in section \ref{334}.

Having referred to the SASI and convective activities in 2D and 3D, we are 
 now ready to perform analysis of the residency and 
 neutrino-heating timescales. First of all, we discuss the residency timescale in
 the next section.

\begin{figure*}[htbp]
    \centering
    \includegraphics[width=.35\linewidth]{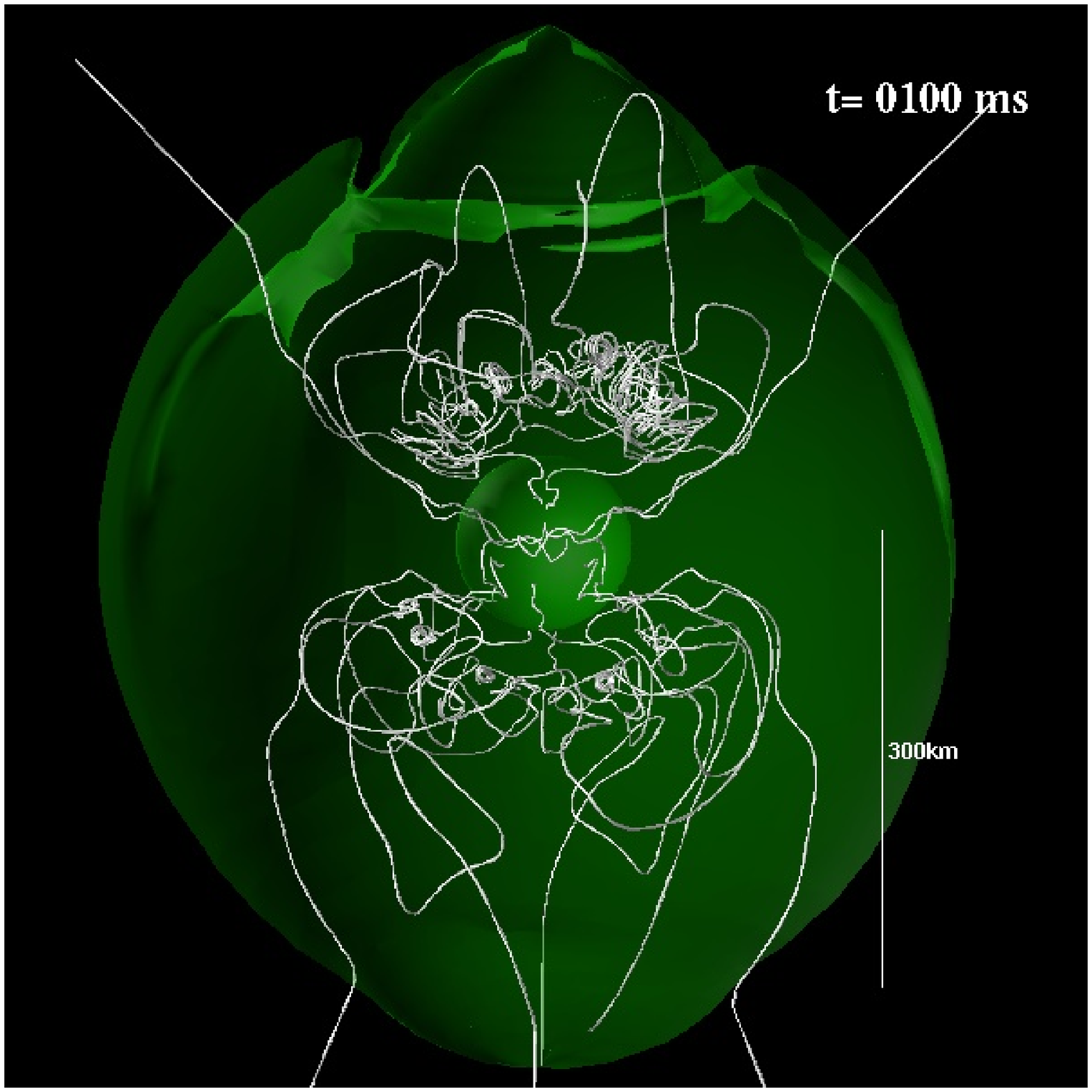}
    \includegraphics[width=.35\linewidth]{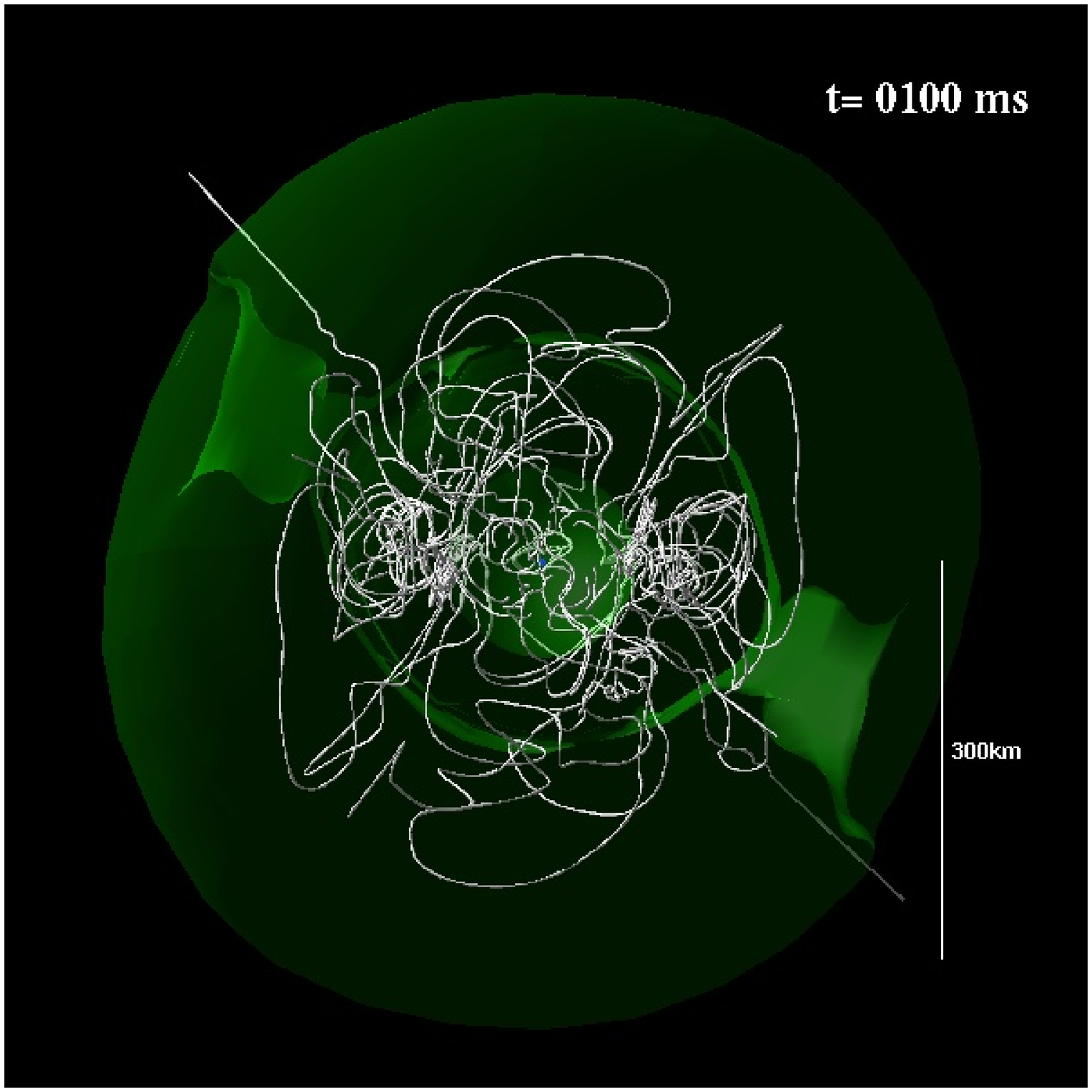}
 \caption{Streamlines of selected tracer particles advecting through the shock wave 
to the PNS for the 3D model. As in Figure \ref{f4}, the left and right panel is for 
the equatorial and polar observer, respectively. Each panel shows several
 surfaces of constant entropy marking the position 
 of the shock wave (greenish outside) and the PNS (indicated by the 
central sphere). The linear scale is given in the right bottom edge 
 of each panel.}
\label{f9}
\end{figure*}

\begin{figure*}[htbp]
    \centering
    \includegraphics[width=.35\linewidth]{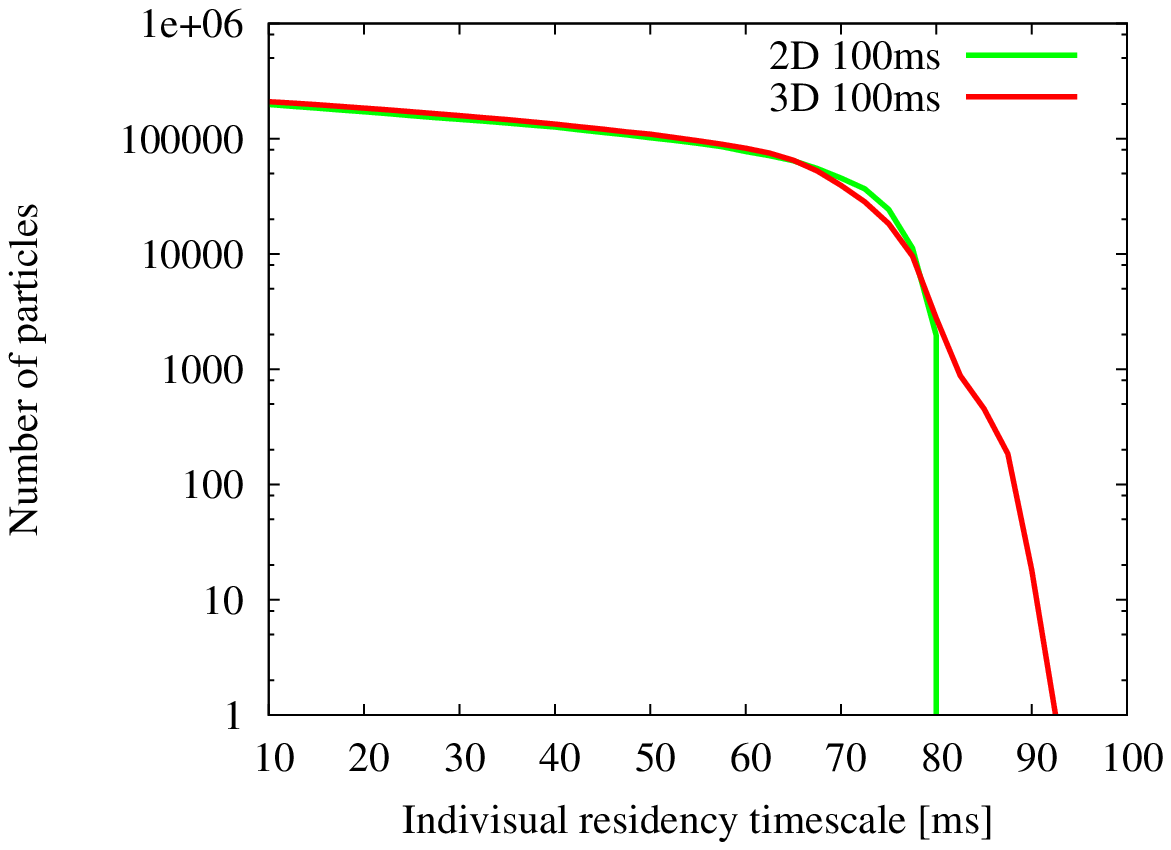}
    \includegraphics[width=.35\linewidth]{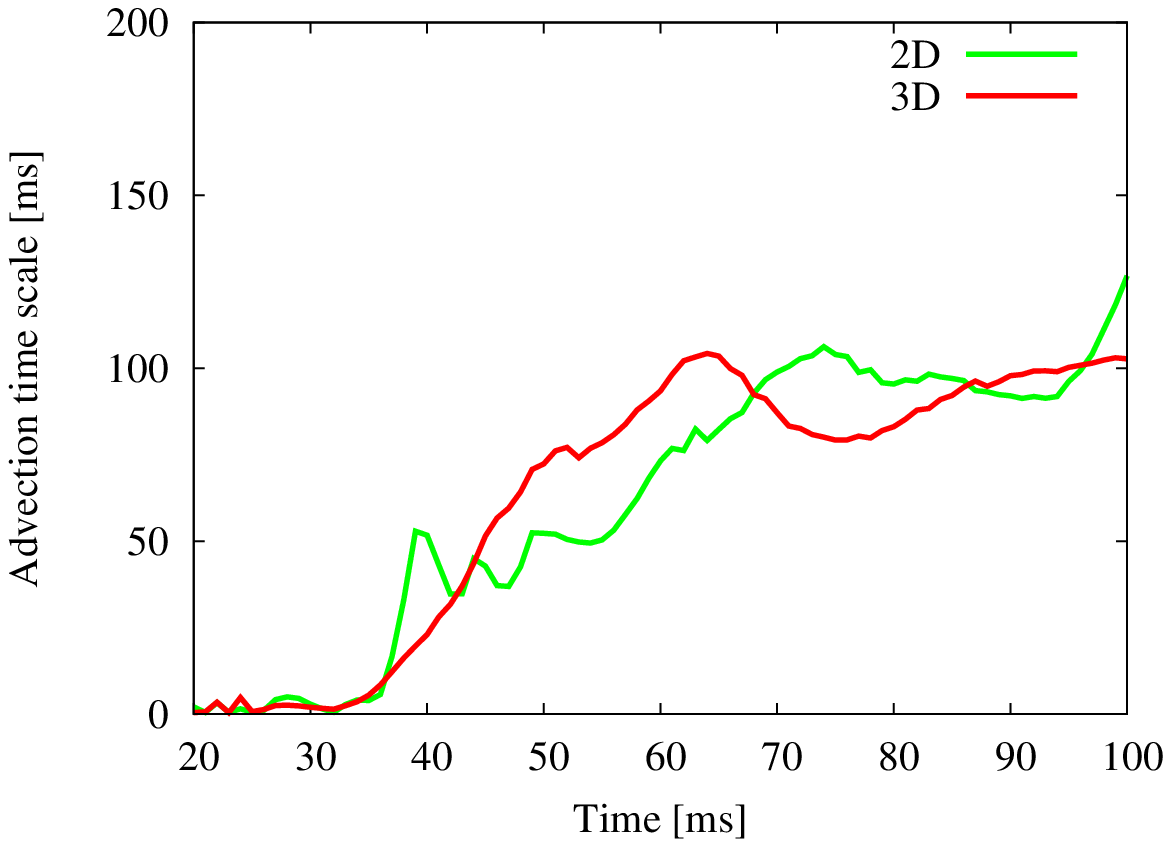}
 \caption{Comparison between our 2D and 3D model at 100 ms after bounce, showing
the number of tracer-particles travelling
 in the gain region as a function of their individual
 residency time (left panel, see text for details),
 and the average advection timescale 
 as a function of the postbounce time (right panel). }
\label{f10}
\end{figure*}

\subsubsection{Residency timescales in 2D and 3D}\label{332}

Figure \ref{f9} depicts the streamlines of tracer particles advecting 
from the outer boundary of the computational domain through the shock 
wave down to the PNS.  The number of the tracer particles that we actually 
injected is $\sim 10^{6}$, however only the trajectories of selected 
particles are shown in Figure \ref{f9} (not to make the figure filled with particles).
 As seen from the left panel, the tracer particles first go down to the 
 shock wave, which is shown by the radial straight lines. Later on, as indicated by the 
tangled streamlines, they experience turbulence before 
 falling to the PNS. The low-modes (here, $\ell =2$) oscillation of the 
accretion shock due to 
SASI activities (as discussed in section \ref{331}, e.g., right panel 
in Figure \ref{f9}) make the residency timescales much longer for multi-D models than 
1D. 
 If the right panel were for 2D models,  the streamlines would 
 be seen as a superposition of circles with different diameters.
In contrast, the non-axisymmetric matter motions can be clearly seen, which is a 
genuine 3D feature. 

The left panel of Figure \ref{f10} compares the number of the tracer particles 
vs. their individual residency timescales between the 2D and 3D model 
(for the same snapshot in Figure \ref{f9}).
 As seen, the maximum residency time is longer for 3D ($t_{\rm res}\sim 92$ ms)
 than 2D ($\sim 80$ ms), which is most likely to be the outcome of the 
non-axisymmetric matter motions in the gain region. As well known, 
the longer residency time is good for producing neutrino-driven explosions 
because of the long exposure to the neutrino heating in the gain region.

The right panel of Figure \ref{f10} shows the comparison of the 
 advection timescale, conventionally employed in literatures (e.g., Equation (4) in 
 \citet{marek}, that is $(R_s - R_g)/|\langle v_r \rangle|$
 with $R_s$, $R_g$, and $\langle v_r \rangle$ 
representing the angle average shock radius, gain radius, and postshock 
radial velocity, respectively. Against our anticipation, the 
 averaged advection timescale is not always longer for 3D.
  Before $\sim 70$ ms after bounce, our 3D model (red line) has generally 
longer advection timescales. However the advection timescale later on 
 can be longer for 2D.

 At around $\sim 70$ ms after bounce, the revived shock wave has already reached 
at a radius of $\sim$ 400 km for 2D and $\sim 320$ km for 3D, respectively
(see the right panel of Figure \ref{f7}). In such a shock expansion phase,
 the definition of the ``advection timescale'' would become rather vague.
 For example, the advection timescale is longer for 2D than 3D at $t=100$ ms 
(right panel of Figure \ref{f10}), however this simply reflects larger shock 
radii for 2D than 3D (e.g., right panel of Figure \ref{f7}).\footnote{
In some sense artificially, this makes the advection timescale 
longer by the larger distances between $R_s$ and $R_g$, however this does not
 evidently mean that the 2D model can gain much more efficient neutrino-heating.} 
 Also in the above residency-time analysis, the longer residency time
 for 2D can be seen around $t_{\rm res} =70-80$ ms in the left panel of Figure \ref{f10} (seen 
 as a dominance of the green line over the red line).
 If the onset time of explosion 
could be much delayed after bounce (such as $\sim$ 600 ms as in \citet{marek}), 
  the advection(or residency)-timescale analysis between 2D and 3D could have been made 
  clearer in the long-lasting bubbling phase.
 In order to see the 3D effects more clearly, we plan to employ 
a more massive progenitor (such as $15 M_{\odot}$) as a follow-up of this study.

\begin{figure*}[htbp]
    \centering
    \includegraphics[width=.35\linewidth]{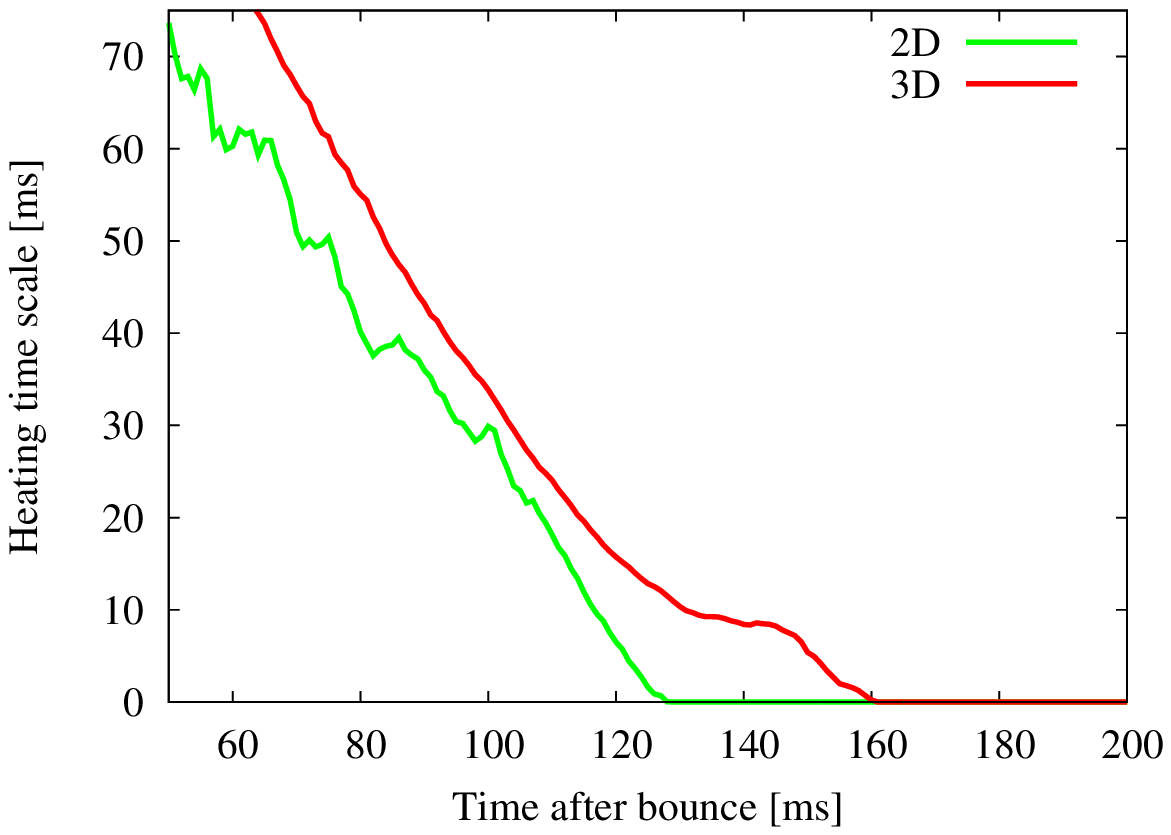}
    \includegraphics[width=.35\linewidth]{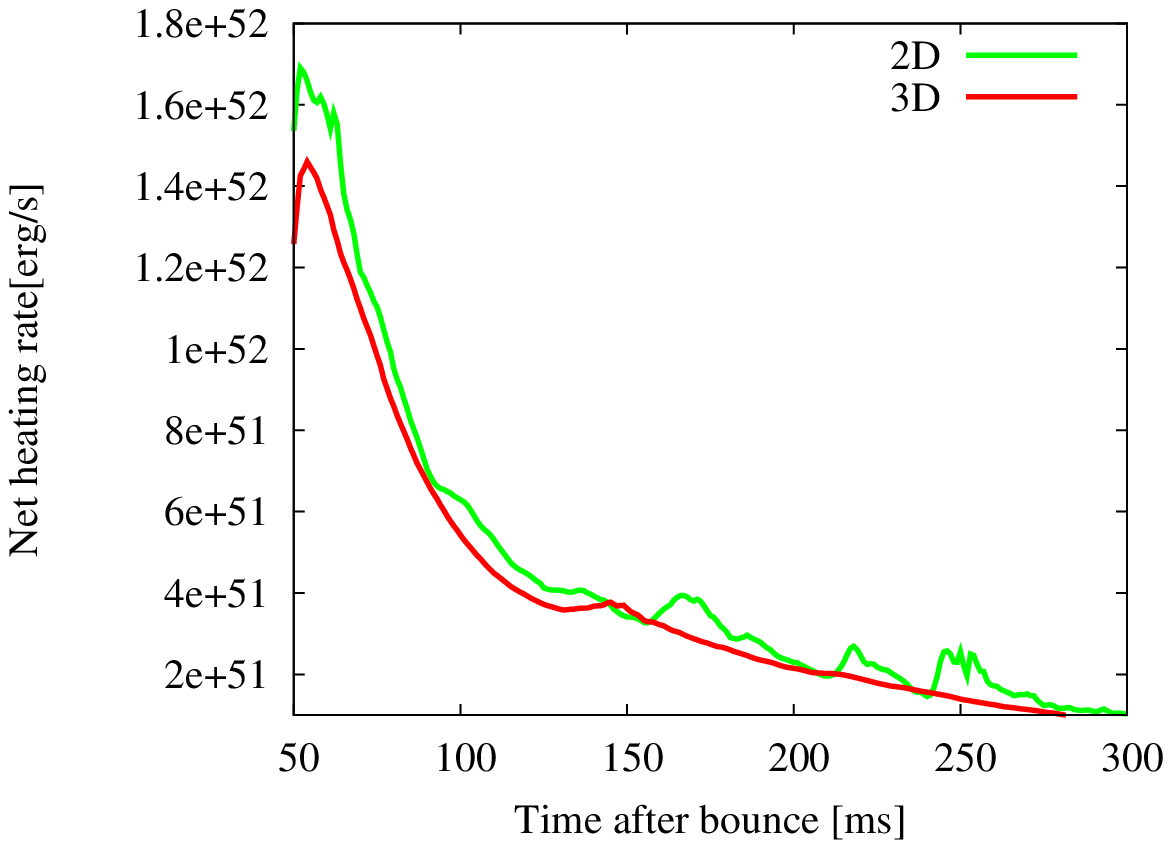}\\
 \caption{Time evolutions of neutrino-heating timescale (left) 
and total net rate of neutrino heating (right) in our 2D (green line) 
and 3D model (red line). }
\label{f11}
\end{figure*}

\begin{figure*}[htbp]
    \centering
  \includegraphics[width=.35\linewidth]{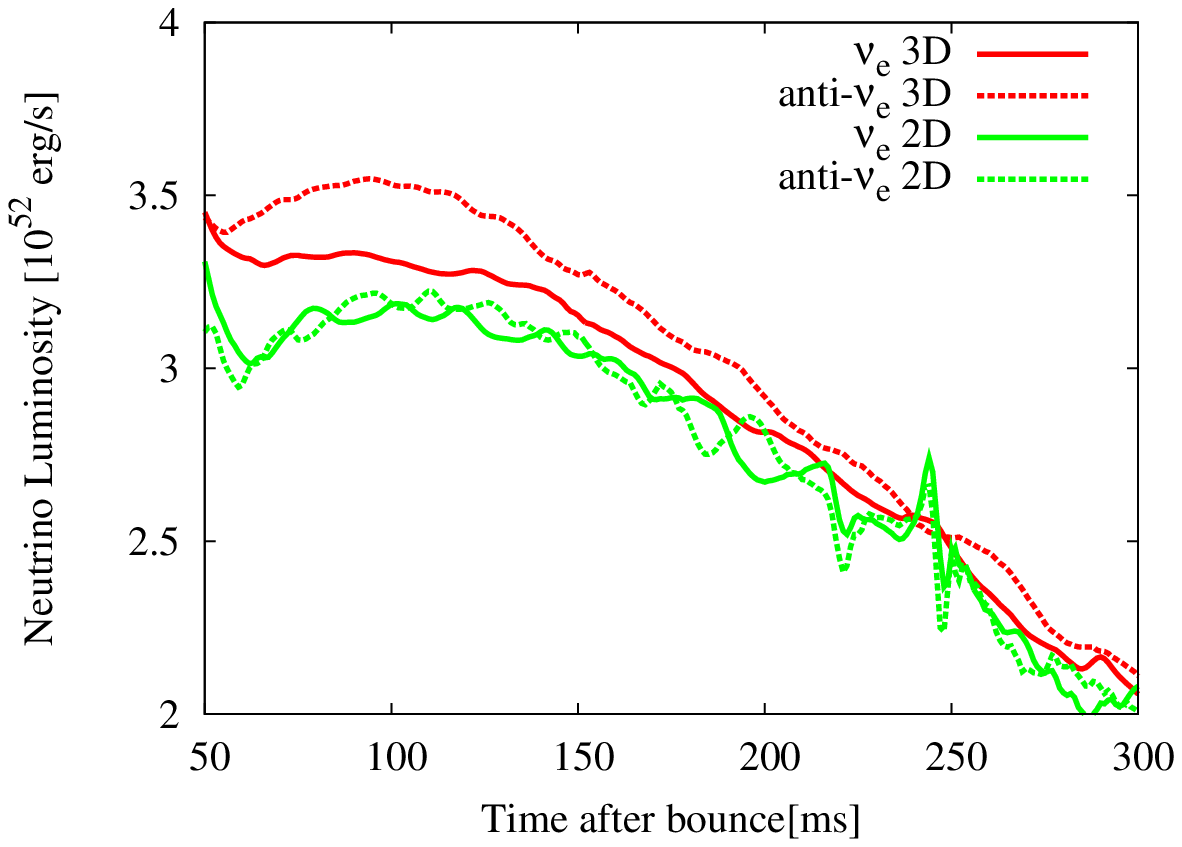}
    \includegraphics[width=.35\linewidth]{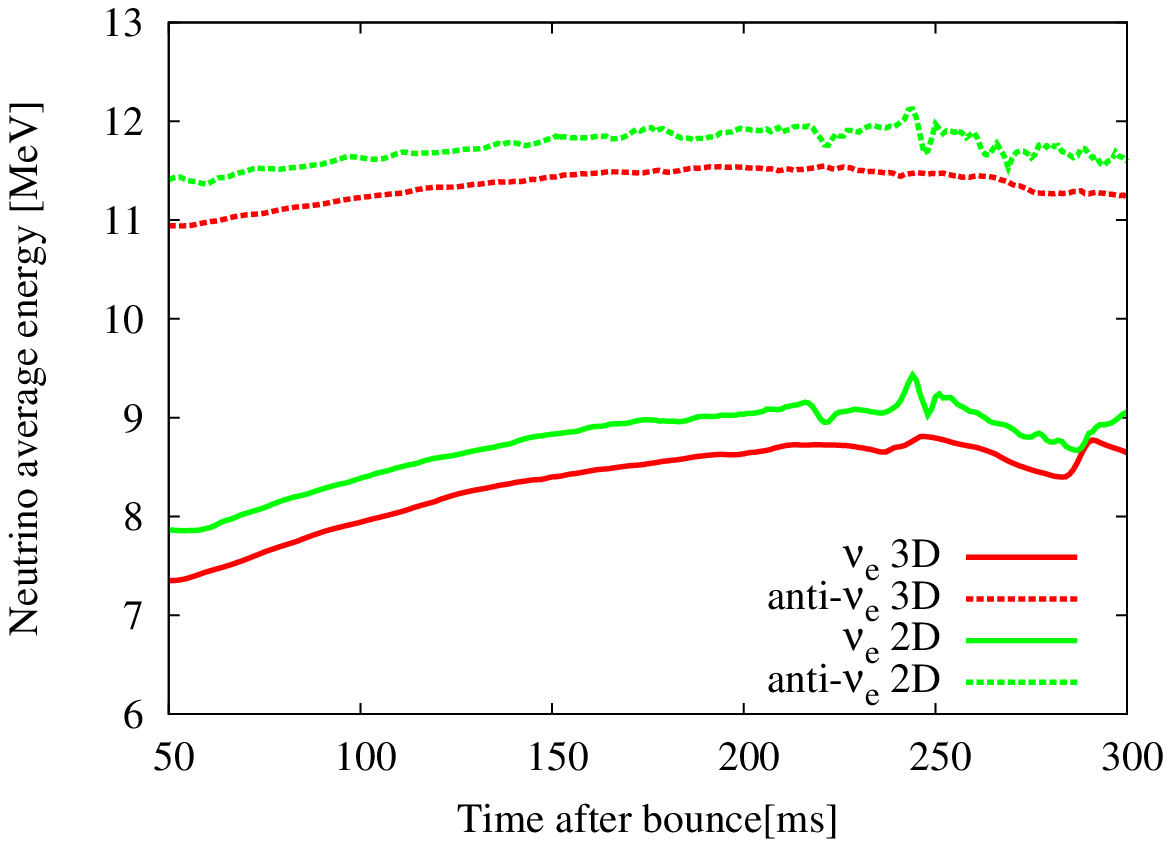}\\
 \caption{Same as Figure \ref{f11} but for luminosities (left) and 
 mean energies (right) of radiated electron ($\nu_e$) or 
 anti-electron ($\bar{\nu}_e$) type neutrinos. }
\label{f12}
\end{figure*}

\begin{figure*}[htbp]
    \centering
    \includegraphics[width=.35\linewidth]{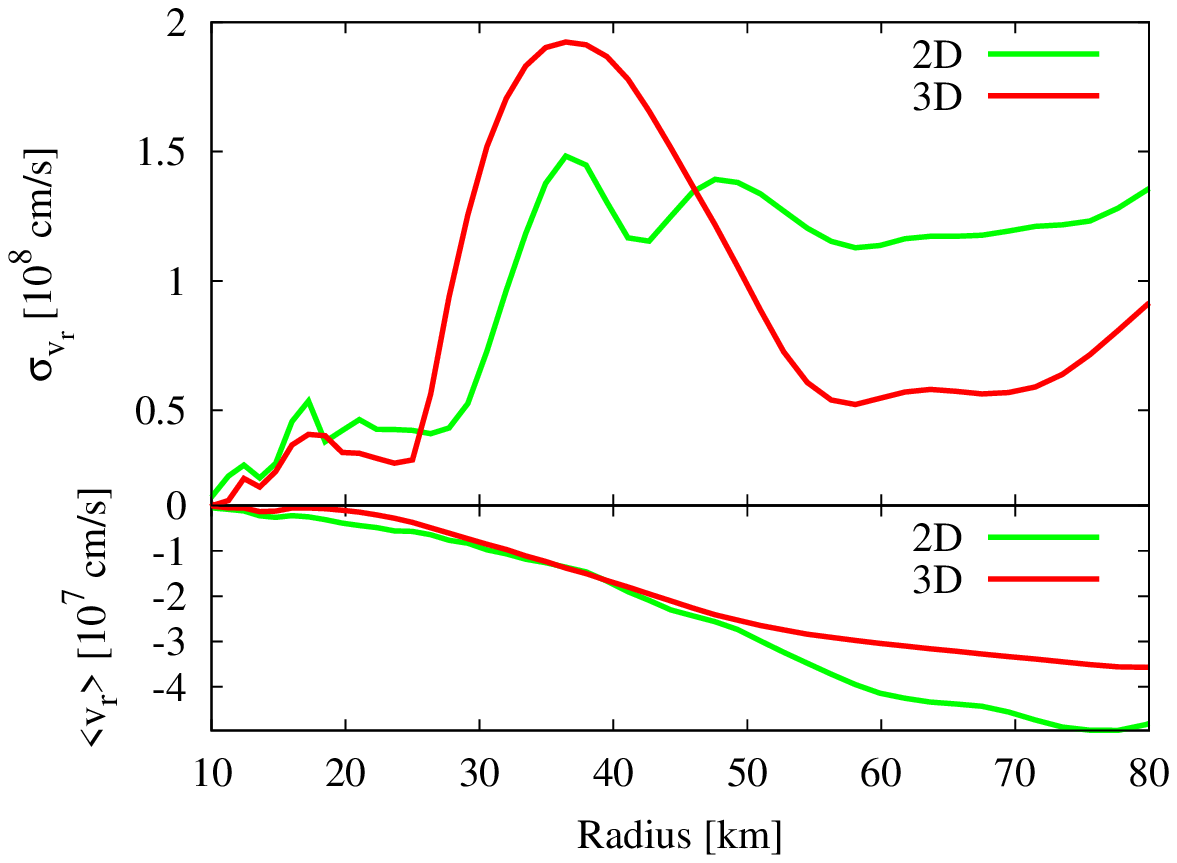}
    \includegraphics[width=.35\linewidth]{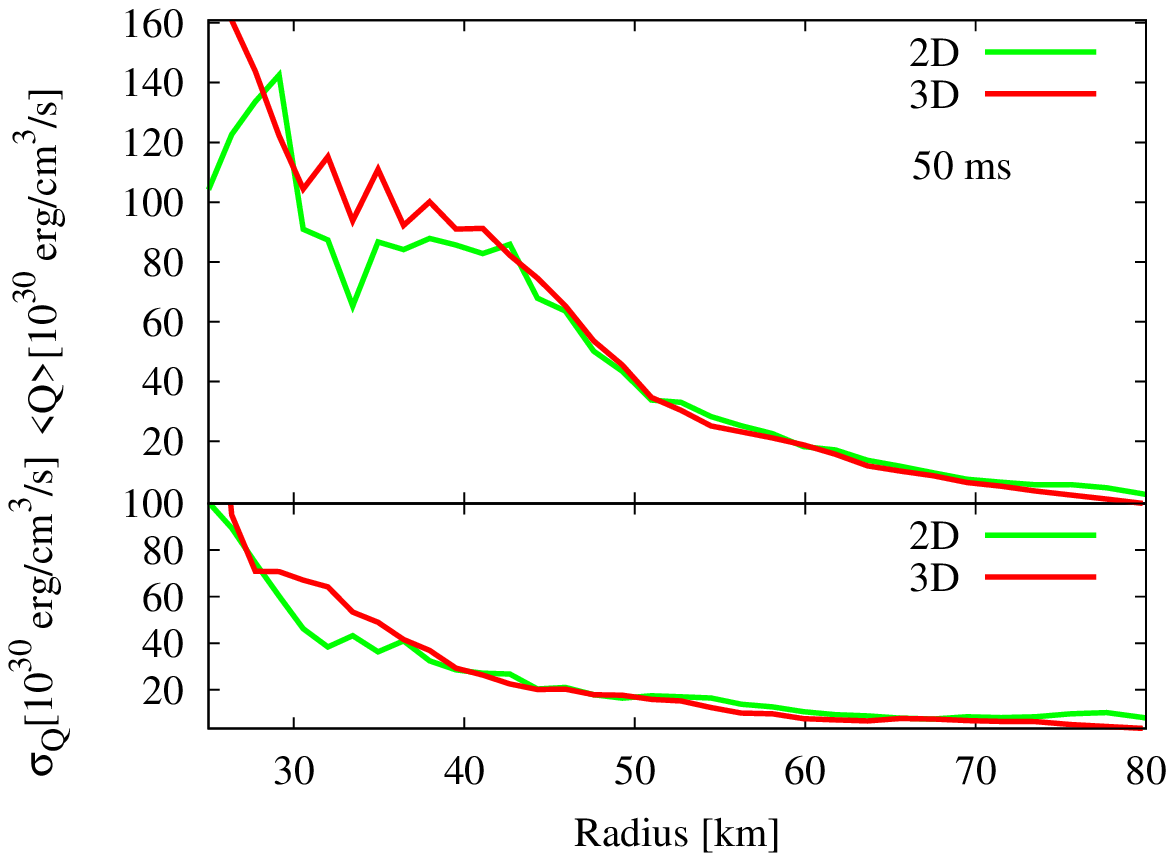}\\
    \includegraphics[width=.35\linewidth]{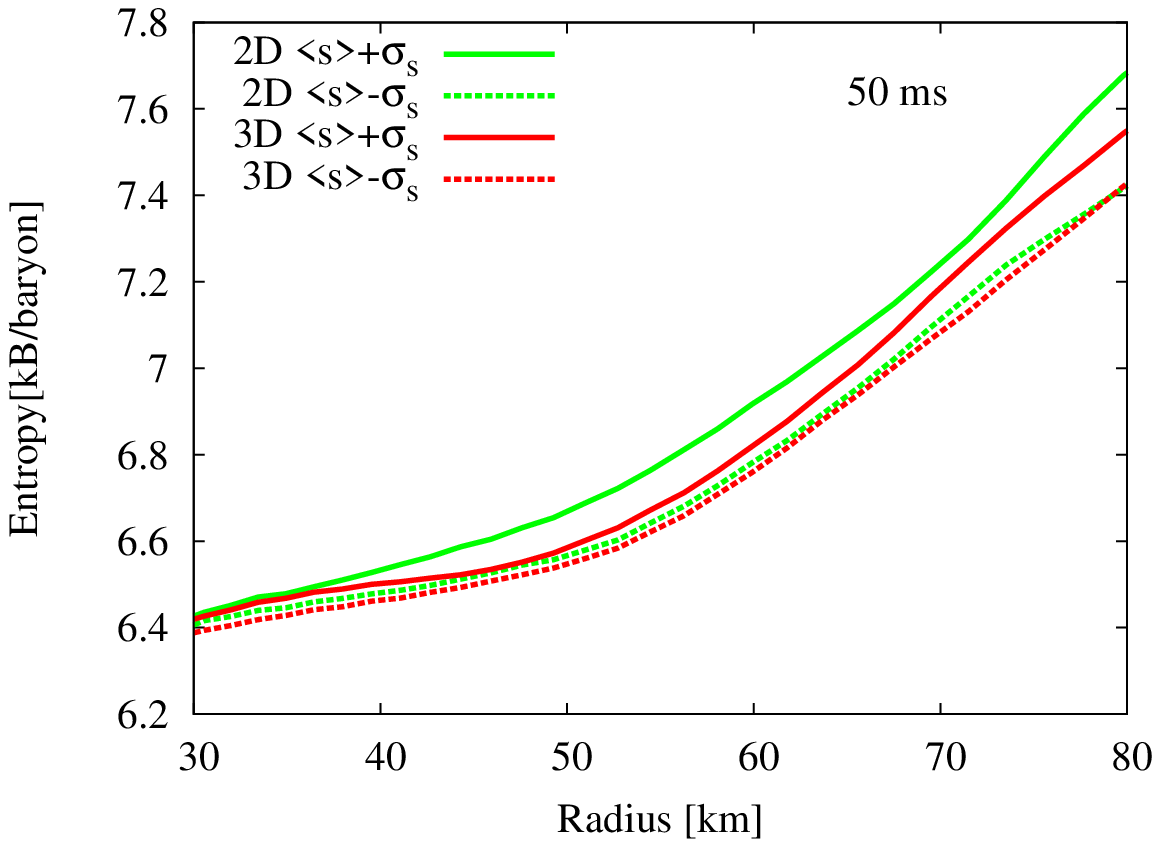}
 \caption{Important quantities for analyzing the properties of neutrino emission
 in multi-D models (see text for more details).
 Top left panel compares the velocity dispersion (top) and the average 
 radial velocity (bottom) between 2D and 3D model. 
Top right panel shows the neutrino cooling rate (top) and
 its dispersion (bottom). The bottom panel compares the profiles 
of maximum and minimum entropy, which is indicated by 
 ($\langle s \rangle +
 \sigma_s$) and ($\langle s \rangle - \sigma_s$), respectively.
 These panels are for $50$ ms after bounce.}
\label{f13}
\end{figure*}

\subsubsection{Neutrino-heating timescales in 2D and 3D}\label{334}

Now we compare the neutrino-heating timescales between our 2D and 3D models.
From the left panel of Figure \ref{f11}, the heating timescale is shown
 to be longer for 3D (red line). As seen from the right panel, this is because the 
 total net rate of the heating rate is generally smaller for our 3D model.
To understand this feature, we analyze the 
 neutrino luminosities ($L_{\nu}$) and mean energies ($\langle \epsilon_{\nu} \rangle$),
  since the neutrino heating rate can be symbolically expressed 
as $Q^{+}_{\nu} \propto L_{\nu}\langle \epsilon_{\nu}^2 \rangle$
 (e.g., Equation (23) in \citet{jank01}).

From the left panel of Figure \ref{f12}, the neutrino luminosities regardless
 of electron or anti-electron type are shown to be generally 
 larger for 3D than 2D. On the
 other hand, the mean neutrino energies are lower for 3D (right panel).
 Although the higher neutrino luminosity is advantageous for producing 
neutrino-driven explosions, the lower neutrino energies predominantly 
 make the heating rate smaller, thus leading to the longer heating timescale in 
our 3D model. 

The higher neutrino luminosity in 3D is due to the stronger convective 
 activities as discussed in section \ref{333}. The left panel of Figure \ref{f13} 
 compares the velocity dispersion (top panel) and the average radial velocity (bottom
 panel) between 2D and 3D. 
 Figure \ref{f13} is at 50 ms after bounce, when 
 the convection and SASI are actively in operation.

 The top left panel in Figure \ref{f13} shows that convective motions are much 
 more vigorous for 3D in a radius of $30- 50$ km 
 (see also the yellowish stripe 
 in Figure \ref{f10b}). The top right panel shows that the 
neutrino cooling rate (top) as well as its dispersion there ($\sigma_{Q}$, bottom panel)
 is larger for 3D than 2D.
 For 3D models computed in this work, these features are generally maintained before
 the revived shock expands further out (typically $\sim $ 100 ms after bounce).
 The bottom panel of Figure \ref{f13} shows that the entropy above $30$ km 
 is generally smaller for 3D (red line) than 2D.
 This is as a consequence of the weaker neutrino-heating in 3D than 2D.
 For the time snapshot in Figure \ref{f13} (at 50 ms after bounce),
 the position of the electron (energy-averaged) neutrinosphere is about 75 km. So 
 the convection deep below the neutrinosphere ($30- 50$ km in radius) 
is the agent to affect the neutrino luminosity and the mean neutrino energy. 
This trend is akin to the one observed in the 2D simulations by \citet{buras06}.

\begin{figure*}[htbp]
    \centering
    \includegraphics[width=.35\linewidth]{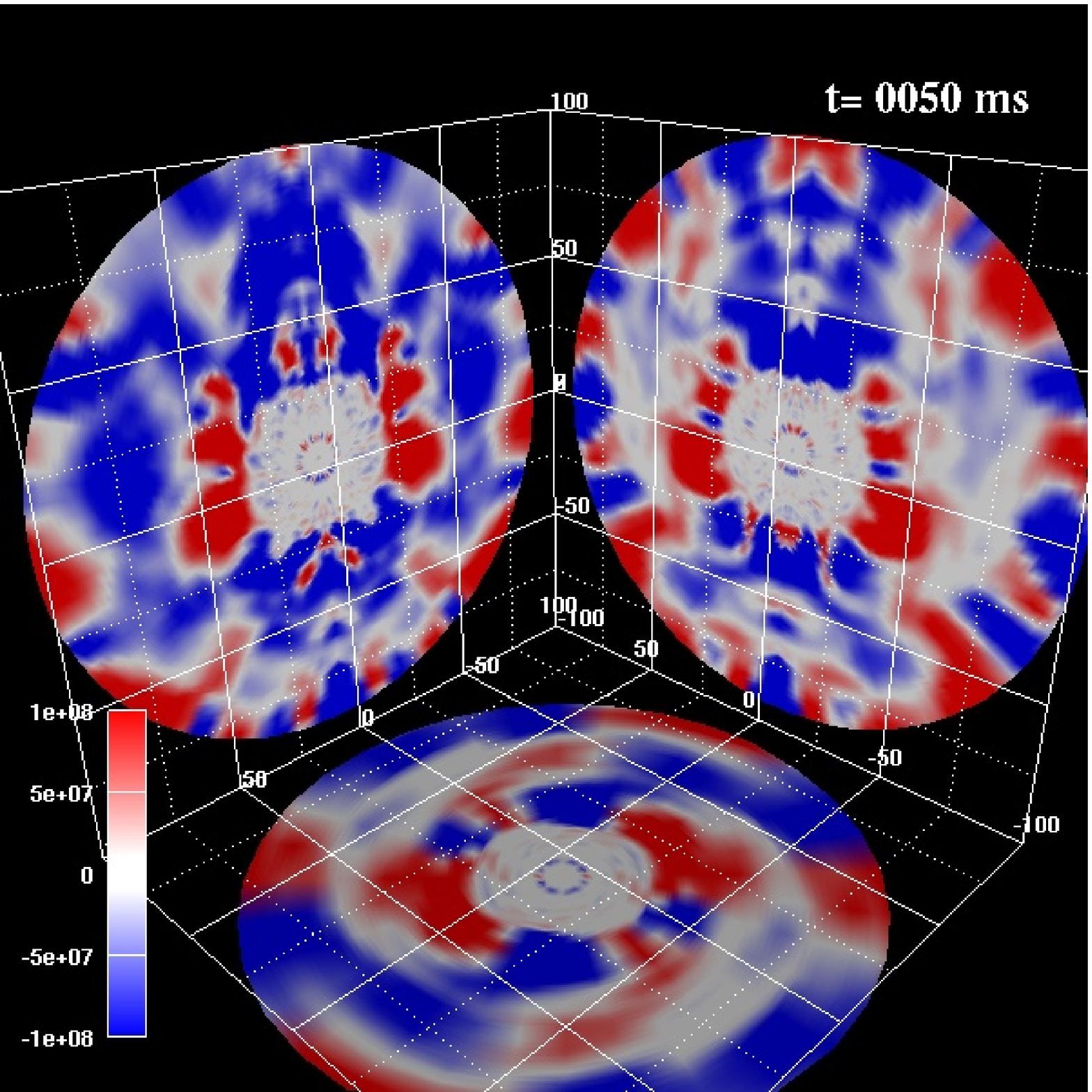}
    \includegraphics[width=.35\linewidth]{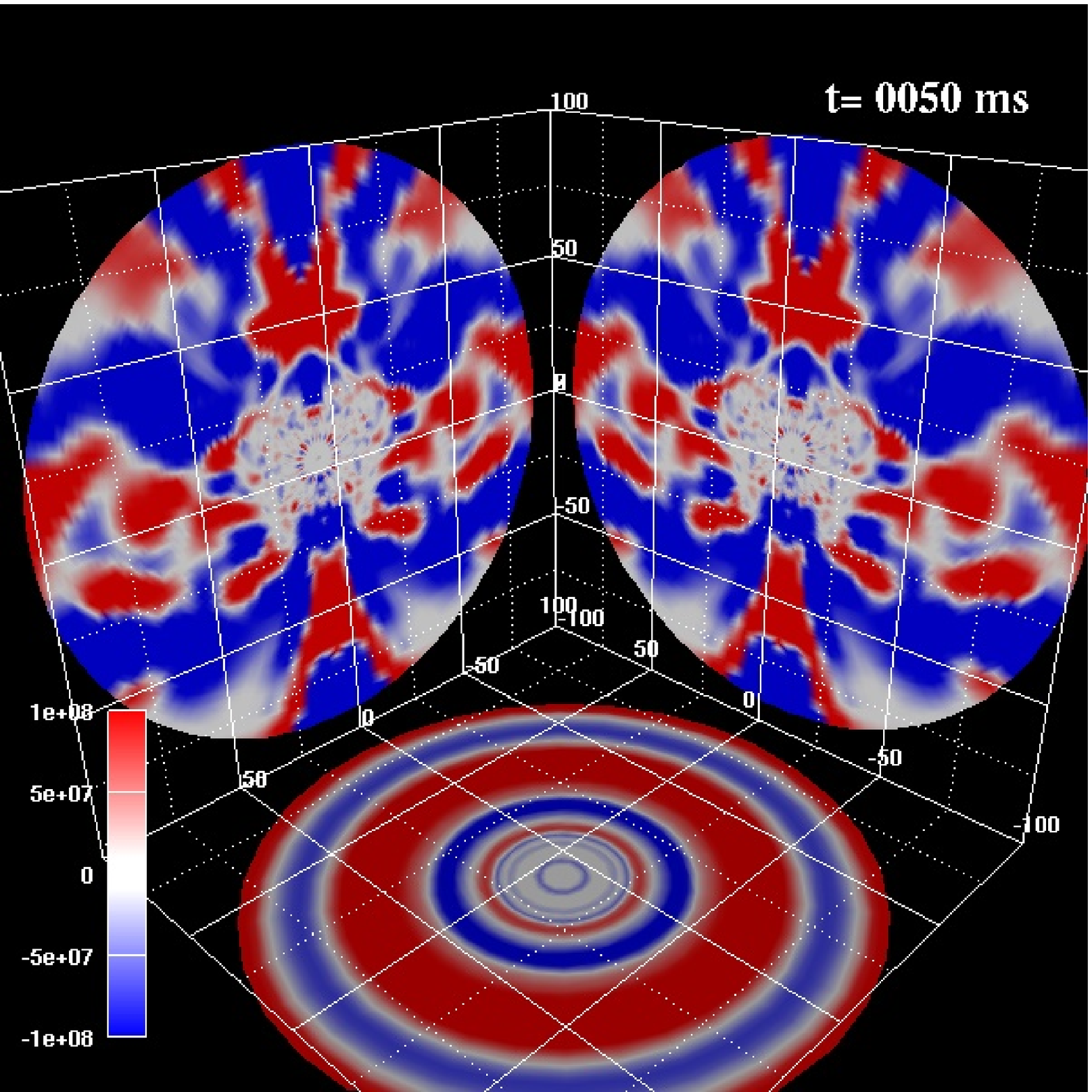}\\
    \includegraphics[width=.35\linewidth]{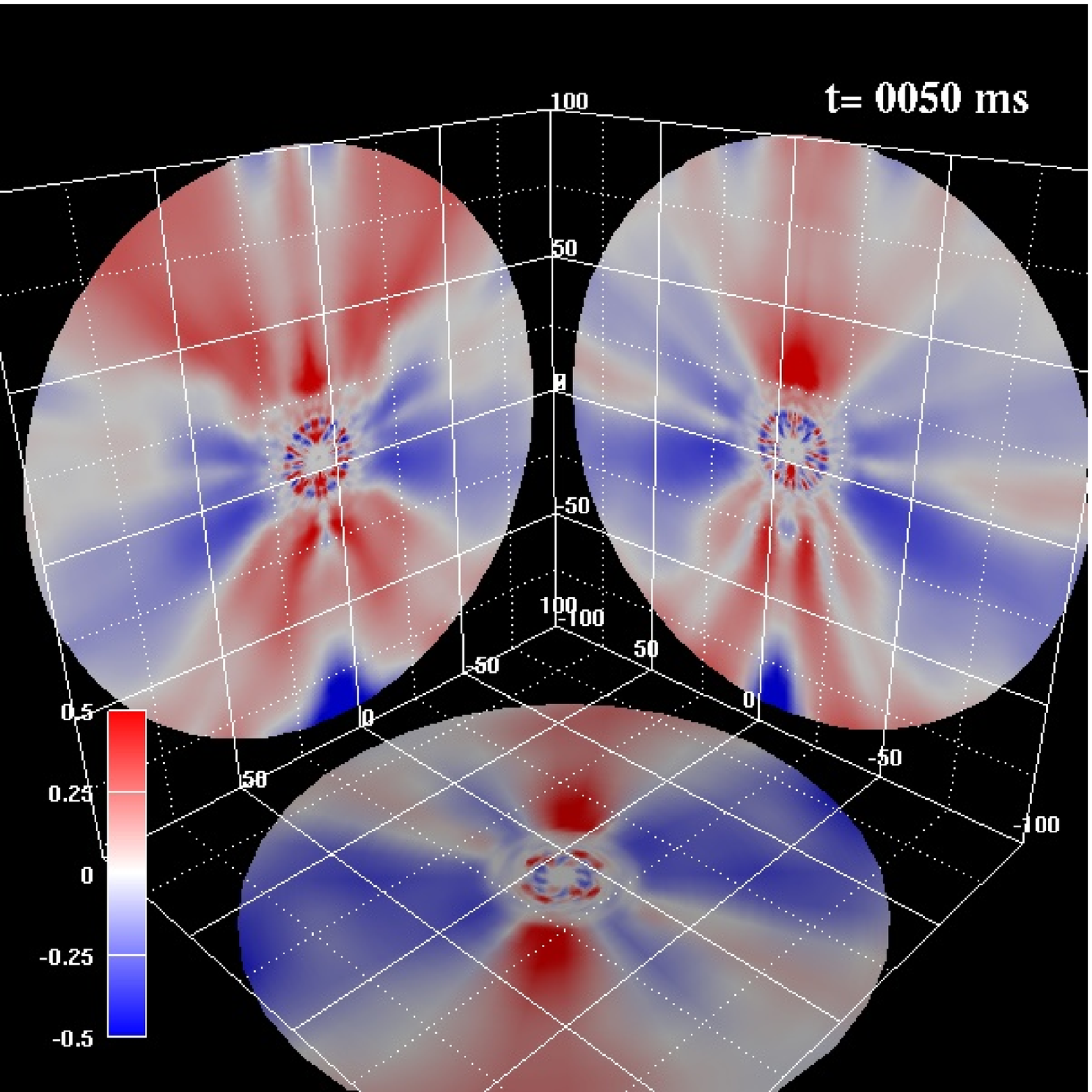}
    \includegraphics[width=.35\linewidth]{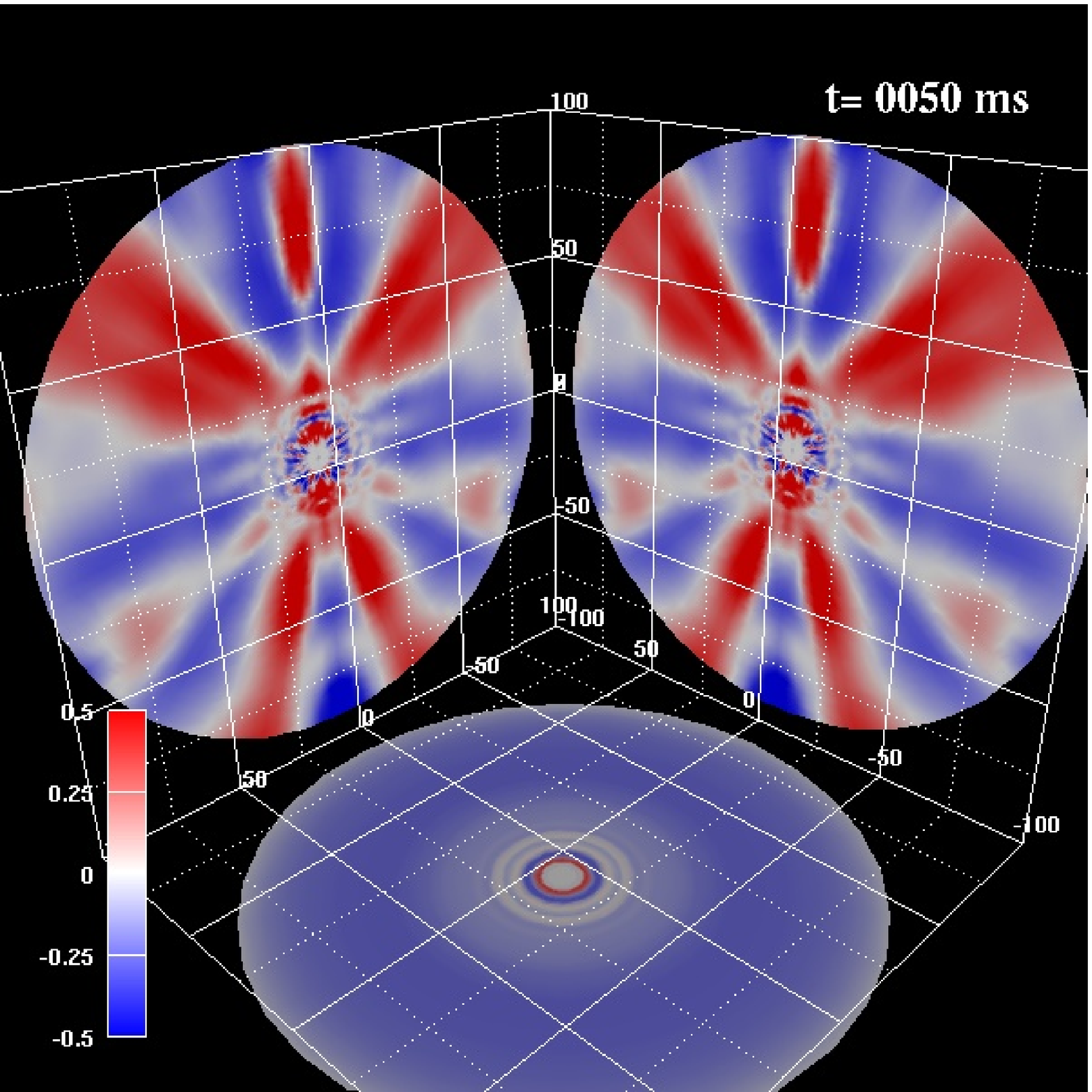}
 \caption{Analysis of flow patterns and matter mixing for our 3D (right) and 2D (left)
 model, respectively.
Top panels show radial velocities, in which bluish and reddish regions correspond 
to downflows and upflows that are distinguished from their local radial velocities 
(negative or positive). Similar to Figure 1, 
 the contours on the cross sections in the 
$x=0$ (back right), $y=0$ (back left), and $z=0$ (back bottom) planes are, 
respectively projected on the sidewalls of the graphs to visualize 
3D structures. Bottom panels show
 the relative angle variation of $Y_{\bar \nu_e}$ ($\delta Y_{\bar{\nu}}$,
 see text for definition). 
In the regions with downflows (bluish in the top panels), the sign of 
 $\delta Y_{\bar{\nu}}$ tends to be positive (colored by red in the bottom panels).
 All the panels are at 50 ms after bounce (same as Figure \ref{f13}).
  }
\label{f16}
\end{figure*}

In Figure \ref{f16}, we proceed to perform a more detailed analysis 
 on matter mixing behind the shock and its impact on the emergent neutrino luminosity.
 Top panels show radial velocities for our 3D (left) and 2D (right) model
 within a radius of 100 km, in which downflows 
and upflows are colored by blue and red, respectively. The central whitish regions
 correspond to the PNS, which is convectively stable (hence, with small radial
 velocities) due to the positive entropy 
 gradient (e.g., section \ref{333} and Figure \ref{f10b}). In the vicinity of 
 the PNS, downflows and upflows are visible near the equator and pole,
 respectively in the 3D model 
(e.g., back left and back right panels in Figure \ref{f16} (top left)).
 The bottom left panel is same as the top left panel but 
 for the normalized angle variation of $\delta Y_{\bar{\nu}}$.
  Here we define the normalized angle variation of quantity $A$ as 
\begin{equation}
\delta A = \left( A - \langle A \rangle \right)/\sigma_A.
\end{equation}
 We focus on anti-electron neutrino ($\bar{\nu}_e$), because the luminosity of 
 $\bar{\nu}_e$ dominates over that of ${\nu}_e$ during the simulation time (e.g.,
 left panel in Figure \ref{f12}). Comparing the top left to the bottom left panel 
 in Figure \ref{f16},
 it can be seen that the positive sign of $\delta Y_{\bar{\nu}}$ (reddish region in the 
 bottom left panel) tends to have a correlation with the downflows (bluish 
 region in the top left panel), which is 
 vice versa for the upflows. This is because material with 
larger $Y_{\bar{\nu}_e}$ in the outer layers is mixed down to the vicinity of the 
PNS that possesses smaller $Y_{\bar{\nu}_e}$ due to convective overturns. 
This can be a possible explanation of the correlation between 
 the gain (loss) in $Y_{\bar{\nu}_e}$ and the upflows (downflows) to the PNS. 
Note that this relation is also visible in our 2D model (right panels).

\begin{figure*}[htbp]
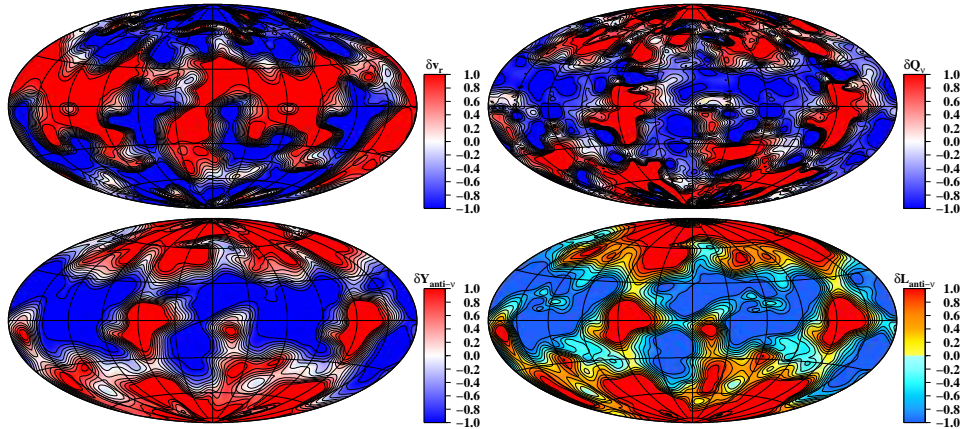

    \centering
    \includegraphics[width=.35\linewidth]{f17a.eps}
    \includegraphics[width=.35\linewidth]{f17b.eps}\\
    \includegraphics[width=.35\linewidth]{f17c.eps}
    \includegraphics[width=.35\linewidth]{f17d.eps}\\
 \caption{The 4$\pi$-maps of various quantities in Figure \ref{f16} 
visualized by the Mollweide projection at a radius 50 km. The top left, top right, and bottom left panel shows 
 the normalized angular variation of the radial velocity ($\delta v_r$), neutrino cooling 
rate ($\delta Q$), and $\bar{\nu}_e$ fraction ($\delta Y_{\bar{\nu}_e}$), respectively. 
The bottom right panel is the normalized angular variation of the (anti-electron)
 neutrino luminosity ($\delta L_{\bar{\nu}_e}$), which is measured at the outer
 boundary of the computational domain.
}
\label{f17}
\end{figure*}

 Figure \ref{f17} depicts angular variations of the flow patterns 
in Figure \ref{f16}. The Mollweide projection (or 4$\pi$-map) of various quantities 
 is taken at a radius of 50 km. From the top left panel ($\delta{v_r}$),
 downflows are shown to flow from the pole (colored by blue), while upflows 
are rather uniformly distributed in the equator (seen like a horizontal red belt).
 From the bottom left panel, the color pattern of blue and red reverses with that 
 of the top left panel. As already mentioned, this reflects the correlation between
 downflows (upflows) and gain (loss) in $Y_{\bar{\nu}_e}$. Reflecting the gain or 
 loss, the neutrino cooling rate ($\delta {Q_{\bar{\nu}_e}}$) has a positive correlation
 with $Y_{\bar{\nu}_e}$ (compare the top right and the bottom left panel). 
The variation in the neutrino luminosity that is measured at the outermost boundary 
of the computation domain (bottom right panel) has a rough positive correlation with 
the neutrino cooling rate (top right panel), which may agree with one's intuition.
 
\begin{figure}[htbp]
    \centering
    \includegraphics[width=.90\linewidth]{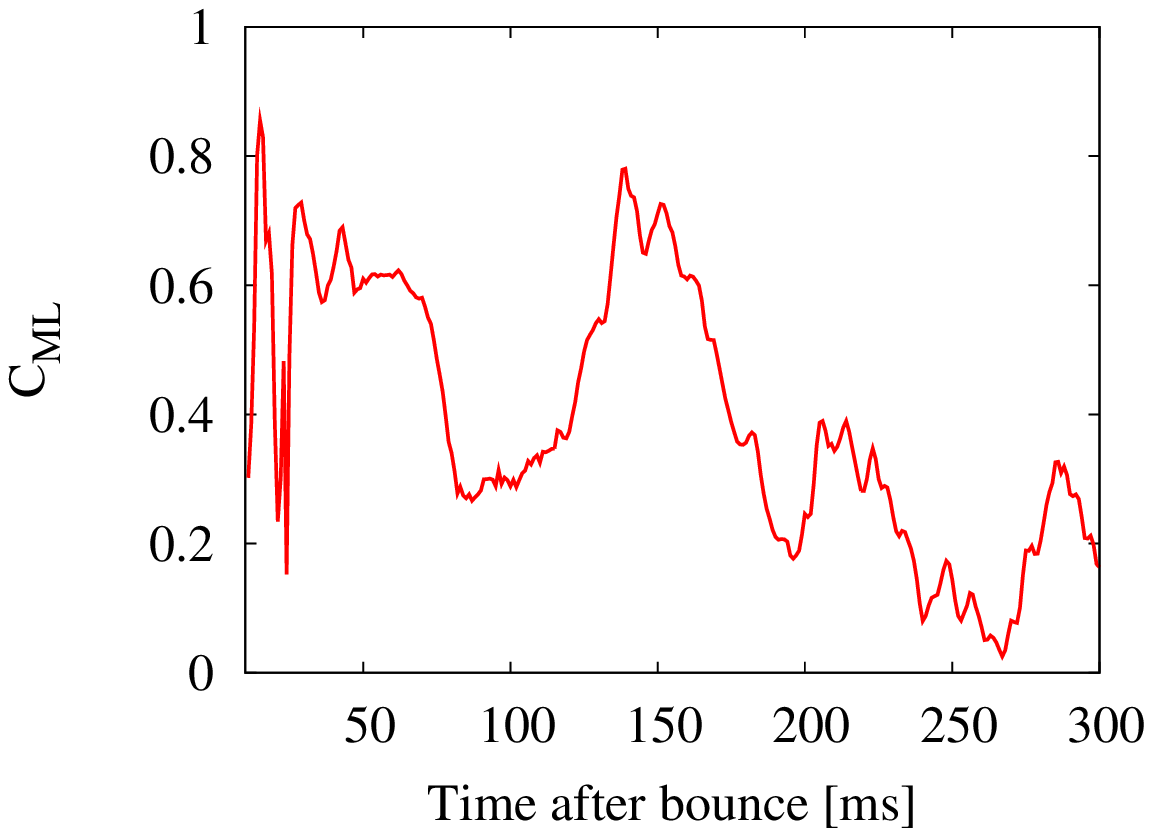}
 \caption{Time evolution of cross-correlation coefficient 
between the mass accretion rate ($\dot{M}$) to the PNS
 and the neutrino luminosity for our 3D model. The coefficient 
  between two variables 
of $A$ and $B$ is defined by $r_{A,B} = \frac{\int \mathrm{d}\Omega
  \left(A- \langle A\rangle \right)(B-\langle B\rangle))}{4\pi}/\left(\sigma_A\sigma_B\right)$.}
\label{f14}
\end{figure}

 Finally we show the cross correlation that we take 
 between the time evolution of the mass accretion rate to the PNS
 and the neutrino luminosity (Figure \ref{f14}).  
As seen, a positive correlation is commonly seen during the simulation time.
 This plot may carry a 
 message that it is important to go beyond the light-bulb scheme, in which 
the input neutrino luminosity is usually kept constant with time 
(e.g., \citet{iwakami1,iwakami2,nordhaus}). To take 
 into account the feedback between the mass accretion and the neutrino luminosity,
 the spectral IDSA scheme, which is beyond the grey transport scheme (e.g., 
 \citet{frye02,fryer04a}), sounds quite 
efficient in the first-generation 3D simulations.

As suggested in the right panel 
 of Figure \ref{f7}, 3D explosions are more easily obtained 
 for models with finer numerical resolutions\footnote{To say something
 very solid on this respect, we apparently 
need to study the effects of numerical resolutions more 
 in a systematic manner. But the result we reported here is sort of
 best what we can do now, which it took more than 4 CPU months by keeping
 the currently available supercomputing facilities at our hand running.}. Our results would indicate
 whether the advantages for driving explosions mentioned above 
 could or could not overwhelm
 the disadvantages should be tested by the next generation 3D simulations 
 with much more higher numerical resolutions. Needless to say, the 3D results
(not to mention 2D results) should depend on the sophistication of the employed 
neutrino transport scheme. Regarding the gravity, we should first go over the monopole 
approximation. This may not be so easy task from a technical point of view,
 because we need to implement a multigrid approach to obtain a high scalability 
in the MPI computing. To go beyond the Newtonian gravity is also a challenging task
 \citep{mueller}. 
 Our 3D results are only the very first step towards a more realistic 3D supernova 
modeling.

\section{Summary and Discussion}
We have presented numerical results on 3D hydrodynamic core-collapse simulations of 
an $11.2 M_{\odot}$ star. By comparing our 1D and 2D results, we have studied
 how the increasing spatial
 multi-dimensionality affects the postbounce supernova dynamics. 
 The calculations were performed with an energy-dependent treatment of 
the neutrino transport based on the isotropic diffusion source approximation scheme.
In agreement with 
 previous study, our 1D model does not produce explosions for 
 the 11.2 $M_{\odot}$ star, while the neutrino-driven 
revival of the stalled bounce shock is obtained both in 2D and 3D models.
 We showed that the SASI does develop in the 3D models, 
however, their saturation amplitudes are generally smaller than 2D. 
By performing a tracer-particle analysis, we showed that 
 the maximum residency time of material 
in the gain region is shown to be longer for 3D due to non-axisymmetric flow motions
 than 2D,
 which is one of advantageous aspects in 3D to obtain neutrino-driven explosions.
 Our results showed that convective matter motions below the gain radius become 
much more violent in 3D than 2D, making the neutrino luminosity larger
 for 3D. Nevertheless the emitted neutrino 
 energies are made smaller due to the enhanced cooling. Our results indicated 
whether these advantages for driving 3D explosions  
could or could not overwhelm
 the disadvantages is sensitive to the employed numerical resolutions.
 An encouraging finding was that the shock expansion tends to become 
 more energetic for models with finer resolutions.
To draw a robust conclusion, 3D simulations 
 with much more higher numerical resolutions and also with more advanced treatment of 
 neutrino transport as well as of gravity is needed.

 Finally we refer to the approximations adopted in this paper.
 As already mentioned, the omission of heavy lepton neutrinos, 
the inelastic neutrino scattering, and the ray-by-ray approach should be 
improved. The former two, should act to suppress the explosion.  
The ray-by-ray approach may lead to the overestimation of the 
directional dependence of the neutrino anisotropies 
(see discussions in \citet{marek}). Although it would be highly computationally 
 expensive, the full-angle transport will 
give us the correct answer (e.g., \citet{ott_multi,brandt}).
Our numerical grid in the azimuthal direction is only 32 to cover 360 degrees.
 Such a low resolution could lead to a large numerical viscosity.
 The numerical viscosity is expected to be large especially in the vicinity of
 the standing accretion shock, which may affect the growth of the SASI. It could also 
 affect the growth of the turbulence in the postshock convectively active regions,
 which is very important to determine the success or failure of the neutrino-driven 
 mechanism. To clearly see these effects of numerical viscosity, 
we need to conduct a convergence test in which a numerical gridding is changed 
in a parametric way (e.g. \citet{hanke11}).

 A number of exciting issues are remained to be studied in our 3D results, 
  such as gravitational-wave signatures (e.g., \citet{kotake_ray,kotake11,ewald11}), 
 neutrino emission and its detectability (e.g., \citet{kistler}), possibility of 3D SASI flows 
generating pulsar kicks and spins \citep{annop}.
 The dependence of progenitors (e.g., \citet{buras06,burrows2}) and 
equations of state (e.g., \citet{marek}) are important to be clarified in 
  3D computations.
 We are going to study these items one by one in the near future. 

 As of July 2011, the $K$ supercompter in Kobe city of Japan 
is ranked as the top on the ``TOP 500 list of 
World's Supercompters''\footnote{http://www.top500.org/}. From early next year,
 we are fortunately allowed to start using the facility for our
 3D supernova simulations. Keeping our efforts to overcome the caveats mentioned 
above, we plan to improve the numerical resolutions as 
 much as possible in the forthcoming run,  
  by which we hopefully gain a new insight into the long-veiled 
 explosion mechanism.

\acknowledgements{
We full-heartedly thank M. Liebend\"orfer for stimulating
 discussions and helpful exchanges in implementing the IDSA scheme. We are also 
 thankful to K. Sato and S. Yamada for continuing encouragements. 
Numerical computations were carried on in part on XT4 and 
general common use computer system at the center for Computational Astrophysics, CfCA, 
the National Astronomical Observatory of Japan.  This 
study was supported in part by the Grants-in-Aid for the Scientific Research 
from the Ministry of Education, Science and Culture of Japan (Nos. 19540309, 20740150,
  and 23540323) and by HPCI Strategic Program of Japanese MEXT.}

\bibliographystyle{apj} 
\bibliography{ms}
\end{document}